\documentclass[twocolumn]{aastex63}
\pdfoutput=1
\usepackage{amsmath, amstext}
\usepackage{mathrsfs}
\usepackage{xfrac}
\usepackage[]{xcolor}

\usepackage[T1]{fontenc}
\usepackage{apjfonts}
\usepackage[figure, figure*]{hypcap}
\usepackage{xspace}
\usepackage{ulem}
\usepackage{enumitem}
\setlist{leftmargin=*}

\newcommand{\msun}{\ensuremath{{\rm M_\odot}}\xspace}

\newcommand{\ak}{\ensuremath{A_{\rm K}}\xspace}
\newcommand{\av}{\ensuremath{A_{\rm V}}\xspace}

\newcommand{\w}[1]{\ensuremath{{\mathrm{W}[{#1}]}}\xspace}
\newcommand{\x}[1]{\ensuremath{{\mathrm{X}[{#1}]}}\xspace}
\newcommand{\N}[1]{\ensuremath{{\mathrm{N}[{#1}]}}\xspace}
\newcommand{\E}[1]{\ensuremath{\times10^{#1}}\xspace}

\newcommand{\htwo}{\ensuremath{{\rm H}_2}\xspace}
\newcommand{\nht}{\ensuremath{\N{{\rm H}_2}}\xspace}

\newcommand{\dgr}[1]{\ensuremath{\delta_{\rm dgr\ifx&#1&\else\,#1\fi}}\xspace}
\newcommand{\deltagas}[1]{\ensuremath{\delta_{\rm gas\ifx&#1&\else\,#1\fi}}\xspace}
\newcommand{\deltamol}[1]{\ensuremath{\delta_{\rm mol\ifx&#1&\else\,#1\fi}}\xspace}

\newcommand{\xco}{\ensuremath{X_{\mathrm{CO}}}\xspace}
\newcommand{\wco}{\ensuremath{W_{\mathrm{CO}}}\xspace}
\newcommand{\nco}{\ensuremath{{N_{\mathrm{CO}}}}\xspace}
\newcommand{\tex}{\ensuremath{T_{\rm ex}}\xspace}
\newcommand{\tdust}{\ensuremath{T_{\rm dust}}\xspace}
\newcommand{\fdep}{\ensuremath{{\rm \eta}_{\,\text{CO}}}}

\newcommand{\magn}{\ensuremath{\xspace\mathrm{mag}\xspace}}
\newcommand{\wunit}{\ensuremath{\xspace\mathrm{K\ km/s}\xspace}}
\newcommand{\xfunit}{\ensuremath{{\rm cm}^{-2}\ {(\wunit)}^{-1}}\xspace}
\newcommand{\msunpc}{\ensuremath{\msun~{\rm pc}^{-2}}\xspace}
\newcommand{\dgunit}[1]{\ensuremath{\mathrm{\ cm^{-2}\, \magn^{-1}}\xspace}}
\newcommand{\mdgunit}[1]{\ensuremath{\mathrm{\ g\, cm^{-2}\, \magn^{-1}}\xspace}}

\newcommand{\twelveco}{\ensuremath{{}^{12}\text{CO}}\xspace}
\newcommand{\thirteenco}{\ensuremath{{}^{13}\text{CO}}\xspace}
\newcommand{\ceighteeno}{\ensuremath{\text{C}{}^{18}\text{O}}\xspace}
\newcommand{\cotwo}{\ensuremath{\mathrm{CO}_2}\xspace}
\newcommand{\abund}[1]{\ensuremath{[#1]}\xspace}
\let\e\E

\newcommand{\herschel}{{\sl Herschel}\xspace}
\newcommand{\lkhalpha}{LkH$\alpha$ 101\xspace}

\newcounter{paranum}
\newcommand{\parnum}{{\color{blue}\stepcounter{paranum}\theparanum}\ ---\ }
\renewcommand{\parnum }{\stepcounter{paranum}}

\defcitealias{2015ApJ...805...58K}{K15}

\submitjournal{ApJ}
\accepted{October 21, 2020}
\shorttitle{The Cold Depths of California}
\shortauthors{Lewis et al.}

\begin{document}

\title{Probing the Cold Deep Depths of the California Molecular Cloud: \\ The Icy Relationship between CO and Dust}
\author[0000-0001-5199-3522]{John Arban Lewis}
\affiliation{Center for Astrophysics $\vert$ Harvard \& Smithsonian, 60 Garden St, MS 72, Cambridge, MA 02138, USA}
\author{Charles J. Lada}
\affiliation{Center for Astrophysics $\vert$ Harvard \& Smithsonian, 60 Garden St, MS 72, Cambridge, MA 02138, USA}
\author{John Bieging}
\affiliation{Steward Observatory, The University of Arizona, Tucson, AZ 85721}
\author{Anoush Kazarians}
\affiliation{California State Polytechnic University Pomona, 3801 W. Temple Ave., Pomona CA 91768}
\author{Jo\~ao Alves}
\affiliation{University of Vienna, Department of Astrophysics, T\"urkenschanzstrasse 17, 1180 Wien, Austria}
\author{Marco Lombardi}
\affiliation{Department of Physics, University of Milan, via Celoria 16, 20133, Milan, Italy}

\begin{abstract}
We study the relationship between molecular gas and dust in the California Molecular Cloud over an unprecedented dynamic range of cloud depth (\av = 3 -- 60 magnitudes). We compare deep \herschel-based measurements of dust extinction with observations of the \twelveco, \thirteenco, and \ceighteeno J=2-1 lines on sub-parsec scales across the cloud. We directly measure the ratio of CO integrated intensity to dust extinction to derive the CO X-factor at over 10$^5$ independent locations in the cloud. Confirming an earlier study, we find that no single $^{12}$CO X-factor can characterize the molecular gas in the cold ( $\tdust\leq20$) regions of the cloud that account for most of its mass. We are able to derive a single-valued X-factor for all three CO isotopologues in the warm ( \tdust>25 K ) material that is spatially coincident with an \ion{H}{2} region surrounding the star \lkhalpha. We derive the LTE CO column densities for \thirteenco and \ceighteeno since we find both lines are relatively optically thin. In the warm cloud material CO is completely in the gas phase and we are able to recover the total \thirteenco and \ceighteeno abundances. Using CO abundances and deep \herschel observations, we measure lower bounds to the freeze-out of CO onto dust across the whole cloud finding some regions having CO depleted by a factor of $>20$. We construct the first maps of depletion that span the extent of a giant molecular cloud. Using these maps we identify 75 depletion-defined cores and discuss their physical nature.

\end{abstract}


\section{Introduction}\label{sec:intro}

\parnum The bulk of the star forming gas is contained in cold giant molecular clouds (GMCs) made primarily of molecular hydrogen ($\rm H_2$). $\rm H_2$ does not emit at the temperatures in GMCs (T$\sim$10 K). The most commonly used proxy for tracing $\rm H_2$ is carbon monoxide (CO). Therefore, the nature of the relationship between the two molecular species is of fundamental importance for molecular cloud studies. This relationship is usually expressed by the CO X-factor,  $\xco = {\rm \nht}/{\rm \wco}$, the conversion factor between CO integrated intensities and H$_2$ column densities. This empirically derived factor is closely related to the abundance of CO and is very useful in calibrating mass determinations of clouds observed in CO. Although often assumed to be universal, measurements of the X-factor in the Galaxy exhibit a significant (factor of 2) variation between clouds and orders of magnitude variations have recently been found within clouds, presumably due to environmentally dependent variations in CO chemistry (e.g., \citet{2014ApJ...784...80L, 2015ApJ...805...58K}).
Moreover, X-factor measurements are inferred to be a function of gas metallicity \citep{2013ARA&A..51..207B}. Usually measurements of the X-factor in individual GMCs are obtained for the \twelveco J=1-0 transition in the low column density regions of the clouds  (i.e., A$_V$ $\lesssim$ 4-5 mag) where \wco %
grows linearly with extinction, which itself tracks the total hydrogen column density in a cloud \citep{2018MNRAS.474.4672L}.
More recently advances in observational capabilities that enable deep extinction measurements of clouds have led to studies that are beginning to extend our knowledge of the relationship between CO and H$_2$ to deeper and deeper cloud layers \citep{2006A&A...454..781L,2013MNRAS.431.1296R,2015ApJ...805...58K}.

\parnum Combining sensitive K-band extinction measurements with CO observations, \citet[hereafter, K15]{2015ApJ...805...58K} were able to extend measurements of the X-factor over a large portion of the California Molecular Cloud (CMC) to cloud depths of A$_V$ $\sim$ 35 magnitudes, a factor of 2 deeper than the previous very deep surveys of the Perseus cloud, Pipe Nebula and Orion A \& B clouds (\citet{2008ApJ...679..481P}, \citet{2009A&A...493..735L}, \citet{2013MNRAS.431.1296R}).
Performing measurements on sub-parsec scales \citetalias{2015ApJ...805...58K} found that no single X-factor could characterize all the gas in the cloud and hypothesized that this was the result of the severe depletion of CO onto dust grains over most of the cloud surface.
However they were able to measure a unique value of the X-factor in hot (\tex > 18 K), undepleted,  molecular gas that is coincident with the only  \ion{H}{2} region in the cloud, NGC 1579. This \ion{H}{2} region is located in L1482 a dark cloud in the southern portion of the CMC \citep{1962ApJS....7....1L, 2009ApJ...703...52L}. \citetalias{2015ApJ...805...58K} found that the X-factor varied with \tex and was influenced by the star forming environment. This variation with \tex was also reflected in abundance variations throughout the region with the abundances decreasing with increasing \av in the colder gas away from the \ion{H}{2} region. Though unable to perform a detailed comparison with the dust temperature, the authors concluded that the abundance variations were due to desorption/depletion processes.

\parnum  Here we significantly extend the work of \citetalias{2015ApJ...805...58K} in three ways. First, we employ Herschel observations to provide dust column densities over an unprecedented dynamic range (A$_V$ $=$ 3 - 60 magnitudes) of cloud depth (a factor of 2 deeper than \citetalias{2015ApJ...805...58K} with increased fidelity). Second, we extended the area of the cloud surveyed in CO by a factor of 3 to cover larger regions with less active star formation \citep{2017A&A...606A.100L}. Third, we use the Herschel maps of dust temperature to directly test the hypothesis that depletion is responsible for the observed variations in the X-factor and CO abundances throughout the cloud. We use the deeper extinction and dust temperature maps, and our more expansive CO observations to derive column densities and measure abundances for \thirteenco and \ceighteeno in order to probe variations in the relationship between CO and molecular hydrogen and more quantitatively evaluate the role of depletion has in producing these variations.

 \parnum Our paper is organized as follows. In \S\ref{sec:data} we present our CO survey and briefly discuss the \herschel dust extinction and temperature maps. In \S\ref{sec:datanalysis} we describe the creation of our moment and column density maps. In \S\ref{sec:results} we discuss what we find in examining the \wco vs. \av and \nco vs. \av relations and present our measurement of the \thirteenco and \ceighteeno abundances in the CMC. In \S\ref{sec:depletion} we map the CO depletion in California and use it as a novel method for detecting cold cores. Finally we summarize our findings in \S\ref{sec:summary}.

\section{Observations and Data Reduction}\label{sec:data}
\begin{figure*}[ht]
    \plotone{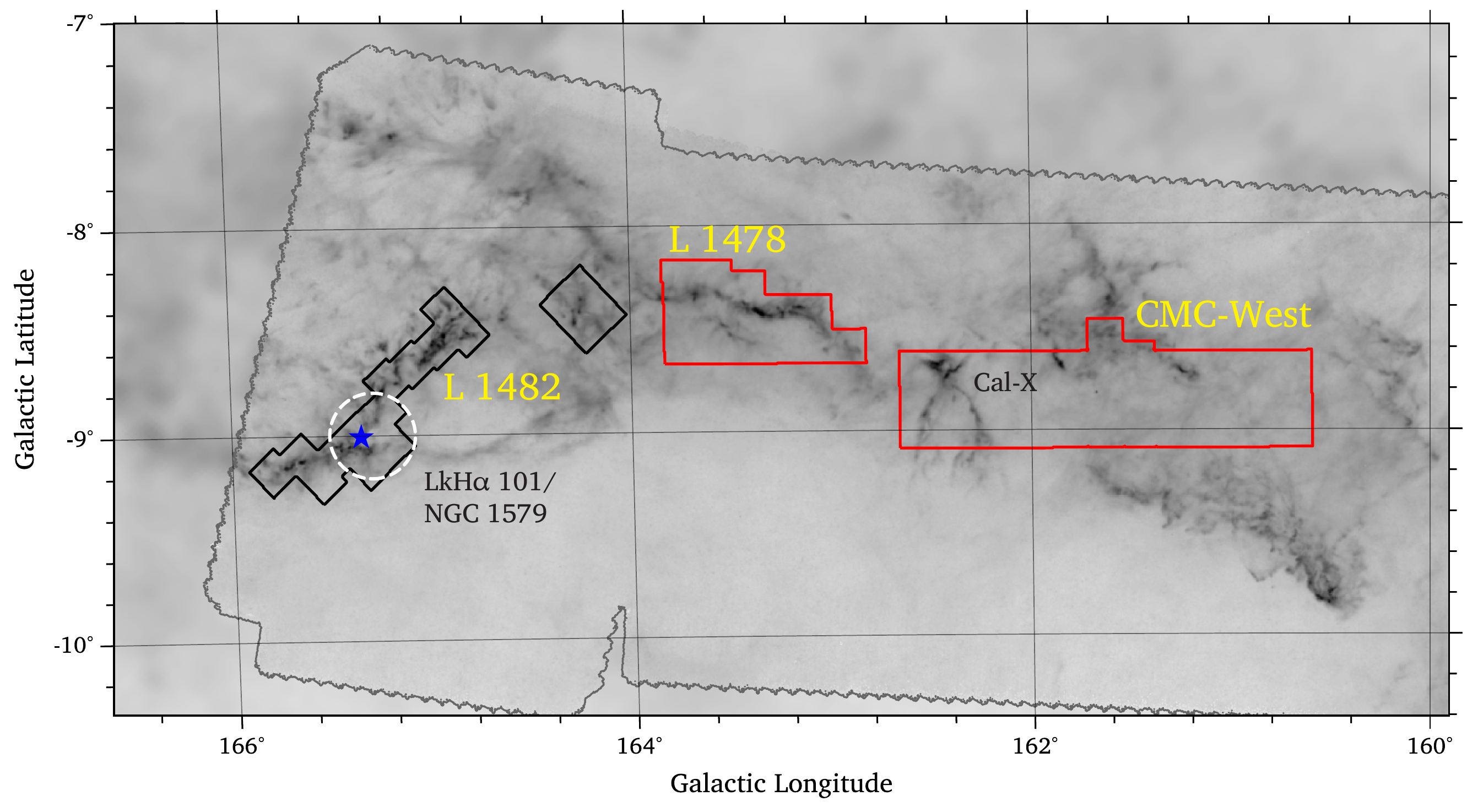}
    \caption{CO survey outlines over \herschel+{\sl Planck} dust maps. The CO survey boundary is outlined in black and red, with the black outlined covering L1482, the southeastern extent of the cloud, being initially presented in \citetalias{2015ApJ...805...58K} and the red outlines indicating the new data presented in this paper (L1478, and CMC-West). \twelveco and \thirteenco were observed in all 3 regions, while \ceighteeno was not observed in CMC-West. The jagged black boarder marks the extent of the \herschel dust data \citep{2017A&A...606A.100L}, with {\sl Planck} data used to fill in outside that. The \herschel data continues further west about $1\deg$. than shown on the map. The CO survey boarder was chosen to cover the majority of the \av$>$3 mag dust. Two subregions of interest are shown. The dashed white circle shows the extent of the \ion{H}{2} region (NGC 1579) associated with the star (shown in blue) \lkhalpha. In CMC-West we show the Cal-X feature described in \citet{2017ApJ...840..119I}. }
    \label{fig:survey}
\end{figure*}

\parnum For this paper we use observations obtained with the Arizona Radio Observatory (ARO) 10 m Heinrich-Hertz Submillimeter Telescope (SMT) and the {\sl Herschel Space Observatory}.

\subsection{\herschel dust opacity and temperature maps}\label{sec:herschel}
\parnum We make use of the \herschel dust opacity and temperature maps  from \cite{2017A&A...606A.100L}. The CMC was observed in the "Auriga-California on \herschel\ program" \citep{2013ApJ...764..133H}, with the PACS and SPIRE instruments. PACS 160 \micron\ and SPIRE 250, 350, and 500 \micron\ maps were convolved to the resolution of the 500 \micron\ map ($\rm FWHM_{500\micron}=36\arcsec$). We briefly describe the data and the method we used to derive the \htwo column density and dust temperature.
In \citet{2017A&A...606A.100L} dust opacity ($\tau_{850}$) and dust temperature ($\tdust$) were derived by fitting the \herschel spectral energy distribution (SED) for each pixel with a modified blackbody,
\begin{equation*}
F_\nu =\tau_{850} \left(\frac{\nu}{\nu_0}\right)^{\beta_d}	\frac{2 h \nu^3}{c^2}\frac{1}{{\rm e}^{h\nu/k\tdust}-1}
\end{equation*}
where, $\nu_0 = 353\ \text{GHz}$, the dust $\beta_d$ is fixed locally from {\sl Planck} maps and $\tau_{850}$ (the opacity at 850 microns) and $\tdust$ are free parameters. Here \tdust\ is an {\it effective dust temperature} since it averages over the temperature profile along the line-of-sight.

\parnum The conversion from $\tau_{850}$ to \htwo\ column density (\nht) is done by comparing the \herschel\ $\tau_{850}$ maps to extinction maps of the CMC derived using the NICEST method \citep{2009A&A...493..735L}. They found a linear relation between $\tau_{850}$ and \ak\ for this cloud,
\begin{equation}
\ak = \gamma\, \tau_{850} + \delta,
\end{equation}
where $\gamma=3593\rm\ mag$,
 which is similar to the value measured previously in the Orion B cloud \citep{2014A&A...566A..45L} and Perseus \citep{2016A&A...587A.106Z}. We convert from K-band extinction to V-band extinctions to facilitate comparisons to previous work. \ak\ is converted to \nht assuming an extinction law $\ak/\av = 0.11$ and a dust-to-gas ratio 
 $\nht/\av = 9.4 \times 10^{20}\, \text{cm}^{-2}\, \text{mag}^{-1}$  \citep{1978ApJ...224..132B,2002ApJ...577..221R}.
The final maps are degraded to $38\arcsec$ resolution for comparison to the CO data. The corresponds to 0.08 pc at distance to the CMC (450 pc, \citet{2009ApJ...703...52L}). The total (atomic + molecular) mass is found by integrating the total gas surface density ( $\Sigma_{gas}$) over the area of the cloud,
$ {\rm M}_{tot} = \int \Sigma_{gas}\, {\rm d}S ,$ where $\Sigma_{gas}/\ak =\ 183\, {\rm M}_\odot\, {\rm pc}^{-2}\, {\rm mag}^{-1}$ \citep{2006A&A...454..781L,1979ARA&A..17...73S}.
The total mass contained in the \herschel\ map is
 $5.52\E4\ {\rm M_\odot}$, \citep{2017A&A...606A.100L}. Approximately 20\% of the total cloud mass, $M = \rm 1.07\E4 M_\odot$, is covered by our CO survey, which is discussed below.

\subsection{Carbon Monoxide Data}\label{sec:CO}
\parnum We observed the $J=2-1$ transition of \twelveco (230.528 GHz), \thirteenco (220.339 GHz) and \ceighteeno (219.560 GHz) with the {\sl Heinrich Hertz Submillimeter Telescope} (SMT) using a prototype ALMA band 6 dual-polarization sideband separating receiver in combination with the 0.25 MHz -- 256 channel filterbank as the backend (0.25 MHz  $\sim0.32\rm\ km/s$ at 230 GHz). We used two observing setups - (1) with \twelveco in the upper-sideband and \thirteenco in the lower-sideband and (2) another with \twelveco in the upper-sideband and \ceighteeno in the lower-sideband. This improves the signal-to-noise of the \twelveco data.
CMC was observed over multiple days during the November 2012 - April 2013 observing season. Those segments are shown as outlines overlayed on \textsl{Planck}+\herschel dust map in Fig. \ref{fig:survey}. CMC West was not observed in \ceighteeno.
The survey boundaries were designed to cover regions with $\av\gtrsim 3\ {\rm mag}$ in the \cite{2009ApJ...703...52L} extinction maps.

\parnum We observed in 'on-the-fly' (OTF) mode \citep{2007A&A...474..679M} - raster scanning $10\arcmin\times10\arcmin$ tiles at a rate of 10\arcsec\ $\rm s^{-1}$, sampling spectra every 0.1 s.
The raw OTF data were put onto a 10\arcsec grid using custom data reduction scripts in the GILDAS CLASS software package \citep{2018ssdd.confE..11P} and then had a linear baseline subtracted. Those reduced data were then exported as tiles into MIRIAD, where adjacent tiles were combined into a single map, with overlapping regions being combined using the rms weighted average. The final three contiguous regions are shown outlined in Fig. \ref{fig:survey}. From east-to-west they are L1482, L1478, and the West cloud.

\begin{figure*}[!ht]
    \centering
    \includegraphics[width=\textwidth]{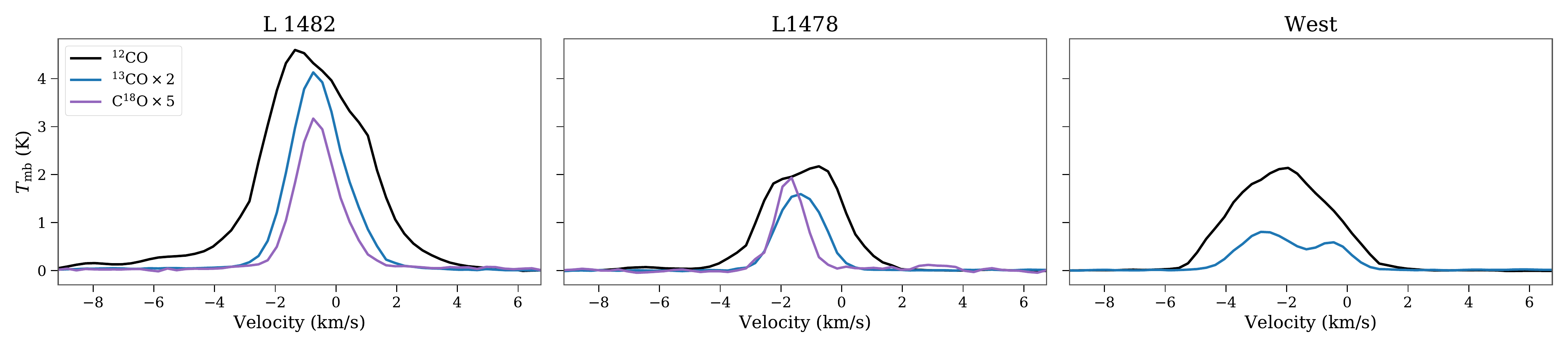}
    \caption{Average \twelveco,\thirteenco, and \ceighteeno spectra for each region of the California molecular cloud for pixels with SNR>3 for \twelveco and \thirteenco, and SNR>5 for \ceighteeno.}
    \label{fig:allspec}
\end{figure*}

\parnum We calibrated the intensity using periodic observations of the CO bright source W3(OH) (more details on the calibration method can be found in \citet{2010ApJS..191..232B}). Since the beam and velocity resolution for each CO isotopologue are all slightly different, the maps were convolved to the resolution of the \twelveco\ map (38\arcsec)
and regridded in velocity to 0.3 km/s channels to facilitate comparison. The average RMS noise per 0.3 km/s channel we achieve is 0.11 K for \ceighteeno, 0.13 K for \thirteenco, and 0.11 K for \twelveco.
The average spectra for each region are shown in Fig. \ref{fig:allspec}. The reduced maps are publicly available on the Harvard Dataverse \citep{DVN/FTOHSO_2020}\footnote{\url{https://doi.org/10.7910/DVN/FTOHSO}}.%

\section{Data Analysis}\label{sec:datanalysis}

\subsection{CO Moment Maps}\label{sec:momentmaps}

\parnum  We derived the line parameters (integrated intensity, central velocity, and velocity dispersion) using moment analysis:
$M_0 = \sum T_{\rm mb}\ {\rm \Delta v}$, $M_1 = \frac{\sum T_{\rm mb}\, {\rm v}\ {\rm \Delta v}}{M_0}$, $ M_2 = \frac{\sum T_{\rm mb}\, ({\rm v -}M_1)^2\ {\rm \Delta v}}{M_0},$ where $\rm \Delta v$ is the channel width. $M_0$ is the integrated intensity (\wco), $M_1$ is the intensity weighted velocity centroid, and $\sqrt{M_2}$ corresponds to an intensity weighted velocity dispersion $\sigma_{\rm v}$. Moments are sensitive to noise in the spectrum, so it is important to carefully select the channels over which the integrals are performed. We use custom velocity windows for every spatial location - that is for every spectrum - in our map. Windows are created by locating the channels which contain emission in a smoothed version of the data cube. The window width and position in velocity are chosen based on the line width and line central velocity and is optimized to capture all the line emission. This method is a large improvement over using a single velocity window per isotopologue per region in the cloud. Data outside the velocity windows are masked, and moments are calculated on the masked data cube.

\parnum Moments calculated using these customized narrow windows return line parameters that are consistent with the values returned from a gaussian fit. As a test, we fit single gaussians to a set of \thirteenco lines that were characterized by a single velocity component. We found the average difference between the velocity dispersion derived from the moment analysis and the one derived from gaussian fitting was $(\sqrt{M_2}-\sigma_{gauss}) \sim -0.03 \pm 0.07$. Similar levels of agreement are seen for \twelveco and \ceighteeno. In general, our lines are not simple gaussians, especially for \twelveco emission, so gaussian fitting is not an appropriate method for deriving line parameters over the whole cloud. The results in this paper rely primarily on the $0^{\rm th}$-moment or integrated intensity so, in any event, our findings are quite insensitive to the details of setting the integration window. For our results, we only use pixel where the signal-to-noise is >3 for \twelveco and \thirteenco and >5 for \ceighteeno.

\parnum  Maps of the integrated intensity for all three lines are shown in Appendix \ref{sec:appwco} for L1482, L1478, and CMC-West, the portions of the CMC that were surveyed.

\subsection{Column Density}\label{sec:seccolumndensity}
\parnum We derived column densities following the method used in \citetalias{2015ApJ...805...58K}, which we describe in Appendix \ref{sec:methnco}. We differ from their calculation by calculating the partition function numerically to the J=100 term rather than using an approximate form. The noise in our data limits the minimum column density we can detect to $\gtrsim 10^{14}{\rm\ cm^{-2}}$ for both isotopologues.

\section{Results and Discussion}\label{sec:results}

\subsection{The \wco -- \av relation and \xco }\label{sec:wcoav}

\begin{figure}[tbp]
\centering
{\includegraphics[width=\columnwidth]{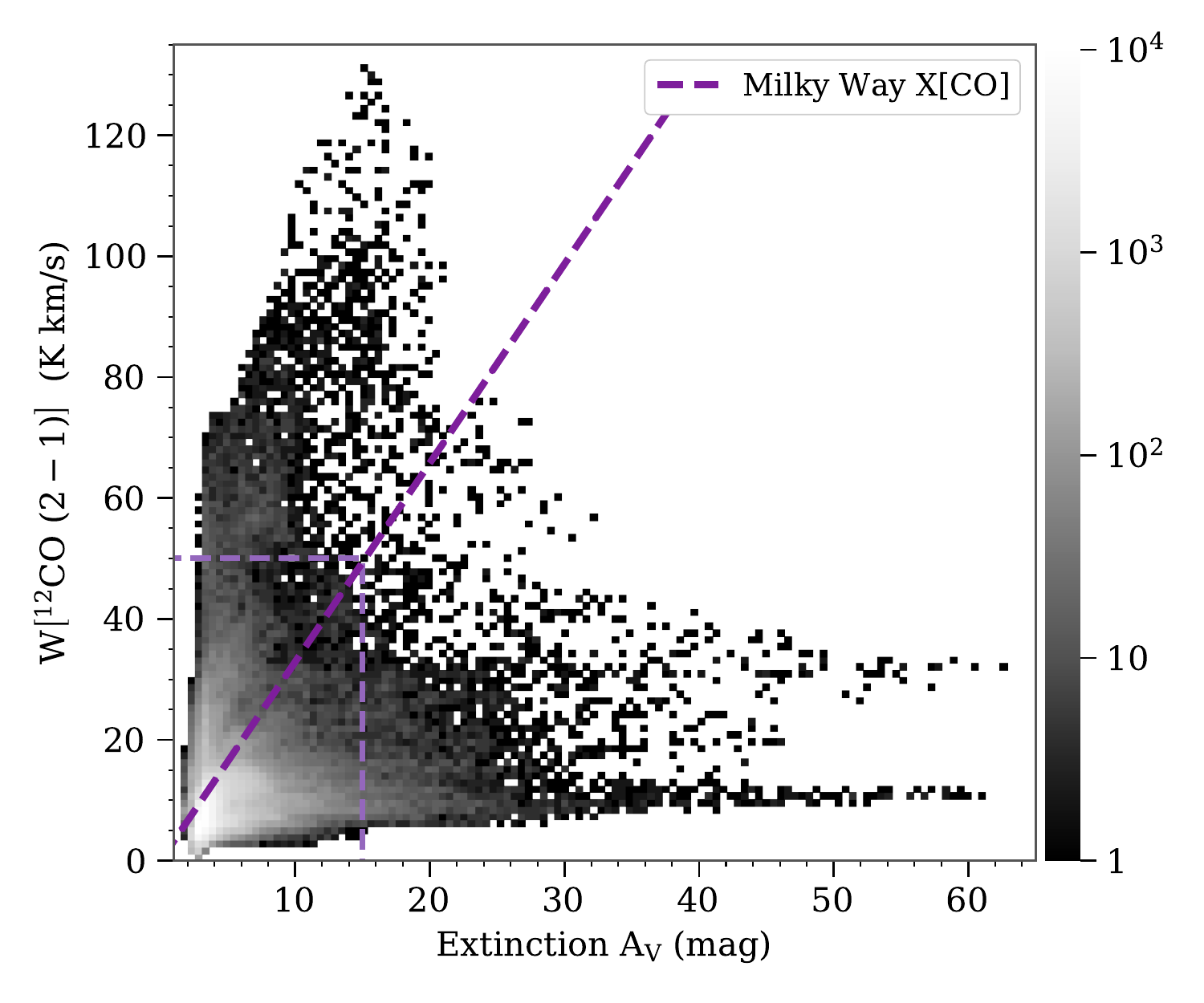}}
\caption{ \label{fig:wcoavall} Binned 2D histogram for \w{\twelveco\ (2-1)} vs \av for the entire California cloud where \w{\twelveco} is detected at greater than 3$\sigma$ ($\gtrsim 1 \text{K}$). The { gray dashed line} is the standard relation for the adopted MW X-factor ($2\times 10^{20}$\xfunit) which is 3.29 \wunit $\rm\ mag^{-1}$. The blue histogram shows the number of pixels in a binned pixel of the plot on a log scale. The black dashed box shows the region containing the bulk of data range covered by previous studies \citep[e.g.,][]{2008ApJ...679..481P,2013MNRAS.431.1296R}. }
\end{figure}

\begin{figure}[tbp]
\includegraphics[width=\columnwidth]{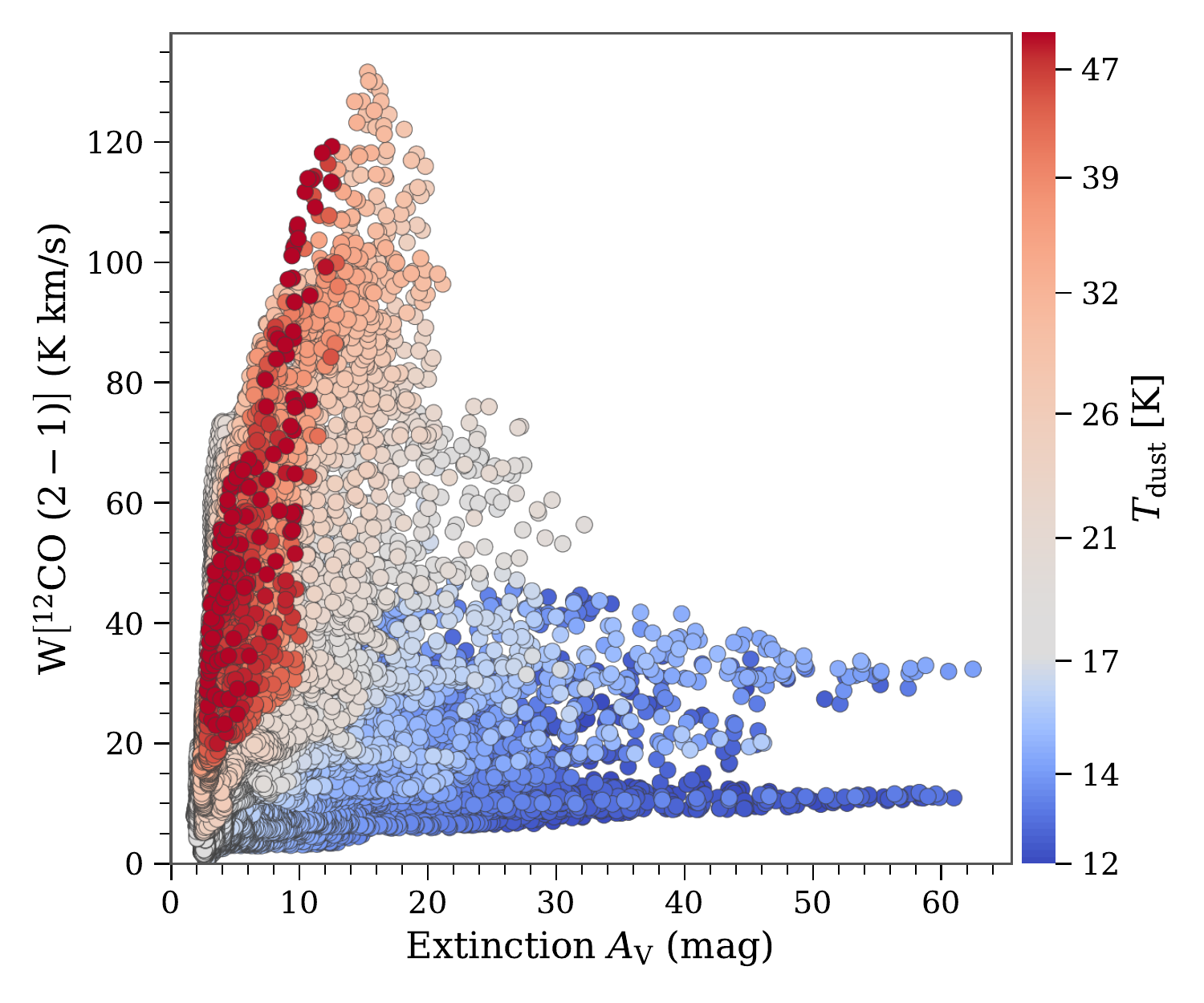}
\caption{  \label{fig:wcoavtemp}
\w{\twelveco(2-1)} vs \av for the entire California cloud where \w{\twelveco} is detected at greater than 3$\sigma$ ($\gtrsim 1\, \text{K}$). The points are colored by their \herschel-derived dust temperature ($\tdust$). The points are plotted so that hotter points are plotted over cooler points.}
\end{figure}

\parnum  On a plot of \wco vs \av, a single valued  X-factor is represented by a straight line through the origin. The \mbox{2-D} distribution of \w{\twelveco\ (2-1)} vs \av for the entire CMC is shown in Fig. \ref{fig:wcoavall}. We cover a large range in extinction (\av = 1.8 - 62.4 {\rm mag}) and \twelveco integrated intensity (\wco=.7 - 132 \wunit). The black box shows the region of parameter space containing the bulk of the data range covered by previous studies of this relation \citep[e.g.,][]{2008ApJ...679..481P,2013MNRAS.431.1296R}. The relation corresponding to the canonical Milky Way X-factor $\x{\twelveco\ (2-1)}_{\rm,MW} =2.9\times10^{20}\ \xfunit$ is  plotted for comparison. We converted the nominal $J=1-0$ CO X-factor to an X-factor for the $J=2-1$ transition by using \w{\twelveco\,(2-1)}/\w{\twelveco(1-0)} = 0.7 \citep{1995ApJS..100..125S,2010PASJ...62.1277Y}.
The larger dynamic range enabled by use of the Herschel-derived extinction measurements and the increased area of the CMC included in the present survey bring the overall behavior of the W[CO]-A$_V$ relation into sharper focus.

\parnum Confirming the earlier results of \citetalias{2015ApJ...805...58K}, it is clear from Fig. \ref{fig:wcoavall} that the CMC data are not well described by a single X-factor, i.e. there is no single, clear linear relationship present between \twelveco and \av.
Typically, if there is a constant CO abundance in a molecular cloud, a linear rise of W[$^{12}$CO] with A$_V$ is expected at low extinctions  with a break at around A$_V$ $\approx$ 4-5 magnitudes  transitioning to a flat relation at high (i.e., A$_V$ $>$ 10 mag) extinction where the $^{12}$CO emission is expected to be saturated due to high line opacity and no longer sensitive to increasing dust/hydrogen column density. \citep[e.g.;][]{1994ApJ...429..694L,2006A&A...454..781L,2008ApJ...679..481P,2013MNRAS.431.1296R,2018MNRAS.474.4672L,2016MNRAS.456.3596G}.
In the CMC the relation is considerably more complex.
The plot is characterized by three branches. One branch is at low extinction and appears nearly vertical, possibly indicative of a quasi-linear relation between \wco and \av in that part of the diagram. At high extinctions the relation is flat and nearly horizontal, as might be expected if CO emission is saturated. However, it is separated into (at least) two distinct branches that  correspond to two different levels of constant W[$^{12}$CO], one at ~10 K km/s and another at ~35 K km/s.

\parnum Important clues concerning the nature of the \wco-\av relation in the CMC are provided by Fig. \ref{fig:wcoavtemp}. Here we plot the W[\twelveco]--\av relation with the individual data points colored by  \herschel \tdust, ranging from 11.8 K - 74.4 K, plotted so cooler points are beneath hotter ones. In these plots, blue corresponds to cold dust (\tdust<18 K) and red to  hotter dust (\tdust>18 K). The branches are clearly temperature dependent, with the 10 \wunit\  branch associated with the coldest dust, the 35 \wunit\  branch associated with slightly warmer dust, and the vertical branch with dust over $\sim 25\rm\ K$. The cold branches extend to low extinctions beneath the points corresponding to the hot dust in red; however 98\% of the cold pixels are below 40 \wunit.

\parnum Further inspection of the data shows that these branches actually correspond to different regions of the cloud. The nearly vertical branch corresponds to CO associated with hot (\tdust>~25 K) dust located in the vicinity of, and coincident with, the \ion{H}{2} region in L1482. This is the region studied by \citetalias{2015ApJ...805...58K} who found similar behavior in the W[CO]-Av relations with CO excitation temperature, in particular, \citetalias{2015ApJ...805...58K} showed that the CO emission in the vicinity of the \ion{H}{2} region is also characterized by high gas excitation temperatures. The upper horizontal branch in the figure  originates in the cooler, more extended star forming regions in L1482 and from within the Cal-X region in CMC-West. The lowest ($\sim$10 \wunit) horizontal branch in the Fig. \ref{fig:wcoavtemp} consists of the coldest  (\tdust<~14K) material which we find to mostly originate in the L1478 and CMC-West regions of the CMC.

\parnum  This complex structure is quite different from what is often seen in other clouds, where the CO emission at low extinctions (i.e., 1 $\leq$ A$_V$ $\leq$ 3-4 magnitudes) more or less linearly increases with A$_V$, and then appears to flatten at higher (A$_V$ $>$ 4 magnitudes) extinctions, though previous studies rarely extend much into the high extinctions \citep{1982ApJ...262..590F,1994ApJ...429..694L,2006A&A...454..781L,2013MNRAS.431.1296R,2015ApJ...805...58K,2018MNRAS.474.4672L}. Indeed, none of the previous studies come close to achieving the dynamic range in cloud depth (i.e., 3 $\leq$ A$_V$ $\leq$ 60 magnitudes) provided by the \herschel observations in our study of the CMC. These extremely deep observations, obtained on sub parsec spatial scales and across a range in cloud environments, indicate that expectations of a simple relationship between CO emission and hydrogen column density and, consequently the existence of single empirical X-factor describing all material across the cloud, may be unrealistic.

\subsection{The X-factor for CO J=$2\to1$}\label{sec:xfactor}

\begin{deluxetable}{lcc}[t]
\tablecaption{Global \twelveco X-factor}
\label{tab:xcoall}
\tablewidth{0pt}
\tablecolumns{3}
\tablehead{\colhead{Method} & \colhead{\x{\twelveco\ (2-1)}} & \colhead{K15.}}
\startdata
$\langle\nht/\w\twelveco\rangle$        & 5.05 $\pm$ 1.80 & 4.4\\
$\langle\nht\rangle/\langle\w\twelveco\rangle$ & 4.18 $\pm$  1.83 &  3.6 \\
Linear fit through origin & 2.98 $\pm$ 1.74 \\\hline
{\bf Average} &  {\bf 4.1$\pm$0.8}  & \omit
\enddata
\tablecomments{Measurements of the global \twelveco (2-1) X-factor for the whole cloud in units of $10^{20}\ \xfunit$. The measurement methods are discussed in  \S\ref{sec:xfactor}. The \citetalias{2015ApJ...805...58K} are scaled by 1.4 to convert them from the original \xco (1-0) to \xco (2-1). }
\end{deluxetable}

\parnum As outlined above, a single X-factor does not characterize all the gas in the cloud. We can measure the X-factor for every pixel by taking the ratio of the \htwo column density maps and the J=2-1 CO integrated intensity maps. In the CMC, we find that \x{\twelveco\ (2-1)} ranges from $0.5 - 52\times 10^{20}\ \xfunit$. However, as previously stated, the region of the CMC that contains hot dust shows a fairly tight linear relationship between \wco and \av, indicative of a well-defined X-factor for at least that region.
Here we measure the CO X-factor for the $J=2-1$ transition, $\xco = \nht/\w{\rm CO\ 2-1}$, using three methods, where $\langle X \rangle=\frac{1}{N}\sum_{i=1}^{N}{X_i}$:
\begin{enumerate}[itemsep=1pt,leftmargin=*]
    \item $\langle\xco\rangle=\langle\nht/\w\twelveco\rangle$, the average of the CO X-factor measured in each pixel, the "per-pixel" X-factor.
    \item $\langle\xco\rangle_{\rm W} = \frac{\langle\nht\rangle}{\langle\w\twelveco\rangle}$, which is the average column density divided by the average CO integrated intensity. It is equivalent to the \wco-weighted average of the per-pixel CO X-factor, $\langle\xco\rangle_W = \frac{\sum{\xco \wco}}{\sum{\wco}}$.
    \item  $\widehat{\xco}$ \- is the solution to a least-squares linear fit through the origin to $\nht = \widehat{\xco}\, \wco$. This is equivalent to the $(\wco)^2$-weighted average of the per-pixel CO X-factor, $\widehat\xco = \frac{\sum{\xco \wco^2}}{\sum{\wco^2}}$
\end{enumerate}
If a single X-factor existed that described the whole cloud, each method would return the same value. To estimate our uncertainty in these values we use the standard deviation for (1) and the \wco and (\wco)$^2$ weighted standard deviation for (2) and (3) respectively. The data noise contributes negligibly being more than an order of magnitude smaller than the (weighted) standard deviation. We use the average
the three methods as our accepted value and use the square-root of the average variance to derive the uncertainty. This uncertainty is a factor of $\sqrt{3}$ larger than the error derived using error propagation and better represents the variation of the X-factor in our data.

\subsubsection{\twelveco J=2-1 X-factor}\label{sec:xfactor12}

\parnum While the large scatter in the \wco--\av relation shows that there is no single-valued X-factor that will work for all individual positions across the cloud, a globally averaged X-factor can still be computed from the data and is useful for comparison with other clouds that are either unresolved or partially resolved. Using the three methods outlined above we found results consistent with those seen in \citetalias{2015ApJ...805...58K}. The three measurements are presented in Table \ref{tab:xcoall} and give an average "global X-factor" value of $(4.1\pm0.8)\e{20}\ \xfunit$, which is twice the Milky Way value. We next focus on the ability to measure the existence of a linear relation and measure the J=2-1 X-factor for all three isotopologues in the warm gas where all share the same \tdust-dependent, nearly linear, branched structure.

\begin{figure}[t]
\centering
\includegraphics[width=\columnwidth]{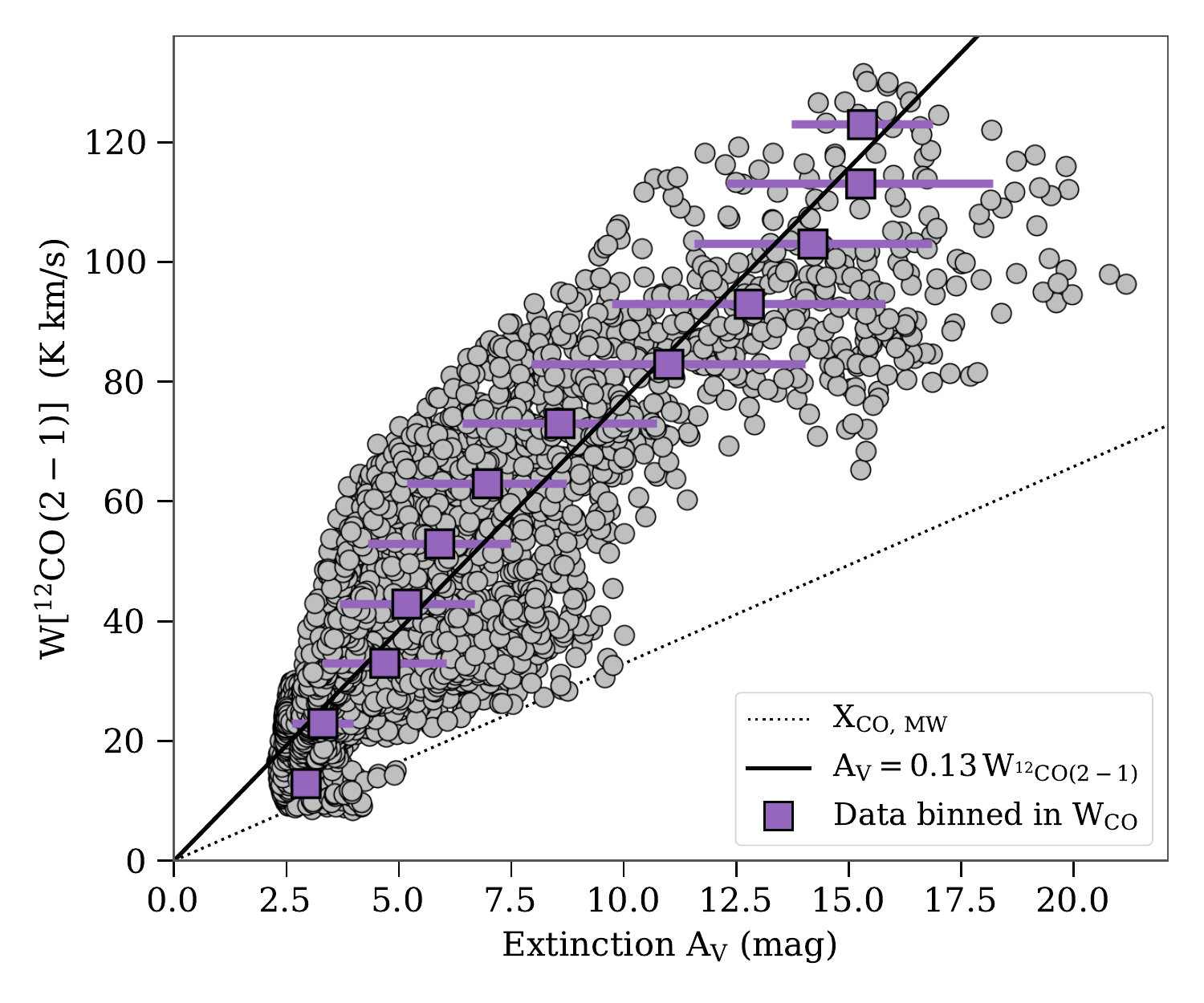}
\caption{\label{wcoavhigh} \w{\twelveco(2-1)} plotted against \av for the CMC  for points with \tdust > 25 K and where \w{\twelveco} is detected at greater than 3$\sigma_{W,12}$. The { black dotted line} is the standard adopted MW X-factor ($2\times 10^{20}$\xfunit). The black solid line is the best fit solution for the data.}
\end{figure}

\begin{deluxetable}{l|CCCC}[t!]
\tablecaption{J=2-1 X-factor for CO Isotopologues\label{tab:xcohot}}
\tablewidth{0pt}
\tablecolumns{4}
\tablehead{\colhead{Method} & \colhead{\x\twelveco} & \colhead{\x\thirteenco} & \colhead{\x\ceighteeno} & \colhead{K15 \twelveco}}
\startdata
$\langle X \rangle$   & 1.36~ (0.40) &  8.99~ (6.61) &  59.7~ (26.2) & 1.8\\
$\langle X \rangle_W$ & 1.27~ (0.36) &  4.67~ (3.22) &  45.1~ (20.8) & \omit \\
$\Hat{X} $            & 1.22~ (0.33) &  3.55~ (1.20) &  36.1~ (13.0) & 2.1 \\\hline
\bf Average & \bf 1.28 \pm  0.36 & \bf 5.74 \pm  4.30 & \bf 46.96 \pm  20.55 & \omit \\
\enddata
\tablecomments{CO X-factor (in units of $10^{20}\xfunit$) derived from pixels containing hot dust (\tdust > 25 K) for the 2-1 transition of \twelveco, \thirteenco, and \ceighteeno in units of $10^{20}\ \xfunit$ derived using the 3 methods we describe in \S\ref{sec:xfactor}. The 1$\rm\sigma$ error is in parenthesis. We adopt the average value for each isotopologue as the accepted values of the X-factor. The error is calculated as the square-root of the average variance (the variance is the square of the  1$\rm\sigma$ error). We converted converted the Kong et al. values to \twelveco (2-1) by scaling with $W_{2-1}/W_{1-0} = 0.7$. }
\tablerefs{\citetalias{2015ApJ...805...58K}}
\end{deluxetable}

\parnum  We measure the \twelveco J=2-1 X-factor in the region of hot (\tdust>25 K) dust  where \w\twelveco and \av are correlated (Pearson $r$ = 0.88) using the 3 methods described above. Those data are shown in Fig. \ref{wcoavhigh} with the data binned along the \wco axis shown in purple. The results are given in Table \ref{tab:xcohot}. The X-factors derived are all consistent with each other within $\sim1 \sigma$ ranging from $(1.2-1.4)\e{20}\ \xfunit$, a little less than 1/2 the Milky Way value. Taking the average of the 3 methods as our X-factor, we find that the \twelveco (2-1) X-factor for \tdust > 25 K is
\begin{equation}
\begin{split}
\x{\twelveco\,(2-1)} &= (1.28\pm 0.36) \\ & \e{20}\ \xfunit.
\end{split}
\end{equation}

\parnum It is interesting to point out here that although the $^{12}$CO gas in Fig. \ref{wcoavhigh} is very optically thick, it exhibits the expected behavior of an optically thin tracer, i.e. \wco increasing with dust/\htwo column density. We find no significant trends in the hot dust in \wco with CO velocity dispersion, $\sigma_{\twelveco}$, or with \tdust  that would provide the increase necessary in the optically thick W[CO] to explain the linear relation in the hot dust pixels. The linear relation seems to be mostly due to a correlation between $T_{\rm  peak} $ and \av in the hot dust as was similarly found in \citetalias{2015ApJ...805...58K}. We can estimate an \textit{effective optical depth}, $\tau_{\rm eff, 12}$, for \twelveco by comparing the \xco one would expect from an optically thin gas to the measured \x\twelveco, $\tau_{\rm eff, 12} = \ln{\left(X_{\rm obs}/X_{\rm thin}\right)}$. The optically thin CO X-factor can be found by dividing equation (\ref{eq:columndensity}) by \nht and assuming $\tau << 1$ to get:
\begin{equation}\label{eq:xcothin}
X_{\rm thin} = C(\tex)/\abund{\twelveco},
\end{equation}
where $C(\tex)$ is as defined in equation (\ref{eq:ctex}), $\abund{\twelveco}$ is the CO:\htwo abundance ratio ${10^{-4}}$,  and for \tex we use mean excitation temperature of the hot dust pixels, $\tex\sim20{\rm\ K}$. We find $X_{12,\, thin} = 5.26\e{18}\ \xfunit$ which gives $\tau_{\rm eff, 12} \approx 3$. \citet{2013ARA&A..51..207B} finds a similar estimate for the optically thin \twelveco X-factor.

\begin{figure*}[tp]
\includegraphics[width=\columnwidth]{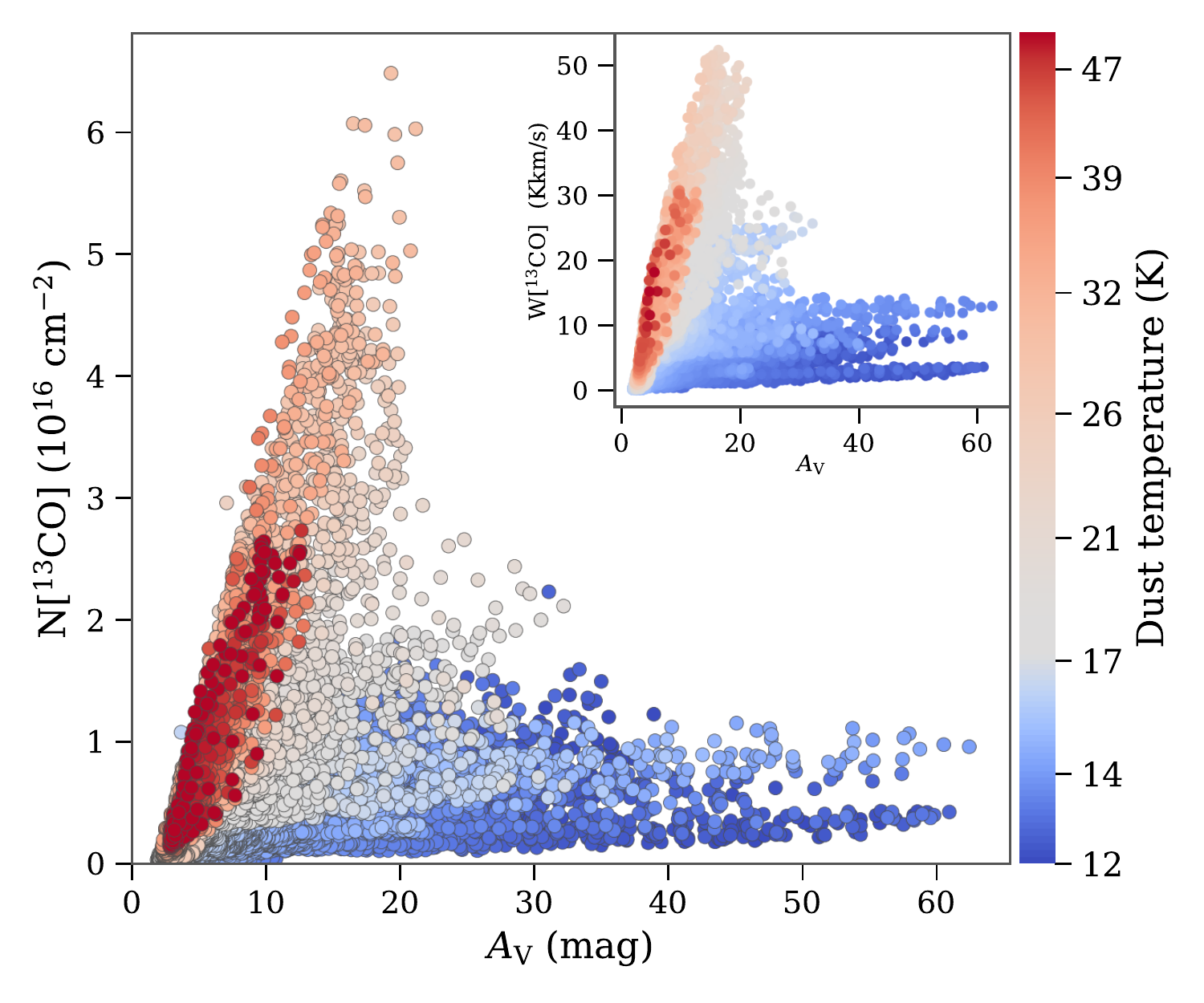}
\includegraphics[width=\columnwidth]{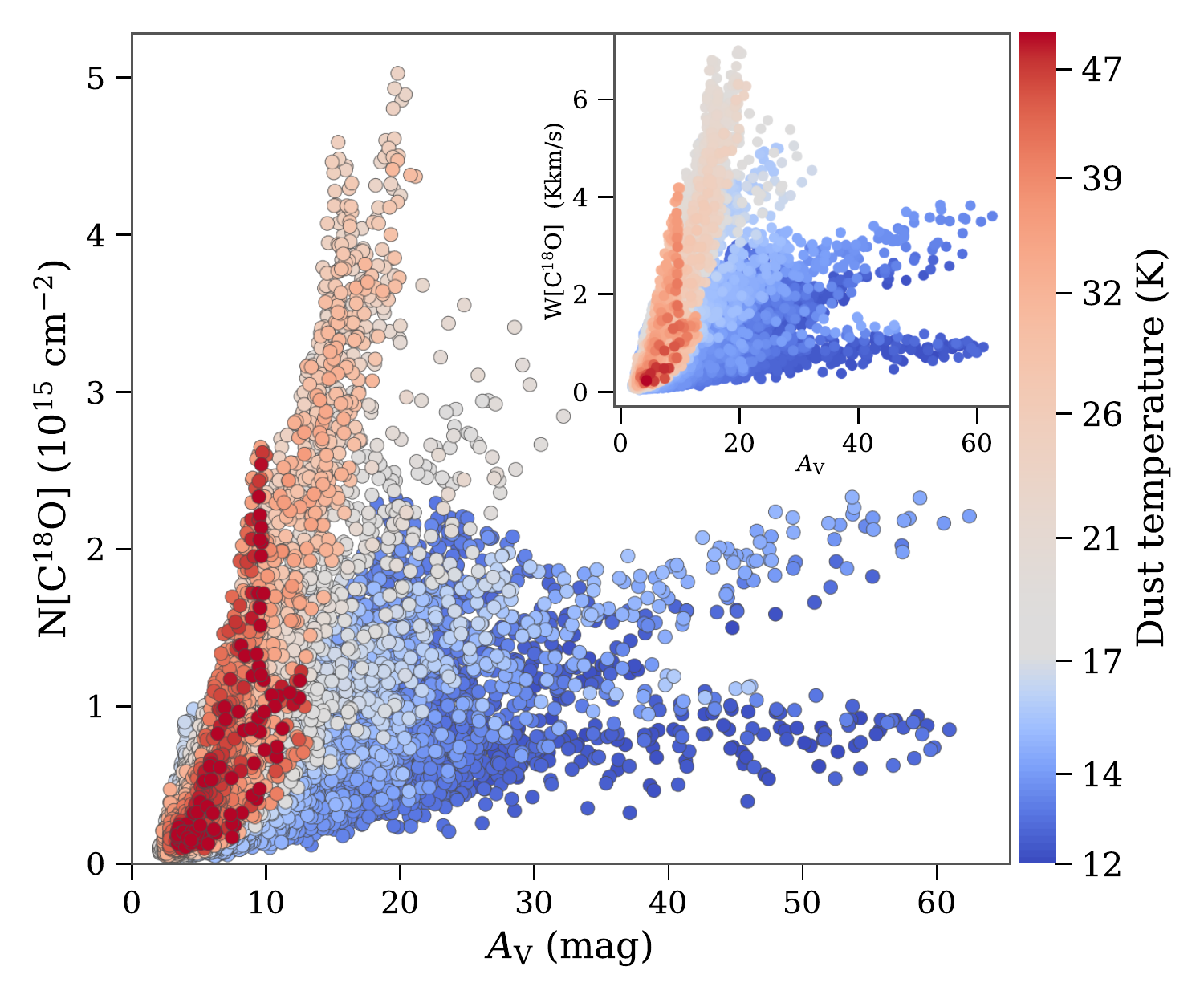}
\caption{\label{fig:ncoav} Plot of the  ({\bf Left}) \thirteenco and ({\bf Right}) \ceighteeno column density vs visual extinction colored by \tdust. The colorbar shows colder pixels as blue, transitioning through gray, and becoming redder as the temperature increases. Data is plotted so that higher temperature points are plotted over cooler points to to emphasize the temperature dependent structure in the plot. The inset shows \wco vs \av for the line with the same coloring scheme.}
\end{figure*}

\subsubsection{\thirteenco \& \ceighteeno J=2-1 X-factor}\label{sec:xfactor1318}
\parnum We measure the X-factors for \thirteenco and \ceighteeno for $J=2-1$ in the same way as done for \twelveco. For both lines, \wco is well correlated with extinction (Pearson $r>0.9$) in the hot dust, indicating the presence of a simple linear relationship with extinction. In the hot dust we find an average \thirteenco X-factor of,
\begin{equation}
\x{\thirteenco\, (2-1)} = (5.74\pm 4.30)\e{20}\ \xfunit,
\end{equation}
and an average \ceighteeno X-factor of,
\begin{equation}
\x{\ceighteeno\, (2-1)} = (4.70\pm  2.05)\e{21}\ \xfunit.
\end{equation}
The error is large here because the linear correlation visible in the hot dust does not pass through the origin like it does for \twelveco.

\parnum Both \thirteenco and \ceighteeno are expected to be optically thin. Assuming a \twelveco abundance relative to \htwo of 1\e{-4}, $\rm ^{12}C/ ^{13}C = 69$, and $\rm ^{16}O/^{18}O = 557 $ \citep{1999RPPh...62..143W} we can apply the same process as above to determine effective optical depths for both lines. For \thirteenco and \ceighteeno we find $\tau_{\rm eff}$ equals 0.69 and 0.31, respectively. If, instead of using \tex = 20 K, we use the \twelveco excitation temperature we derived for each pixel (see Appendix \ref{sec:methnco}) and compare $X_{\rm thin}(\tex)$ (eq. \ref{eq:xcothin}) to \xco in each pixel, we find that on average $\tau \approx 0$. In other words, the X-factors we measure are consistent with those predicted by an optically thin model using the \twelveco excitation temperature and cosmic abundance ratios.

\subsection{N[CO] vs. \av}\label{sec:abund}

\parnum  In Fig. \ref{fig:ncoav} we show the \thirteenco and \ceighteeno column density plotted against extinction with points colored by \tdust. They show the same temperature dependent branched structure as \twelveco, with \ceighteeno showing two very distinct cold branches. Plots of \wco vs \av for these lines are shown inset on the corresponding \nco vs \av figure. The plots of \wco--\av for \ceighteeno and \thirteenco are very similar to the corresponding \nco--\av plots,  suggesting a tight correlation between \wco and \nco. For both lines, the Pearson correlation coefficient between \wco and \nco is $r > 0.9$.
In the optically thin limit equation \ref{eq:columndensity} becomes $N_{\rm CO} = C(\tex) W_{\rm CO}$. The best fit \tex is 7.6 K for \thirteenco and 9.9 K for \ceighteeno, or equivalently
\begin{align}
    \N\thirteenco &= 9.2\e{14}\ \w\thirteenco \label{eq:ctn13}\\
    \N\ceighteeno &= 6.8\e{14}\  \w\ceighteeno. \label{eq:ctn18}
\end{align}This is consistent with the median \tex derived from \twelveco, 7.9 K for the pixels with \thirteenco and 9.9 K for those with \ceighteeno.

\parnum When deriving the column density we also derive the optical depth. In the cold branches the \thirteenco optical depth is $\lesssim 1.5$, while \ceighteeno is optically thin ($\tau_{18} < 0.75$) everywhere. With $\tau<2$, if CO is in the gas phase, we expect to see \wco and \nco increase with \av. Instead \thirteenco shows a distinct flattening in both beyond $\av\sim 10$ and \ceighteeno even shows some evidence for flattening in the coldest branch (the lowest branch on the plot).
The three branches are most distinct in \ceighteeno, with the vertical and upper horizontal branches coming from L1482, and the lowest (and coldest, \tdust$\sim$12-13 K for \av>30 {\rm mag}) originating in L1478. The pixels that make up the upper horizontal branch, are warmer due to being associated with a region with a large number of embedded young stars.

\subsubsection{The \thirteenco and \ceighteeno Abundance}\label{sec:abundmeas}

\parnum  Using our CO and \htwo column density measurements we can measure the \thirteenco:\htwo and \ceighteeno:\htwo ratios, which are the abundances relative to \htwo. Square braces, $[\cdot]$, will be used to denote the abundance relative to \htwo. The abundance relative to H is simply $0.5\times$ the abundance relative to \htwo. The line-of-sight abundance can be determined by simply dividing the column density maps by the \nht map. This {\it in situ} abundance is not constant across the cloud.

\parnum We see in Fig. \ref{fig:ncoav} that again, as with \w\twelveco--\av, only the hottest dust shows a clear linear relation. To measure the true CO abundance, we must limit our measurements to the part of the cloud where CO is likely entirely in the gas phase, namely in the region containing the hot dust.
We measure the total CO abundance by measuring the average abundance, $\left\langle\nco/\nht\right\rangle $, in the hot (\tdust>25 K) dust,
using pixels with signal-to-noise >3 for \thirteenco and >5 for \ceighteeno and \tdust > 25 K, and find
\begin{align}\label{eq:13coabund}
 [\thirteenco]  &= (1.31\pm.89) \E{-6} \\
\label{eq:c18oabund}
 [\ceighteeno]  &= (1.27\pm.59) \E{-7}
\end{align}
These measurements are consistent within the errors with abundances derived from standard atomic abundances ratios assuming all the carbon is in CO\footnote{We use $\rm ^{12}C/H=2\e{-4}$, and adopt the isotopic abundances from \citet{1999RPPh...62..143W} for the local interstellar medium $\rm ^{12}C/^{13}C=69$, $\rm ^{16}O/^{18}O=557 $}:
$\rm [\thirteenco]_{MW}=1.45\E{-6}$, $\rm [\ceighteeno]_{MW}=1.80\E{-7}$. Looking at the hot pixels, we see that a straight line through the data would not intersect the origin. By performing a linear regression to the data we can determine the linear relation between CO and \htwo.
\begin{align}
    \N\thirteenco &=\  3.19\e{15}(\av - 2.50) \label{eq:ncoavfita}\\
    \N\ceighteeno &=\  2.47\e{14}(\av - 3.55) \label{eq:ncoavfitb}
\end{align}
These equations can be combined with equations (\ref{eq:ctn13}) and (\ref{eq:ctn18}) to derive a linear relation for \wco--\av. The \av offset is the extinction below which the measured \thirteenco or \ceighteeno column density is zero.
As mentioned previously, CO has been detected at lower extinctions, below \av $\sim$ 1 mag, in both emission (e.g.; towards the Pipe \citep{2006A&A...454..781L} and Cygnus \citep{ 2016A&A...587A..74S}) and absorption (e.g.; towards O \& B stars, \citep{2007ApJ...658..446B}), so this offset is not a threshold for CO formation.

\begin{figure}[t]
    \centering
    \includegraphics[width=\columnwidth]{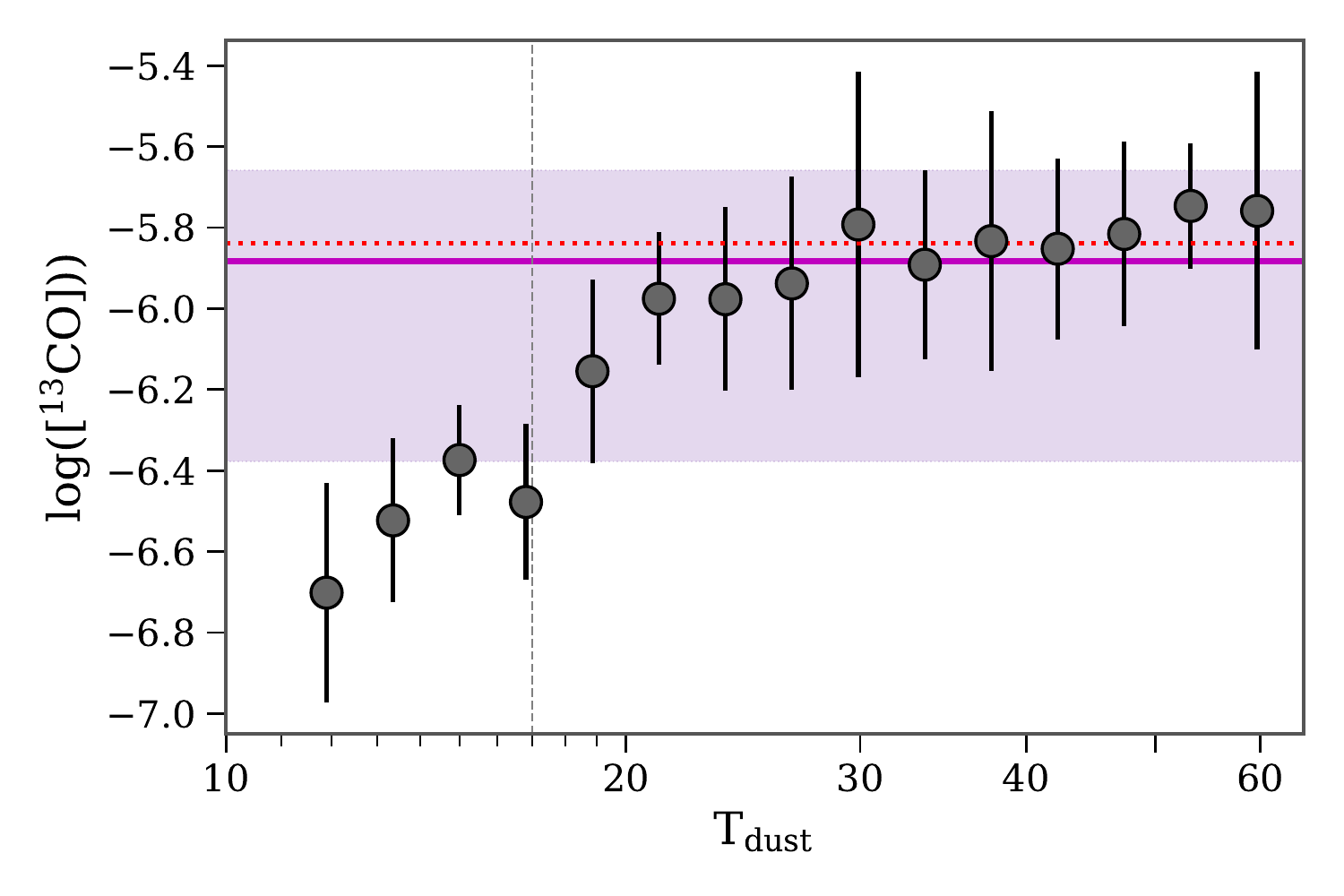}
    \includegraphics[width=\columnwidth]{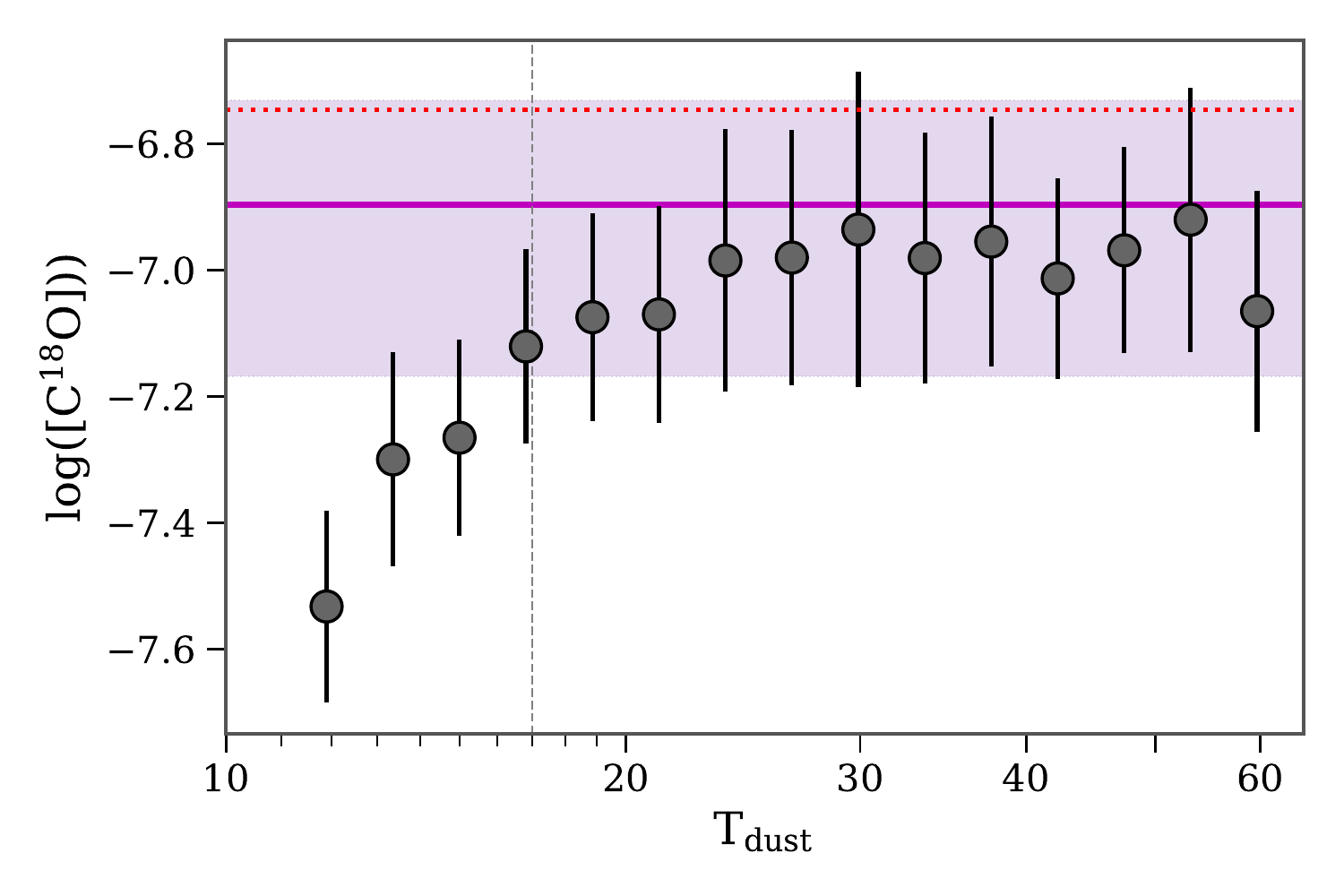}
    \caption{ \label{fig:abundtdust} Binned $\log_{10}$ median abundance for ({\bf Top}) \thirteenco and ({\bf Bottom}) \ceighteeno versus dust temperature. The abundance is binned by \tdust in logarithmic bins from 10 K - 63 K in 0.05 dex steps. The error-bar is the median absolute deviation scaled to the standard deviation. The purple line with the purple shaded region marks the abundance measured in the hot dust and the corresponding $1\sigma$ error. The red dashed line is the cosmic abundance}
\end{figure}

\subsubsection{Abundance versus \tdust}

\parnum  As with \wco vs \av, \nco vs \av shows a continuous transition from the cold horizontal branches to the hot vertical branch, indicating that the abundance may be tied to the dust temperature. Plotting the bin-averaged abundance as a function of \tdust binned in 0.05 dex bins shows us that the abundance is temperature dependent. Fig. \ref{fig:abundtdust} shows the average CO abundance through the cloud binned by \tdust overlaid with the average abundance we derived in the hot dust with the $1\sigma$ range shown in purple and the value derived assuming cosmic abundances as a dashed red line. The abundance clearly increases with temperature and flattens out beyond $\gtrsim 20\ \text{K}$.

\parnum The \thirteenco abundance plateaus to the abundance we measured in the hot dust, which is coincident with its cosmic abundance. \ceighteeno, however, plateaus just short of its cosmic abundance. This is in part because the high temperatures span a large range in extinction, and at low extinction \ceighteeno is selectively photodissociated \citepalias{2015ApJ...805...58K}. Because so much of the high temperature dust is at low extinction, the areas where \ceighteeno can be destroyed could dominate the average and drive the average abundance lower. This would not be much of a problem at low temperatures, since these regions are characterized by much higher \av where the CO is better shielded from far-UV radiation. This effect is seen in photodissociation regions (PDRs), and observationally can be seen as an increase in \abund{\thirteenco}/\abund{\ceighteeno} or \w\thirteenco/\w\ceighteeno \citep{2014A&A...564A..68S} and was observed in the CMC by \citetalias{2015ApJ...805...58K}. Alternatively, the difference between our measured and the cosmic \ceighteeno abundance could be caused by a variation in the $\rm {}^{18}O$ abundance. It is observed to have a large range of values, ranging from 300-600 away from the galactic center \citep{ 2005A&A...437..957P, 2012M&PS...47.2031N}, which would be enough to explain the difference we see.

\parnum The rise of abundance with \tdust is suggestive of CO desorbing from the dust grains in the warm and hot dust, with the inverse relationship being CO depletion onto grains. In Fig. \ref{fig:abundtdust} the transition from a positive slope to a flat one occurs around 16-20 K. This is consistent with the temperature at which CO is expected to sublimate off dust grains \citep{1995ApJ...441..222B,2006A&A...449.1297B}. \citeauthor{1995ApJ...441..222B} shows that the time scale for all of the CO to desorb into the gas phase at \tdust $\sim$ 20 K is of order $100\ {\rm yrs}$. The reverse process, freeze-out or depletion, occurs at similar temperatures with timescale of $\sim 10^{6}$yrs.

\parnum In their study of the L1482, \citetalias{2015ApJ...805...58K} suggested desorption/depletion as the cause of abundance trends with \tex. However, depletion depends on the dust temperature which they were unable to make direct comparisons too at the time. The temperatures we measure in the cold dust are significantly lower than the sublimation/depletion temperature and CO should freeze out onto grains in the cold, dense central regions of the cloud. Similar variations in the nature of the \N\thirteenco--\av relation were found in Orion by \citet{2013MNRAS.431.1296R} who came to the conclusion that depletion was responsible for the flattening of the relationship beyond \av$\sim10\ mag$.

\parnum Our observations of the relationship between CO (\wco, \nco) and the dust point to it being shaped by desorption/depletion processes and are consistent with previous observations of depletion (e.g.; in Orion). We conclude that the spatial variation in CO abundance in the CMC is due in large part to changes in desorption and depletion caused by the dust temperature changing with distance from star forming regions in the cloud.

\section{Depletion}\label{sec:depletion}

\parnum In this section we measure the level of depletion across the CMC and create the first maps of depletion on giant molecular cloud scales.

\subsection{Measuring Depletion}\label{sec:measdepl}

\begin{figure}[t]
    \centering
    \includegraphics[width=\columnwidth]{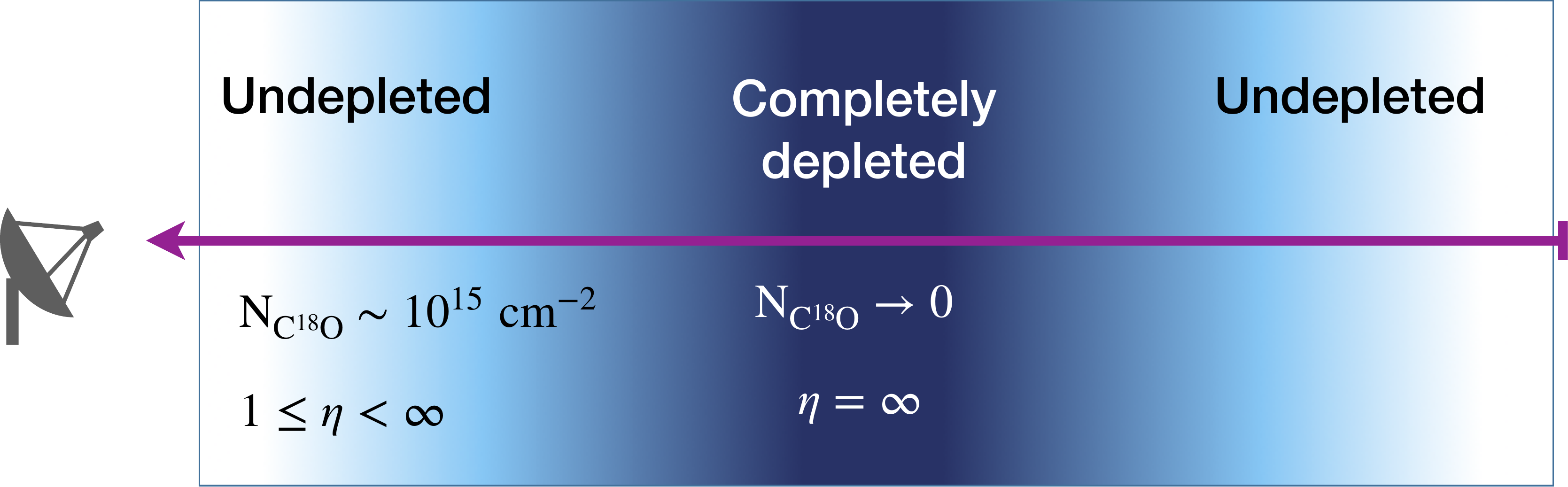}
    \caption{Schematic drawing showing how the depletion factor is an average of the depletions along the line-of-sight and necessarily underestimates the depletion in the center regions.}
    \label{fig:cartoon}
\end{figure}

\parnum The depletion factor, \fdep, is the gas phase abundance at some point in the cloud relative to the true abundance. We measure depletion independent of any outside data. We are using abundances relative to \htwo. The most similar previous study was completed by \citet{2010ApJ...721..686P} in Taurus who relied on relationships between the CO and $\rm CO_2$ ice abundance and \av to derive a total (gas + ice) CO column density which was used to derive a depletion factor - a method we examine later. We bypass needing to estimate the total CO column entrapped in ice by having a direct measurement of the true abundance in the hot dust (\S\ref{sec:abundmeas}). We define the depletion factor the same way \citet{1999A&A...342..257K} do, as a ratio of abundances,
\begin{equation}\label{eqn:depletion}
\fdep = \frac{A_{\rm ISM} }{A_{{\rm gas}} }
               = \dfrac{
\left\langle{N[\text{CO}]}/{\nht}\right\rangle_{\rm\! ISM} }{{\left({N[\text{CO}]}/{\nht}\right)}_{{\rm\! gas}} },
\end{equation}
 where $A_{\rm gas}$ is the abundance in a particular pixel, and $A_{\rm ISM}$ is the true underlying abundance. We will denote depletion factors for \thirteenco and \ceighteeno as $\eta_{13}$ and $\eta_{18}$. \citet{2011ApJ...738...11H} use a very similar method to map depletion in an infrared-dark cloud; however they derive their depletion factor relative to the CO abundance found using cosmic abundance ratios.
  If we were to adopt the same assumption, it would only change the result by a small multiplicative factor. The depletion factor varies from 1 (no depletion) to $\infty$ (completely depleted)\footnote{The depletion factor is related to another common parametrization, also called the depletion factor ($\delta_{\rm CO} = N_{ice}/N_{tot} = 1 - A_{gas}/A_{ISM}$), by $\delta_{\rm CO} = 1 - \frac{1}{\fdep}$ which varies from 0 (not depleted) to 1 (totally depleted)}.


 \parnum It is important to note that what we are measuring is a line-of-sight (LOS) average depletion factor, which is consequently only a lower limit to the peak depletion reached along the LOS. Fig. \ref{fig:cartoon} shows a cartoon of the depletion structure of our molecular cloud. In practice any highly depleted inner region of the cloud will be measured to have a lower depletion factor than it really has due to the measurement being diluted by outer, undepleted layers of the cloud.

\subsection{Depletion Factor vs \av and \tdust}\label{sec:depavtdust}

\begin{figure}[t]
\centering
\includegraphics[width=\columnwidth]{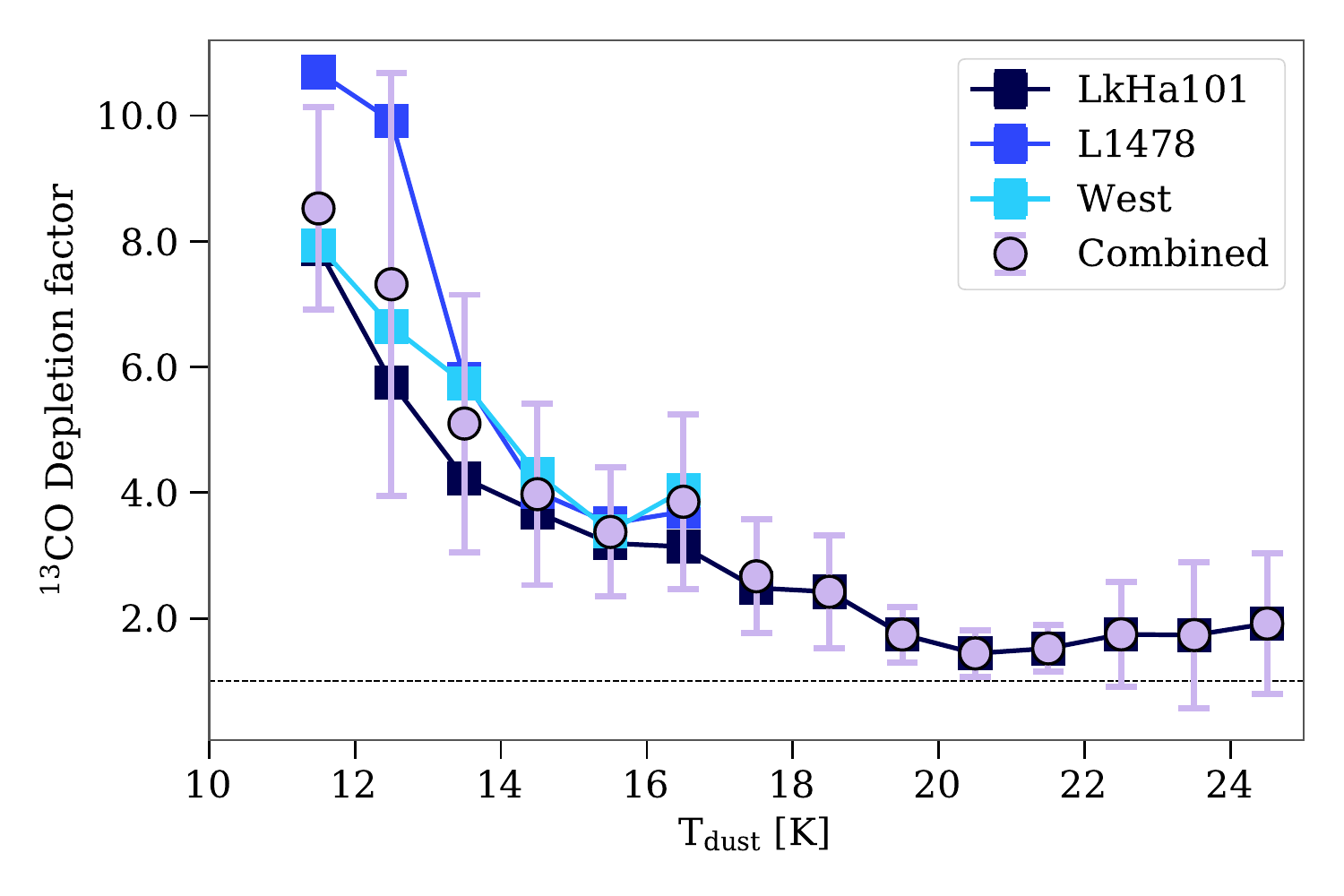}
\includegraphics[width=\columnwidth]{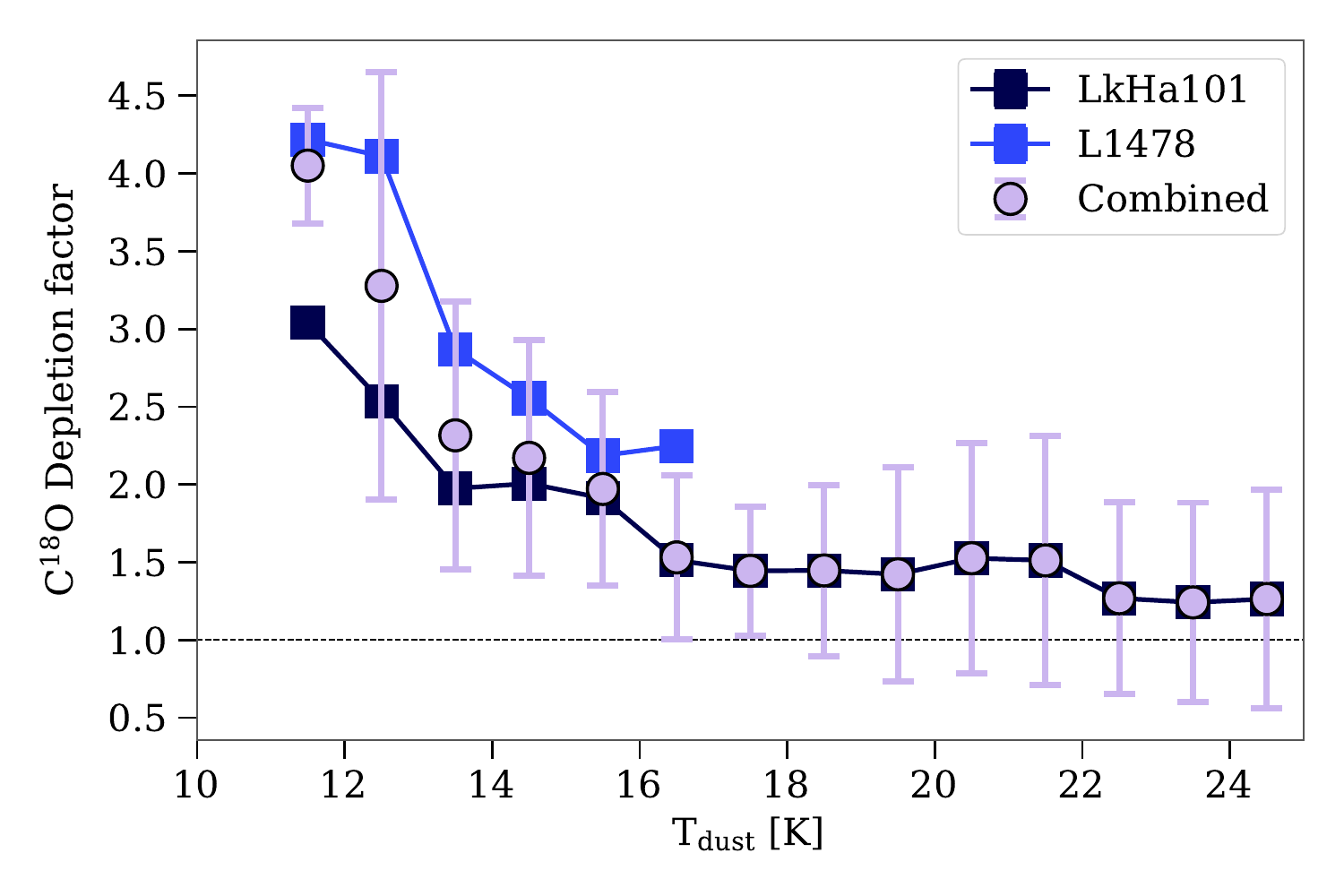}
\caption{Depletion Trends \label{fig:deptrend1} \textbf{Top:} binned measurements of the \thirteenco depletion in each California sub-region. They are binned in 1 K bins for \tdust. The lines with no errors show the binned relationship for each sub-region, and the light-purple points with error-bars is the average of all the data with $1\sigma$ errorbars. The horizontal black dashed line indicates \fdep=1. \textbf{Bottom:} Same as top but for \ceighteeno. }
\end{figure}

\parnum Fig. \ref{fig:deptrend1} shows how the depletion factor varies with the dust temperature, binned in 1 K bins. The relationship for each CMC region is shown with square markers and the average for the whole cloud is shown with circles. There is a large scatter, however, the binned values show some consistency in general shape between regions, though the absolute scale can be different (e.g.; L1478 has higher measured depletion factors than L1482 in \ceighteeno). The curve begins flattening out considerably between 15-18 K, and is quite flat for \tdust> 20 K. This trend is essentially the inverse of that seen in Fig. \ref{fig:abundtdust}. The temperature range at which it transitions from a steep to flat slope, is consistent with the sublimation temperature for CO, $\sim 17\ \text{K}$.

\parnum The binned \fdep--\tdust relationship shows a similar exponential dependence on \tdust as was noticed in \citet{1999A&A...342..257K} in \ceighteeno in the dense filament of IC 5146; however the rise seen in \citeauthor{1999A&A...342..257K} is much shallower, having $\fdep\sim 1.5$ at \tdust= 10 K. This could be due to the fact that the extinction measurements in IC 5146 covered a smaller dynamic range  (1-25 magnitudes) in extinction because of the lower sensitivity of the near-infrared extinction measurements there.

\subsubsection{CO on Ice}\label{sec:ice}

\begin{figure}[th]\label{fig:icyfig}
    \centering
    \includegraphics[width=\columnwidth]{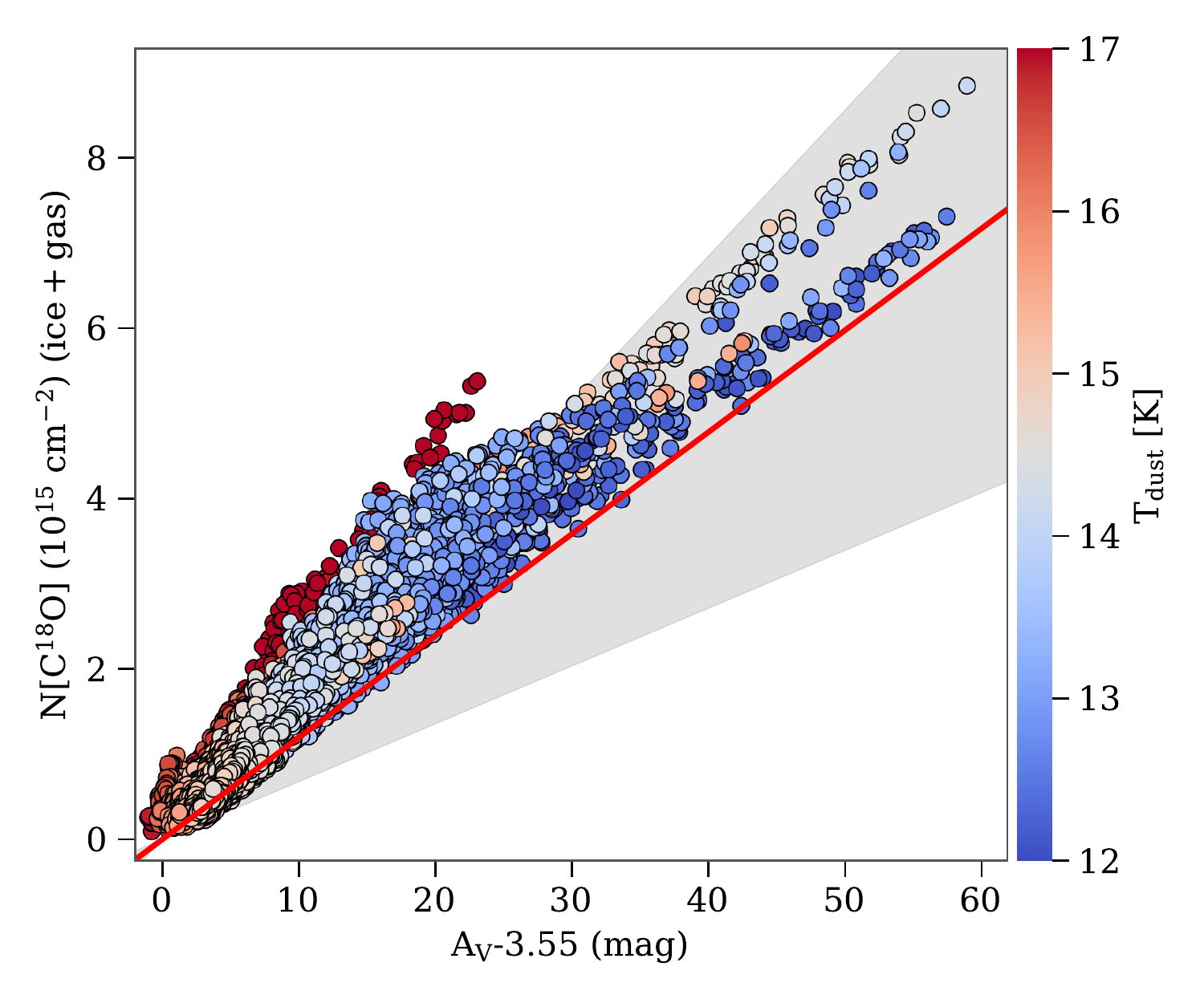}
    \caption{Plot of the total (gas + ice) \ceighteeno column density  plotted against $(\av - 3.55)$ mag. The points are colored by \tdust - note that the scale is different than in Figs. \ref{fig:wcoavtemp} and \ref{fig:ncoav}. The red line shows the \N\ceighteeno predicted using the abundance, $[\ceighteeno] = 1.27\e{-7}$ ignoring the material below the 3.55 mag threshold, $\N\ceighteeno \propto {[\ceighteeno]}(\av - 3.55)$. The gray shaded region shows the $1\sigma$ uncertainty in the \ceighteeno abundance. The total column derived from adding in the contribution from ice is consistent with the total column density derived from the abundance measured in the hot dust. }
\end{figure}

\parnum Depletion results in CO being locked in ices on dust grains. The major reservoirs for CO  in ice are CO ice and \cotwo ice. \cotwo ice is formed from the oxidation of CO on dust grains. By estimating the amount of CO locked away in ice along the LOS, we can determine the total CO column density, $\N{\rm CO}_{\rm total} =\N{\rm CO}_{\rm gas} + \N{\rm CO}_{\rm ice}^{\rm total}$. $\N{\rm CO}_{\rm ice}^{\rm total}$ is the sum of the CO and CO$_2$ ice column densities. Using CO and \cotwo spectral absorption measurements for  sight-lines in Taurus with \av between 5 - 24 mag, \citet{2007ApJ...655..332W} found the following equations describing the relationship between the dust and CO and $\rm CO_2$ ice,
\begin{align*}
    \rm N[\twelveco]_{\rm ice} & = 0.400 (\av - 6.7) \e{17}\ \text{cm}^{-2}\mathbf{,\ \av > 6.7} \\
    \rm N[\twelveco_2]_{\rm ice} & = 0.252 (\av - 4.3) \e{17}\ \text{cm}^{-2}\mathbf{,\ \av > 4.3}.
\end{align*}

\parnum  We convert $\N{\twelveco}_{\rm ice}^{\rm total}$ to $\N{\ceighteeno}_{\rm ice}^{\rm total}$ by scaling it by the cosmic abundance ratio, $\rm ^{16}O/^{18}O = 557$. We add the $\N{\ceighteeno}_{\rm ice}^{\rm total}$ to the $\N{\ceighteeno}_{\rm gas}$ in pixels where $\tdust < 18{\rm\ K}$ to get $\N\ceighteeno_{\rm total}$. In Fig. \ref{fig:icyfig} we plot $\N\ceighteeno_{\rm total}$ against $(\av-A_{\rm V,0})$, where $A_{\rm V,0}=3.55$ is the offset we measured for \N\ceighteeno in eqn (\ref{eq:ncoavfitb}). The data points are colored by the dust temperature along the same line-of-sight. $\N\ceighteeno_{\rm total}$ shows a reasonably tight linear correlation with extinction across the entire extinction range. As a fiducial comparison we also plot in red the line that corresponds to the abundance we measured in the hot dust. The gray shaded region shows the error in the abundance. For this hot material we assume that all the CO is fully evaporated from the grains and thus that this measured abundance represents the true CO abundance in the CMC.

\parnum For \av > 35 magnitudes the relation consists of two branches distinguished by differing dust temperatures. These branches are composed of the same pixels as those that are seen in the CO column density vs extinction plots in Fig. \ref{fig:ncoav}.
The cold lower branch belongs to L1478 and falls along the red line consistent with the predictions derived from the Taurus ice observations. The warmer upper branch comes entirely from the extended regions around L1482. As discussed previously, L1482 has more star formation activity than the area associated with L1478. The warmer dust may indicate that there is likely less CO locked in ice in these regions than in the cold material of L1478 which is apparently well represented by the Taurus-derived relations used to predict the ice abundances. Thus application of these relations to the warm dust likely over predicts $\N\ceighteeno_{\rm total}$ in the warmer regions. However, we note that both branches fall entirely within the shaded region of our uncertainty in the derived abundances.

\begin{figure*}[t]
    \centering
    \includegraphics[width=\columnwidth]{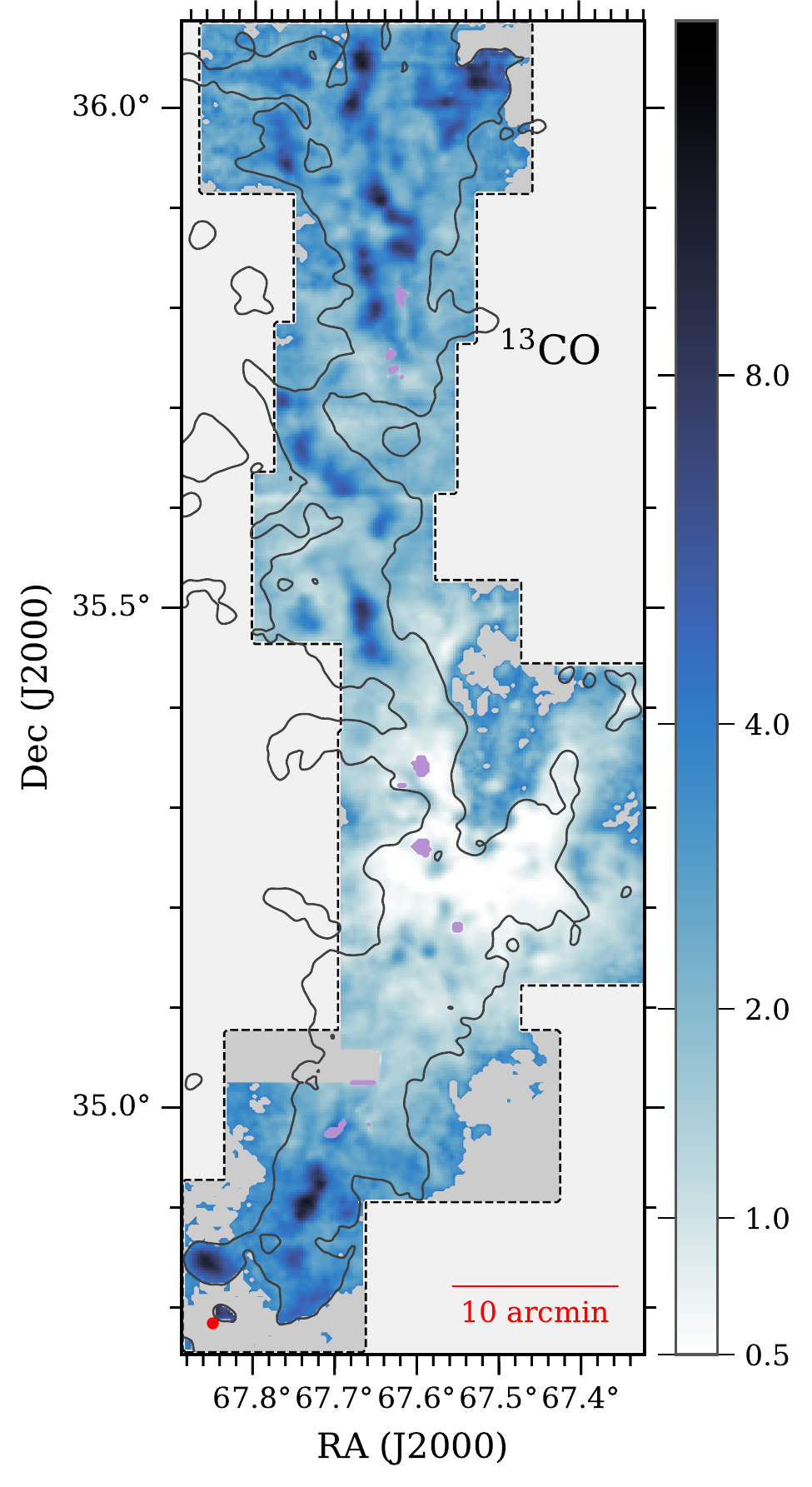}
    \includegraphics[width=\columnwidth]{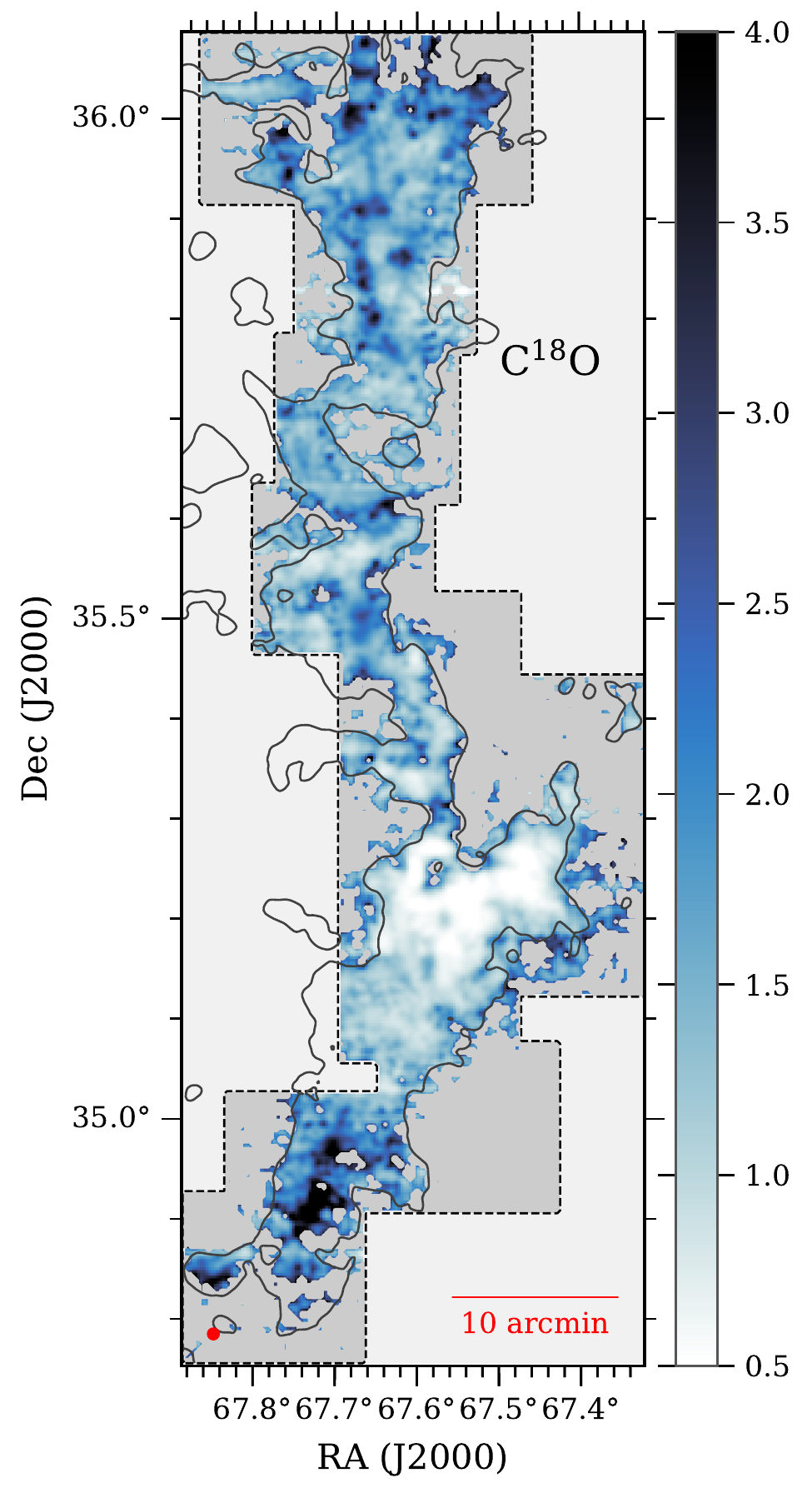}
    \caption{CO depletion maps for CMC-East (Left: \thirteenco, Right: \ceighteeno). The colormap shows the highly depleted regions as dark blue, and the black contour is \av = 5 contour. The dark gray shading marks where there was not sufficient signal-to-noise to detect emission, and the purple patches show where we were unable to calculate column densities due to the optical depth being undefined (see Appendix \ref{sec:methnco}). The map for \ceighteeno has been smoothed to 47.5\arcsec for display only. The black dashed line indicates the survey boundaries. The red circle shows the map resolution (\thirteenco, $38\arcsec$ and \ceighteeno $47.5\arcsec$). }\label{fig:13deplkha}\label{fig:18deplkha}
\end{figure*}

\begin{figure*}[ht]
    \centering
    \includegraphics[height=2.4in]{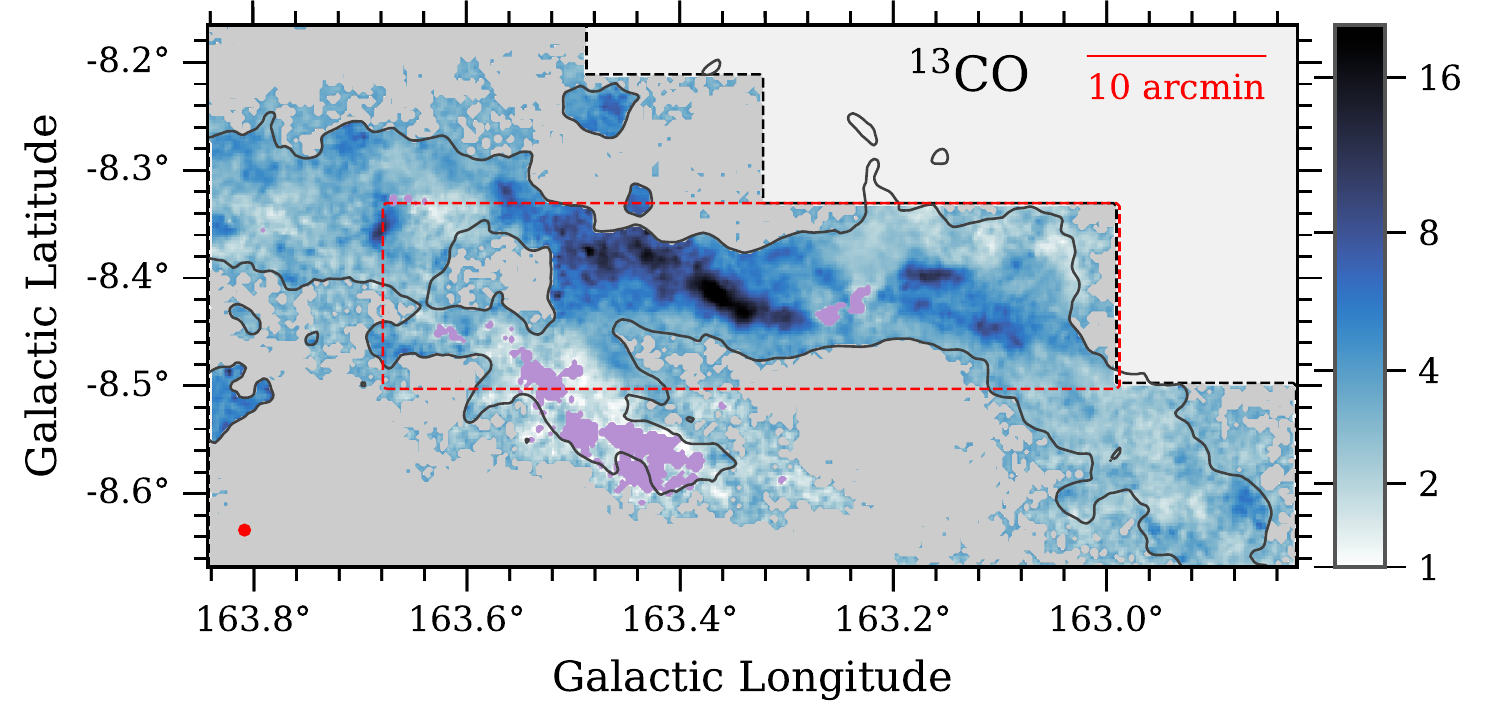}
    \includegraphics[height=2.4in]{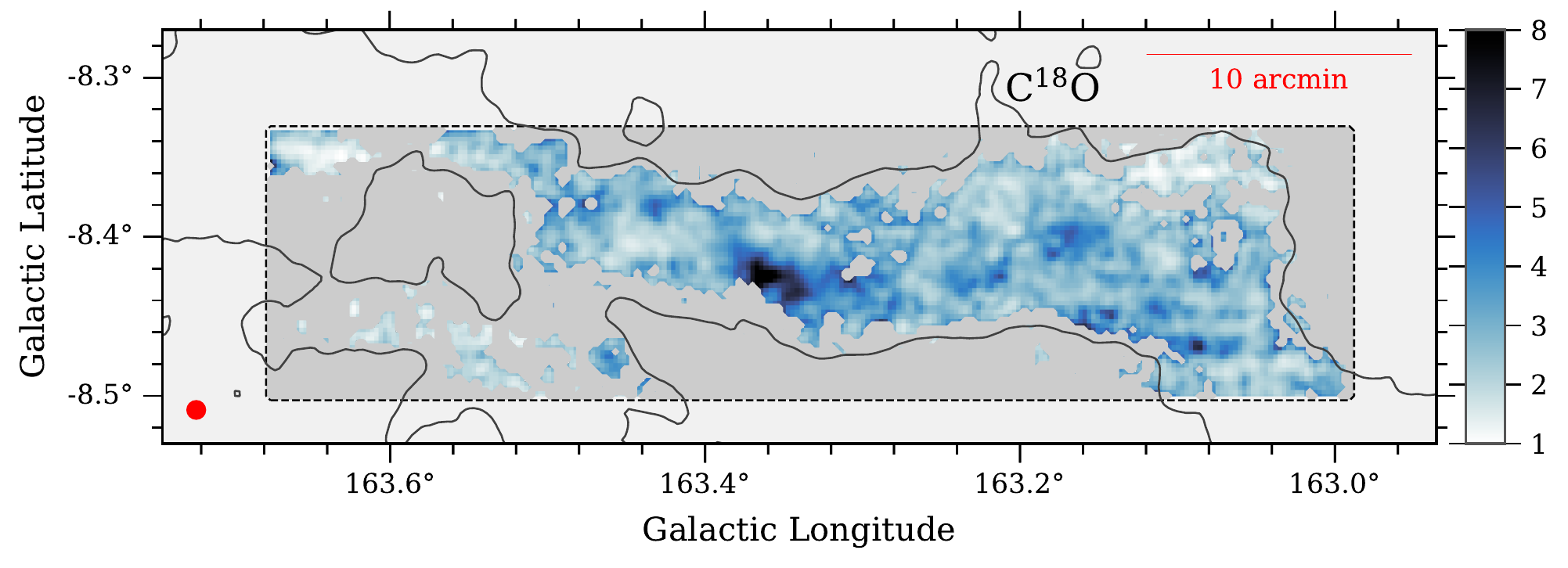}
    \caption{CO depletion maps for L1478 (Top: \thirteenco, Bottom: \ceighteeno). same color scheme as Fig. \ref{fig:13deplkha}. The red dashed box in \thirteenco map shows boundary of the \ceighteeno survey below it. }
    \label{fig:13depl1478}\label{fig:18depl1478}
\end{figure*}

\begin{figure*}[ht]
    \centering
    \includegraphics[height=2.4in]{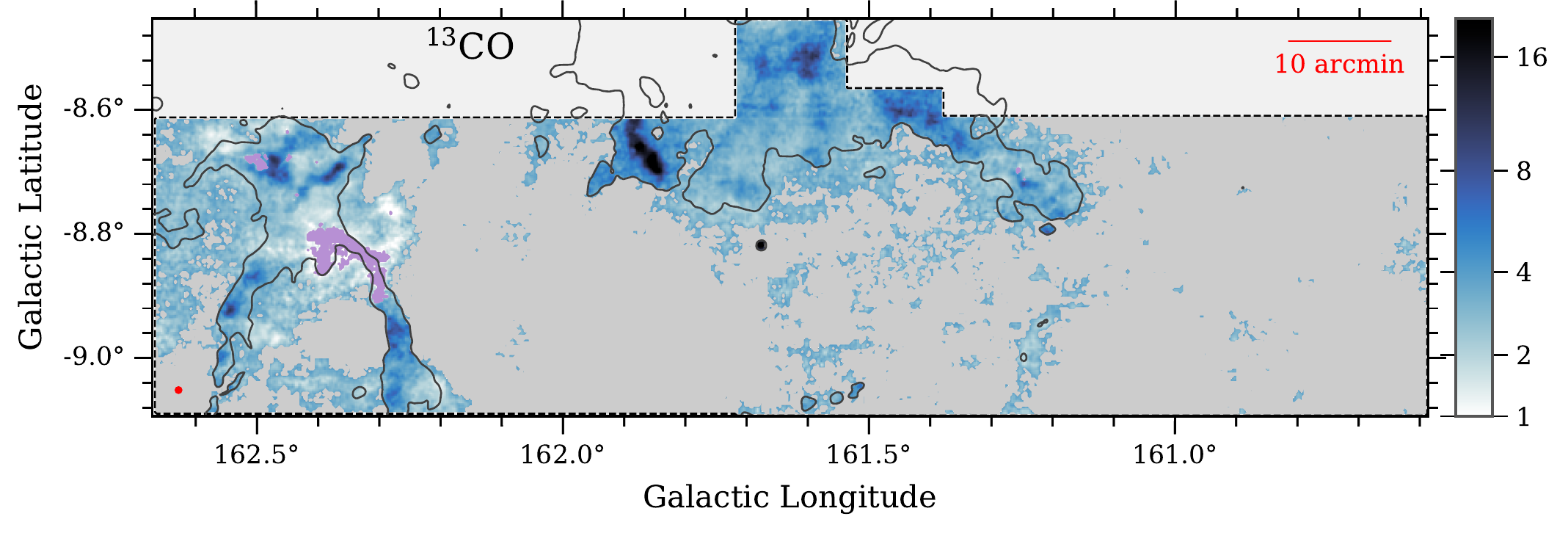}
    \caption{\thirteenco depletion map for CMC-West. Same color scheme as Fig. \ref{fig:13deplkha}. CMC-West was not observed in \ceighteeno.}
    \label{fig:13deplwest}
\end{figure*}

\subsection{Large Scale Depletion Maps}\label{sec:depletionmaps}

\parnum We present the first wide-field, high resolution maps of CO depletion across a significant portion of a single GMC. In figures \ref{fig:13deplkha},\ref{fig:13depl1478}, \& \ref{fig:13deplwest}, we show our cloud scale maps of the \thirteenco and \ceighteeno depletion factors, $\eta_{13}$ and $\eta_{18}$, respectively. The figures show depletion as the colormap with the \av = 5 {\rm mag} contour overlaid. The purple regions blank out  places where column densities were unable to be derived due to the fact that  the \thirteenco peak intensities are larger than those of  \twelveco (see Appendix \ref{sec:methnco}). The \ceighteeno depletion factor maps have been spatially smoothed (for clarity). We measure depletion factors ranging from 0.5 - 25 over the extent of the cloud. Depletion factors with values less than one arise entirely within the area coincident with the \lkhalpha cluster and \ion{H}{2} region.

\parnum The maps show a great deal of spatial variation in the depletion factor. In regions where both \ceighteeno and \thirteenco were observed, the extent of  $\eta_{13}$ is greater than that of $\eta_{18}$ due to stronger \thirteenco lines on average. $\eta_{13}$ also tends to be larger  than $\eta_{18}$ (note the difference in the scale of the colorbar). Both molecules reveal clear peaks in the distribution of depletion factors across the regions. There seems to be a good general, but not perfect,  correspondence between the peaks seen in the \thirteenco and \ceighteeno depletion factors.

\subsection{Depletion cores}\label{sec:depletioncores}

\parnum We perform a search for cores in the depletion map using a dendrogram analysis. A good introduction to the method and its usefulness can be found in \citet{2008ApJ...679.1338R}. In short, dendrograms search for structure from the highest values down, breaking the image in to leaves (independent peaks/structures) and branches (collections of leaves and other branches) forming a tree-like representation of the structures in the data. The leaves form the list of peaks from which we select the cores. We use \texttt{astrodendro}\ \citep{ 2019ascl.soft07016R} for our analysis. We identify cores only in the \thirteenco depletion map because it has better coverage and sensitivity than the \ceighteeno depletion map. Regions with no data, or no column density measurement (i.e.; the pink regions in figures \ref{fig:13deplkha}-\ref{fig:13deplwest}) were excluded. The input parameters for \texttt{astrodendro} are: \texttt{min\_npix}=25 ($2\times$ beam area), \texttt{min\_delta}=0.5, \texttt{min\_value}=1, and $\rm\langle\eta_{13}\rangle$ > 2. This gave an initial list of 398 depletion peaks.

\parnum For a depletion peak to be to be identified as a core it can't be on the map edge, at least 70\% of its pixels must have $\av\geq0.5 \magn$, it must have peak $\eta_{13,peak}>3$, and $\eta_{13,peak} - \langle\eta_{13}\rangle>1$. That final criteria removes sources which don't have a significant local peak. After this filtering 82 sources remained in our list. After examining this list we further manually removed 7 false detections.
Our method is tuned to select objects we can confidently call cores and this exercise comes at the expense of a more complete sample. We identify 75 cores - 28 in the Southeast cloud which is associated with L1482, 20 in L1478, and 27 in West. Our depletion core catalog is presented in Table \ref{tab:coresshort}.

\parnum In Fig. \ref{fig:fullcoremap} we show the contours for the cores overlaid on \herschel extinction, and in Appendix \ref{sec:appcores} we focus on each individual region and show the core's position represented by the best fit ellipse. The best fit ellipse is found using the moment method described in \citet{2006PASP..118..590R} which is implemented in \texttt{astrodendro}.

\begin{deluxetable*}{ccccccccccccc}[t]\label{tab:coresshort}
\tablecaption{Catalog of Depletion Cores in the California Molecular Clouds}
\tablehead{\colhead{Name} & \colhead{Right Ascension} & \colhead{Declination} & \colhead{Radius [pc]} & \colhead{$M_{\rm core}$} & \colhead{$\tdust$} & \colhead{$\ak$} & \colhead{V$_{\rm cen}$} & \colhead{$\sigma_{\rm 13}$} & \colhead{$\sigma_{\rm 18}$} & \colhead{$\langle\eta_{13}\rangle$} & \colhead{$\langle\eta_{18}\rangle$} & \colhead{{\it Zhang et al.}} \\
&  &  & \colhead{pc} & \colhead{$M_\odot$} & \colhead{K} & \colhead{$\magn_{\rm K}$} & \colhead{km/s} & \colhead{km/s} & \colhead{km/s} &  &  & }
\startdata
L1482-014 & 4:31:04.28 & 35:57:40.08 & 0.24 & 30.6 & 14.4 & 0.86 & -0.86 & 0.52 & 0.30 & 3.89 & 5.28 & 188 \\
L1482-015 & 4:30:30.71 & 35:57:08.77 & 0.08 & 6.3 & 13.8 & 1.25 & -0.97 & 0.59 & 0.32 & 3.56 & 1.80 & 192 \\
L1482-016 & 4:30:38.33 & 35:58:29.51 & 0.07 & 5.9 & 13.5 & 1.53 & -1.05 & 0.60 & 0.55 & 4.43 & 2.21 & 185 \\
L1482-017 & 4:30:42.69 & 36:00:17.49 & 0.08 & 10.1 & 13.4 & 1.97 & -1.11 & 0.53 & 0.38 & 6.80 & 3.72 & 183 \\
L1482-018 & 4:30:15.92 & 36:00:15.72 & 0.10 & 13.0 & 13.3 & 1.96 & -0.70 & 0.62 & 0.36 & 5.57 & 3.03 & 181
\enddata
\tablecomments{Abbreviated catalog of CMC depletion cores. The full catalog can be found at the end of this paper in Table \ref{tab:cores}. The radius is the beam-deconvolved radius of the source, $r_d$, as defined in the text. \tdust and \ak are the average \herschel dust temperature and extinction respectively. V$_{\rm cen}$ is the central velocity of the average \thirteenco line in the core. The last column is the core number for the best-matched dust core from Zhang et al. (2018, Table 3)}
\end{deluxetable*}

\begin{figure*}
    \centering
    \includegraphics[width=2\columnwidth]{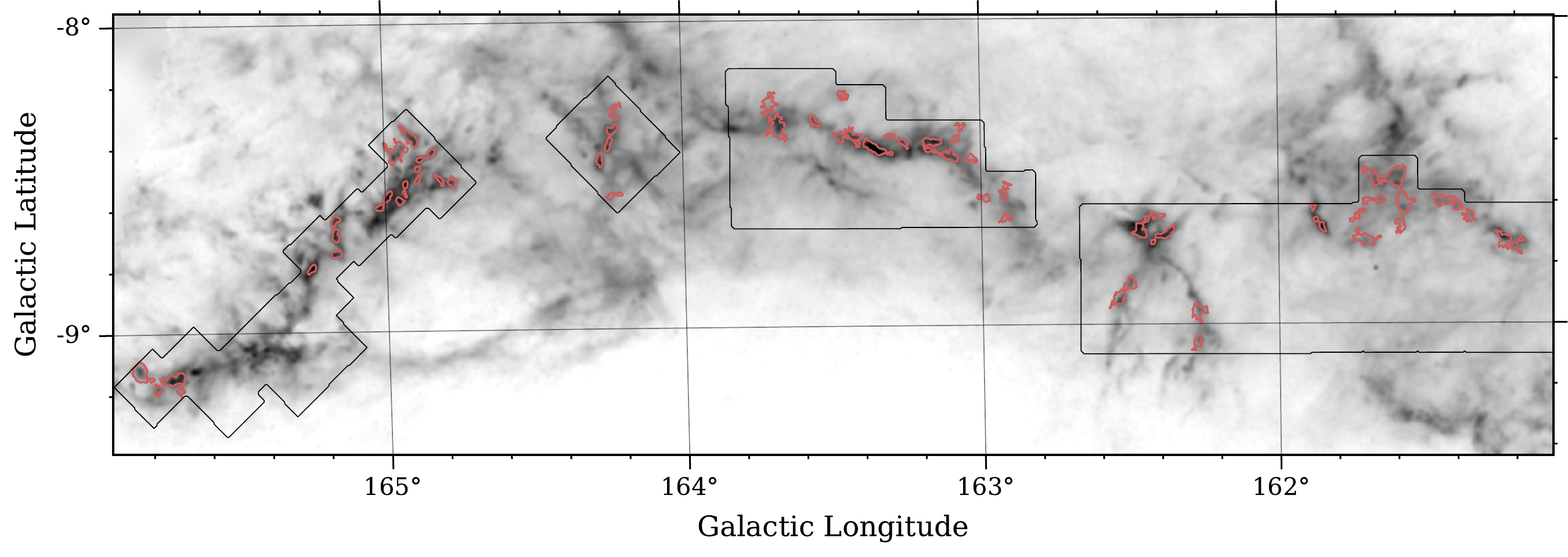}
    \caption{The core contours produced by \texttt{astrodendro} for our catalog of cores overlaid on the \herschel dust map in grayscale. For more detail, see the plots of each region in Appendix \ref{sec:appcores}}
    \label{fig:fullcoremap}
\end{figure*}

\subsubsection{Core Properties}

\begin{figure}
    \centering
    \includegraphics[width=\columnwidth]{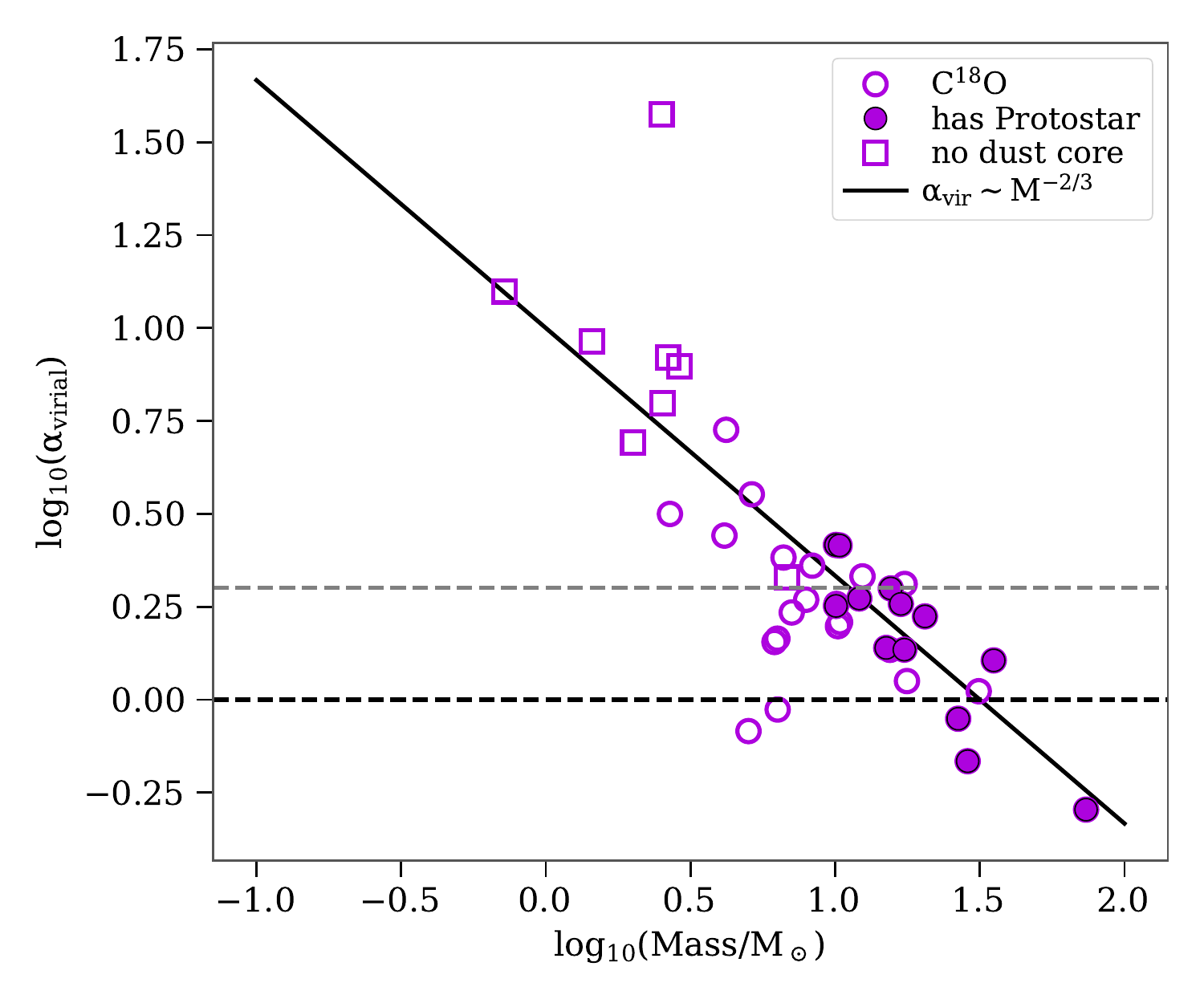}
    \caption{\label{fig:virial} The log of the virial parameter, $\alpha_{\rm virial}$ plotted against log-mass for \ceighteeno depletion cores. Open and filled are are the starless and protostellar cores respectively. The squares indicate depletion cores which are not associated with a dust core in \cite{2018A&A...620A.163Z}. The black solid line shows the $\alpha \sim M^{-2/3}$ relation from  \citep{1992ApJ...395..140B} for pressure bound cores. The black dashed line is $\alpha_{\rm virial}=1$ (cores below this are virialized), and the gray dashed line is $\alpha_{\rm virial}=2$ (cores below this are bound). The $\sim 1/3$ of cores which are bound are also the most massive.}

\end{figure}
\parnum We derived a suite of properties for the cores and list them in Table \ref{tab:coresshort}.
 The core mass is defined as
$$M_{core} = \int \Sigma_{gas}\, {\rm d}S,$$
where $\Sigma_{gas} = 183\,\ak\  \msunpc\,\magn_{\rm K}^{-1} $. To account for the unrelated mass contributed by the large scale structure in the cloud, we subtract the 1 pc-scale structure from our extinction map using a python re-implementation of the \textsc{FINDBACK} algorithm from the \textsc{CUPID} package \citep{2007ASPC..376..425B}, and integrate over the background-subtracted map to measure the core mass. The derived masses range from 0.38 - 73 $\rm M_\odot$ with a mean of 10 \msun. The core radius is simply $r=\sqrt{\rm Area/\pi}$, and the radius deconvolved with the telescope beam is
$$r_d = \sqrt{(\text{Area}/\pi) - \left({\rm FWHM}_{\rm beam}/{2}\right)^2}.$$
We use $r_d$ as the core radius for our analysis.
The derived radii range from 0.04 - 0.25 pc (0.04 pc corresponds to the physical radius of a core with our minimum allowed area). From these we estimate the core number density  $n= 3 M_{core}/(4\pi r_d^3 m_H)$. The density spans ${10^3} - 10^5{\rm\ cm^{-3}}$ with $\langle\log n\rangle = 4.2$. With a mean dust temperature in the cores of  $\langle \tdust \rangle \sim$ 15 K, we expect to see the onset of significant depletion (\fdep$\sim$ 5) around $n\sim 10^4{\rm\ cm^{-3}}$ \citep{1995ApJ...441..222B}, consistent with the mean density in the cores. We also calculate the average spectrum for each core to measure the velocity dispersion using the $2^{\rm nd}$ moment of the spectrum, as described in \S \ref{sec:momentmaps}. Since the cores are defined to be peaks in the depletion maps and most of the observed CO emission likely arises from outer undepleted layers of cloud (Fig. \ref{fig:cartoon}), the velocity dispersion we derive is likely an upper limit to the true velocity dispersion within the core.

\parnum We compare the position of the cores with the YSO catalog from  \citet{2017A&A...606A.100L} , which merged  \citet{2014ApJ...786...37B} and \citet{2013ApJ...764..133H} and and updated the YSO classifications, to identify which cores contain a protostar (Class 0/I) or a disk (Class II). The protostars are shown alongside the cores in Figs. \ref{fig:coremaps}-\ref{fig:coremaps2}. We find that  16/75 of the cores are associated with a protostar (14) or disk (6). Several cores are associated with multiple YSOs. We consider cores with a Class 0/I source a protostellar core, and find that a large fraction of the cores are starless. The protostellar cores tend to be more massive have slightly higher densities than the starless cores, otherwise their dust temperature and size cover a fairly similar range as the starless cores considering only 20\% of our cores are protostellar.

\parnum  As our depletion cores are generally well correlated with extinction peaks (Figs. \ref{fig:fullcoremap}, \ref{fig:coremaps}-\ref{fig:coremaps2}), it is of interest to compare our depletion cores with the \herschel dust core catalog of \citet{2018A&A...620A.163Z}. Out of 300 dust cores in that catalog, 180 are contained within our survey boundaries. We match objects in the two lists using \texttt{skyellipse} cross-matching in \texttt{TOPCAT} \citep{2005ASPC..347...29T} and found that roughly half (i.e., 48/75) of our depletion cores are associated with dust cores identified by Zhang et al. Only 26\% (48/180) of the Zhang et al. dust cores match with our depletion cores. We list the matched dust cores in Table \ref{tab:coresshort}. We found that within our matched sources, the depletion core mass and radius are larger than the cross-matched dust core mass and radius. This is likely because our core identification method can produce larger cores since it tends to merge any adjacent structures that are not sufficiently higher than the background. Unmatched depletion cores are less massive and warmer than depletion cores with matches despite maintaining similar levels of measured depletion. None of the unmatched depletion cores are associated with protostars.

\parnum Some insight into the nature of the depletion cores might be gained from a virial analysis using \ceighteeno to estimate the velocity dispersion within the cores. However, because these lines arise in the outer layers of the cores (see Fig. \ref{fig:cartoon}), estimates of the velocity dispersions are necessarily upper limits. Nonetheless, we use \ceighteeno because it probes deeper layers of the clouds than \thirteenco and is optically thin over the entire area it was observed. Approximately 1/2 of the \thirteenco depletion cores have been observed in \ceighteeno. We perform a simple virial analysis to examine the boundedness of these cores recognizing that the velocity dispersions we use and the virial parameters we derive are only upper limits to the true values in the depleted cores. We compute the virial parameter as %

\begin{equation}\label{eq:virial}
    \alpha_{virial} \leq \frac{5 \sigma^2 r_d}{{\rm G} {\rm M_{core}}}
\end{equation}
where $r_d$ is the deconvolved radius, $\rm M_{core}$ is the core mass, and $\sigma$ is the 1D velocity dispersion.
The velocity dispersion $\sigma$ is  the combination of the thermal and turbulent velocity dispersions, $\sigma^2 = c_s^2 + \sigma_{NT}^2$, where $c_s^2$ is the sound speed or thermal velocity dispersion (\citep{1992ApJ...395..140B}). Using the dust temperature as a proxy for the gas temperature, the velocity dispersion is ,
\begin{equation}
\sigma^2= \frac{k \tdust}{\mu m_{\rm H}} +
    \sigma_{\rm 18}^2 -  \frac{k \tdust}{m_{18}} ,
\end{equation}
where $\mu = 2.33$ is the mean molecular weight corrected for helium, $m_{18}$ is the mass of the \ceighteeno molecule, and $\sigma_{\rm 18}$ is the \ceighteeno velocity dispersion. Like the velocity dispersion, \tdust is also an upper limit as it includes warmer material along the line of sight. If we assumed a central core temperature of $\tdust\sim 10{\,\rm K}$ (\citet{2014A&A...562A.138R} measured 9.3 K for B68), instead of using the mean dust temperature that we measure, $\sigma^2$ would decrease on average by $\sim$10\%. We plot the \ceighteeno virial parameter against mass in Fig. \ref{fig:virial}. The filled and open symbols correspond to depletion cores with and without protostars respectively. The squares corresponds to depletion cores which do not have a match in \citeauthor{2018A&A...620A.163Z}. The black dashed horizontal line is $\alpha_{\rm virial} = 1$ and the gray dashed line is $\alpha_{\rm virial}=2$. Cores are virialized if $\alpha_{\rm virial}<1$ and bound for $\alpha_{\rm virial}<2$. Approximately 1/2 (23/41) of the \ceighteeno cores are gravitationally bound  ($\alpha_{18}<2$) including almost all the protostellar cores. We can derive a lower limit for the virial parameter by assuming that a core's non-thermal linewidth is $\sim$sonic towards its center. Thus the the velocity dispersion that goes into (\ref{eq:virial}) is $\sigma^2 \approx 2 c_{s}^2$. With these assumptions, we find that all of the cores are bound with $\alpha_{\rm virial, sonic} < 1$.

\parnum
The black solid line in Fig. \ref{fig:virial} shows the predicted relationship  (i.e., $\alpha \sim M^{-2/3}$) between mass and $\alpha_{virial}$ for {\it virialized} clouds when the surface (i.e., external pressure) terms are included in the virial equation  (\citet{1992ApJ...395..140B}).
Similar behavior in this relation has been reported for \ceighteeno and NH$_3$ observations of the dense core populations in the Pipe (\citet{2008ApJ...672..410L}, and Orion A  (\citet{2017ApJ...846..144K}) molecular clouds. We can infer from the behavior of the cores' virial parameters that, even if some are not gravitationally bound, the depletion cores in the CMC are likely pressure confined. Since the virial parameter calculated here is an upper limit, it is likely that more of the cores will be shown to be bound if an undepleted species (e.g., N$_2$H$+$) is used to trace the very dense gas in the central regions of the cores. Future observations of additional molecular tracers (e.g., NH$_3$, N$_2$H$+$, HCN, etc.) with higher angular resolution are desirable to obtain  a deeper and more complete understanding of the chemistry and kinematics of this interesting population of cores.

\section{Summary \& Conclusions}\label{sec:summary}

\parnum In this paper we investigate the relationship between molecular gas and dust  over an unprecedented dynamic range of cloud depth (A$_V$ = 3 -- 60 magnitudes) within a single giant molecular cloud. We build on and significantly extend the earlier study of the California Molecular Cloud by \citet{2015ApJ...805...58K} by acquiring extremely deep measurements of dust extinction toward the cloud using the \herschel satellite and by enlarging, by a factor of three, the area (1 sq. deg.) of the cloud surveyed on sub-parsec spatial scales in  \twelveco, \thirteenco, and \ceighteeno J=2-1 lines using the {\sl Heinrich Hertz Submillimeter Telescope}. We directly compare CO integrated intensities with extinction to derive the CO conversion or X-factors for each isotopologue across the cloud on sub-parsec spatial scales. We derive LTE \thirteenco and \ceighteeno column densities and compare them with extinction measurements to derive the abundances of these two CO isotopologues. We compare these results to the dust temperature distribution to investigate CO depletion. We summarize our main results as follows:

\begin{enumerate}[itemsep=1pt,leftmargin=*]
\item No single \twelveco X-factor can describe all the gas in the cloud, confirming the earlier study of \citetalias{2015ApJ...805...58K}. We find the \twelveco X-factor to vary by two orders of magnitude through the cloud becoming essentially infinite at high extinctions in the regions with the coldest dust. We find the variations to be both spatially and temperature dependent.

\item Similar to \citetalias{2015ApJ...805...58K} we find that in the hot dust (\tdust>25 K) dust  we are able to measure single-valued \mbox{CO 2-1} X-factors for \twelveco, \thirteenco, and \ceighteeno that are valid to $\av \sim 20\ \magn$, well beyond the point were \twelveco is expected to saturate. This hot dust is confined to pixels that are coincident with the \ion{H}{2} region associated with the \lkhalpha cluster and is likely heated by the exciting stars in the cluster. We assume that all the CO in this region is in the gas phase and therefore we adopt the following conversion factors for regions with undepleted CO:
\begin{align*}
{\rm X}_{12,2-1} &= (1.28\pm0.36)\e{20}\ \xfunit \\
{\rm X}_{13,2-1} &= (5.74\pm4.30)\e{20}\ \xfunit \\
{\rm X}_{18,2-1} &= (4.70\pm2.05)\e{21}\ \xfunit
\end{align*}
In many applications it is more convenient to express the CO conversion factor with respect to cloud mass instead of column density. The corresponding values of $\alpha_{\rm CO(2-1)}$\footnote{$\alpha_{\rm CO} = 2\mu\xco $, converted to units of $\msunpc (\wunit)^{-1}$, such that $2\e{20}\ \xfunit \to 4.39\  \msunpc (\wunit)^{-1}$} are listed below:
\begin{align*}
{\rm \alpha}_{\rm 12,2-1} &=  2.81\ \msunpc (\wunit)^{-1}\\
{\rm \alpha}_{\rm 13,2-1} &=  12.6\ \msunpc (\wunit)^{-1} \\
{\rm \alpha}_{\rm 18,2-1} &=  103\ \msunpc (\wunit)^{-1}
\end{align*}
For comparison, the values of \xco and $\alpha_{\rm CO}$ for the J=1-0 transition of \twelveco  are $2\e{20}\ \xfunit $ and $ 4.39\  \msunpc (\wunit)^{-1}$, respectively.
\item The \thirteenco and \ceighteeno abundances are found to vary with dust temperature. At high dust temperatures (\tdust $\gtrsim$ 18 K) the abundance is approximately constant while for lower temperatures the abundance decreases with decreasing temperature.
By 25 K all of the CO is in the gas phase, and so we measure the total CO abundance using only the pixels with \tdust > 25 K:
\begin{align*}
\thirteenco:\htwo &= (1.31\pm.89) \E{-6} \\
\ceighteeno:\htwo &= (1.27\pm.59) \E{-7}
\end{align*}
Both measurements are in reasonable agreement (within $1\sigma$) with the abundances derived from cosmic abundance ratios.

\item We conclude that the variations in gas abundances (and \xco) with dust temperature are due to depletion and desorption processes occurring in the cloud. When we correct for the expected contribution of ices to the total (ice+gas) CO column density we recover the CO gas phase column density consistent with the abundances we measure in the undepleted gas. This suggests that total (gas+ice) CO abundance in the CMC is relatively constant across the cloud.

\item Combining the CO and extinction measurements we measure depletion factors for both \thirteenco and \ceighteeno and construct the first maps of CO depletion across the extent of an entire GMC.

\item The depletion maps show structure whose peaks we identify as depletion cores. In most cases these correspond to cores in the extinction maps, although the boundaries of the depletion cores are more clearly defined than those of the extinction cores. We produce a catalog of these cores and list their basic properties. The cores range in mass from 0.38 - 73 \msun, with a characteristic mass of 10 \msun. Their sizes range between 0.04 and 0.25 pc.
These depletion defined cores show evidence of being pressure confined. Those that contain protostars appear to be gravitational bound.

\end{enumerate}

\acknowledgements
We gratefully acknowledge Shuo Kong and Liz Lada for assistance in the data acquisition and Tom Dame for useful discussions. We thank the anonymous reviewer for comments which improved the content of this paper. We used data from the Heinrich Hertz Submillimeter Telescope which is operated by the Arizona Radio Observatory as part of the Steward Observatory at the University of Arizona. This research made use of astrodendro, a Python package to compute dendrograms of Astronomical data (http://www.dendrograms.org/). ; Astropy a community-developed core Python package for Astronomy \citep{ 2013A&A...558A..33A,  2018AJ....156..123A}; APLpy, an open-source plotting package for Python (Robitaille and Bressert, 2012; Robitaille, 2019).

\facilities{Heinrich Hertz Submillimeter Telescope, Herschel Space Observatory}

\software{Astropy \citep{2013A&A...558A..33A, 2018AJ....156..123A}, GILDAS/CLASS \citep{ 2005sf2a.conf..721P, 2018ssdd.confE..11P, 2013ascl.soft05010G}, CUPID package \citep{2007ASPC..376..425B}, MIRIAD \citep{ 1995ASPC...77..433S}, TOPCAT \citep{2005ASPC..347...29T} }

\clearpage
\appendix

\section{Integrated intensity maps\label{sec:appwco}}
Here we show the maps of integrated intensity for \twelveco (2-1), \thirteenco (2-1), and \ceighteeno (2-1) for the California Molecular Cloud. In the following maps integrated intensity is displayed as grayscale with contours. The black dashed line shows the survey boundary. \ceighteeno was only observed in L1482 and L1478 regions. We see that the integrated intensities of the various CO isotopologues show similarity in structure. In each map the beam size is shown as a red circle. We estimate the noise in our integrated intensity maps, by taking the rms of all the pixels outside the integration mask. This gives a map of noise across the observed area. The average rms scaled to the width of the largest window for each line is $0.25\ \wunit$, $0.20\ \wunit$, and $0.19\ \wunit$ for \twelveco (2-1), \thirteenco (2-1), and \ceighteeno (2-1) respectively.

\begin{figure}[ht]
    \centering
    \includegraphics[height=2.5in]{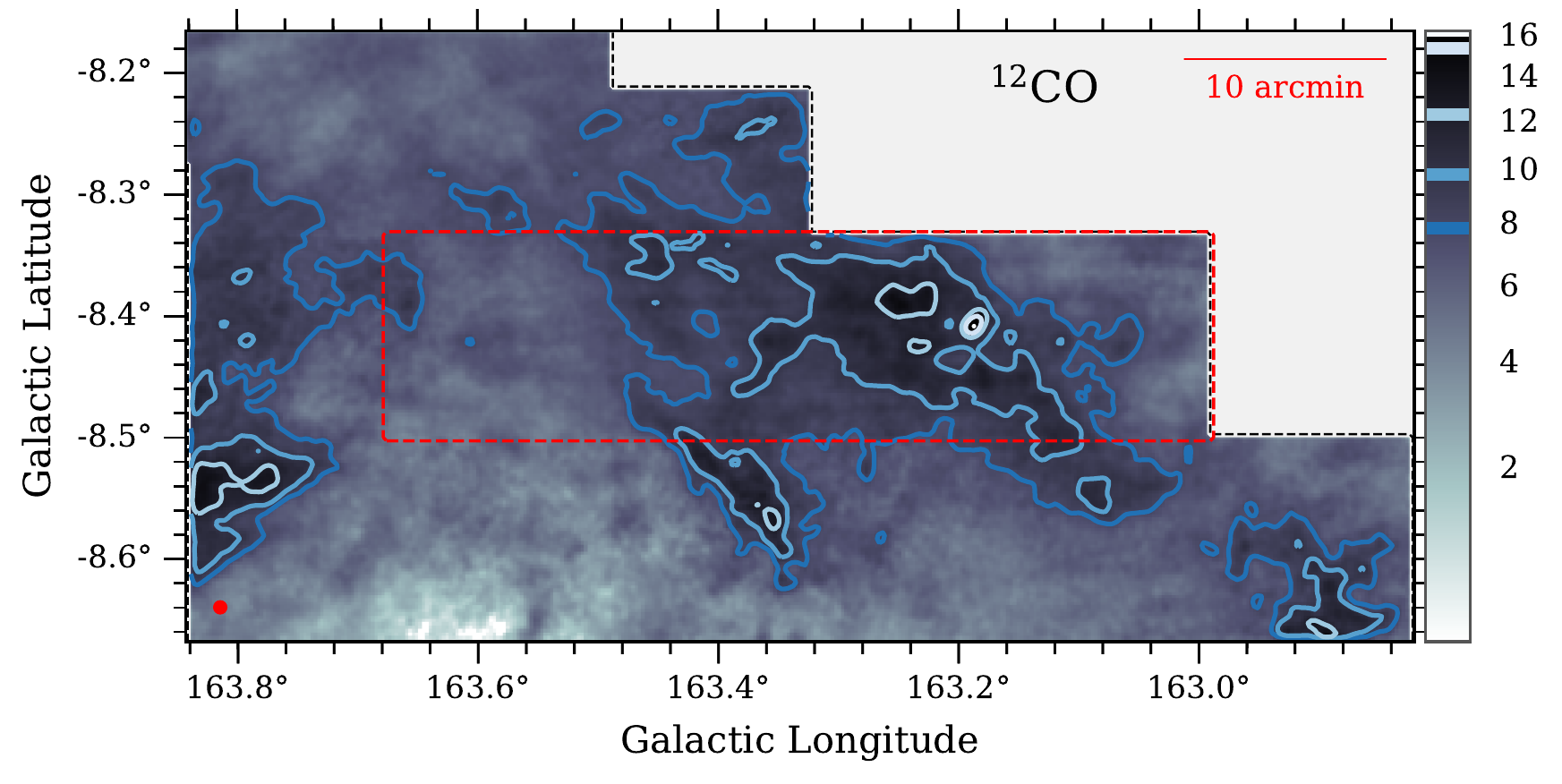}
    \includegraphics[height=2.5in]{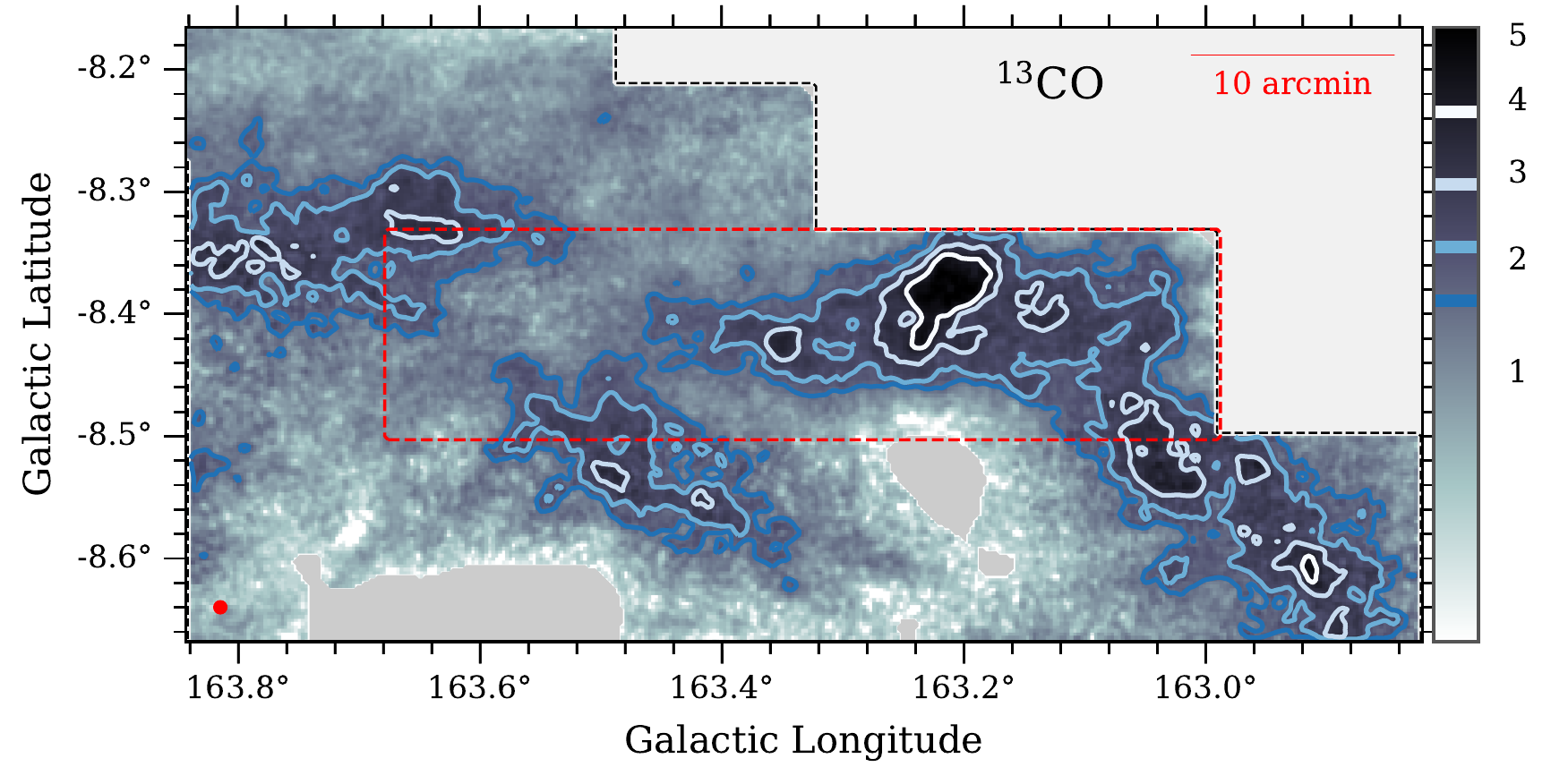}
    \includegraphics[height=2.5in]{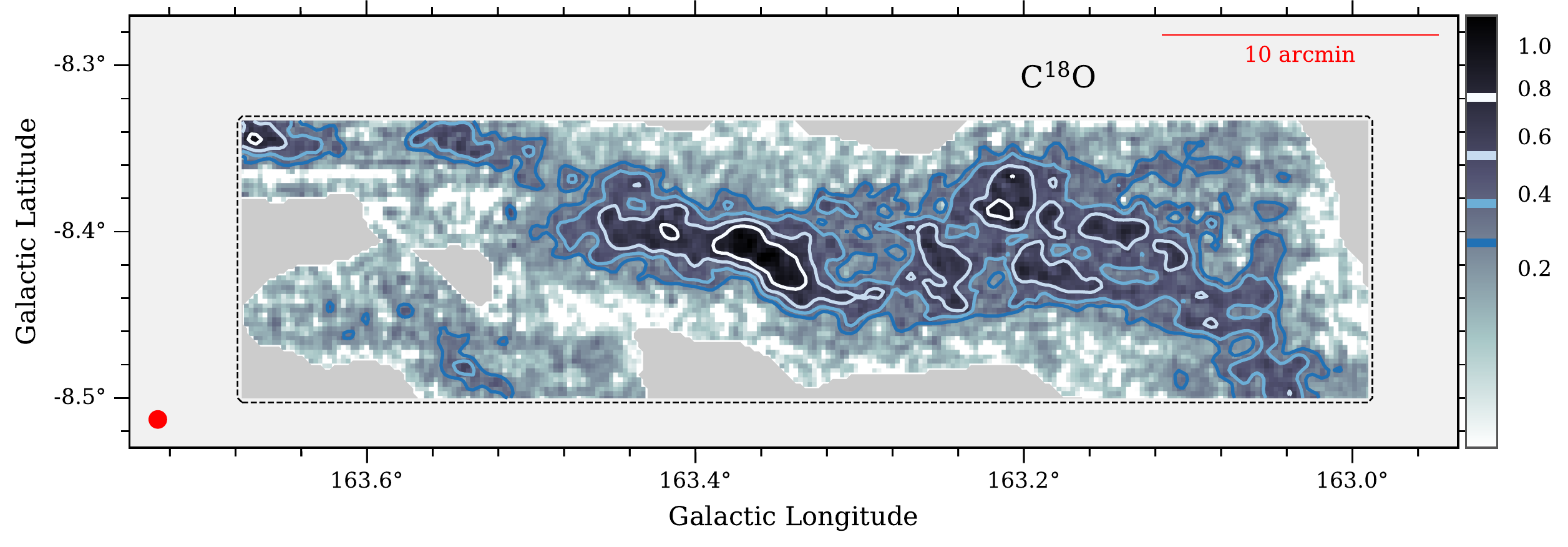}
    \caption{CO integrated intensity maps (Top:\twelveco, Middle:\thirteenco, Bottom:\ceighteeno) for the L1478 region in the CMC. The black dashed line indicates the survey boundaries, and the red dashed line visible in the \twelveco and \thirteenco plots indicates the extent of the \ceighteeno survey. The survey resolution (38\arcsec) is shown as a red circle in the lower left hand corner. The grayscale is the total integrated intensity in K km/s. The contours from light to dark are for - \twelveco: (6.8, 8.3, 9.6 12.0, 14.9) K km/s; \thirteenco: (1.1, 1.8, 2.5, 3.5, 5) K km/s;  \ceighteeno (0.16, 0.31, 0.47, 0.74, 1.0) K km/s. }
    \label{fig:wcol1478}
\end{figure}

\begin{figure}[ht]
    \centering
    \includegraphics[height=2.5in]{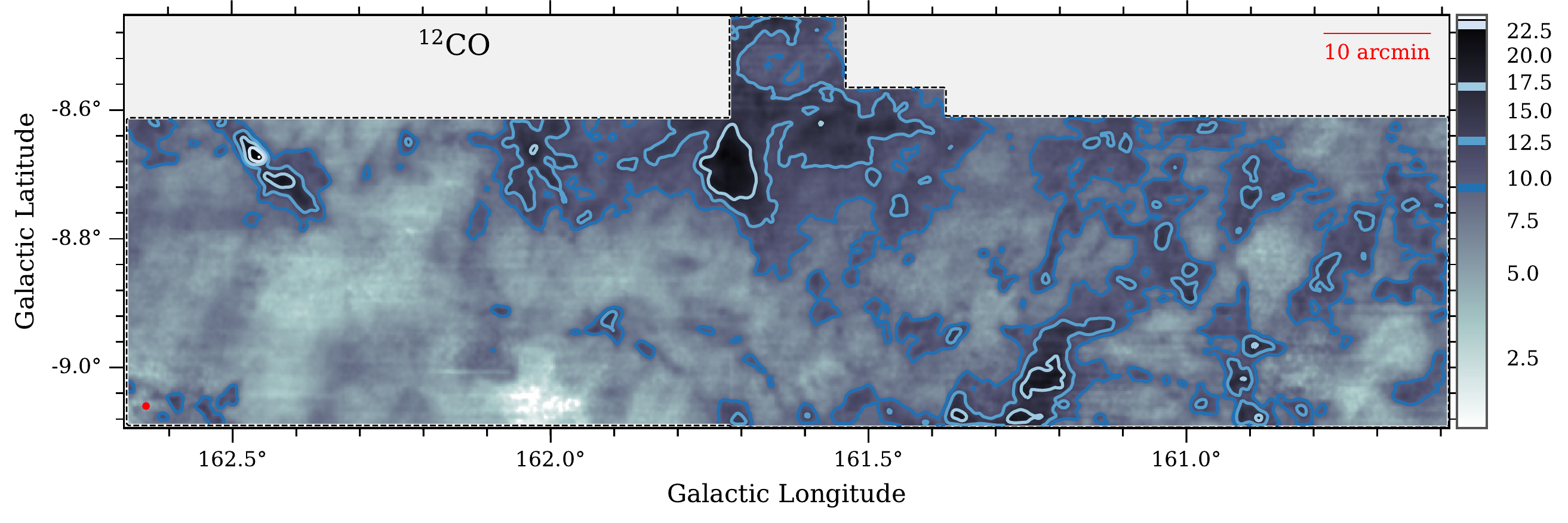}
    \includegraphics[height=2.5in]{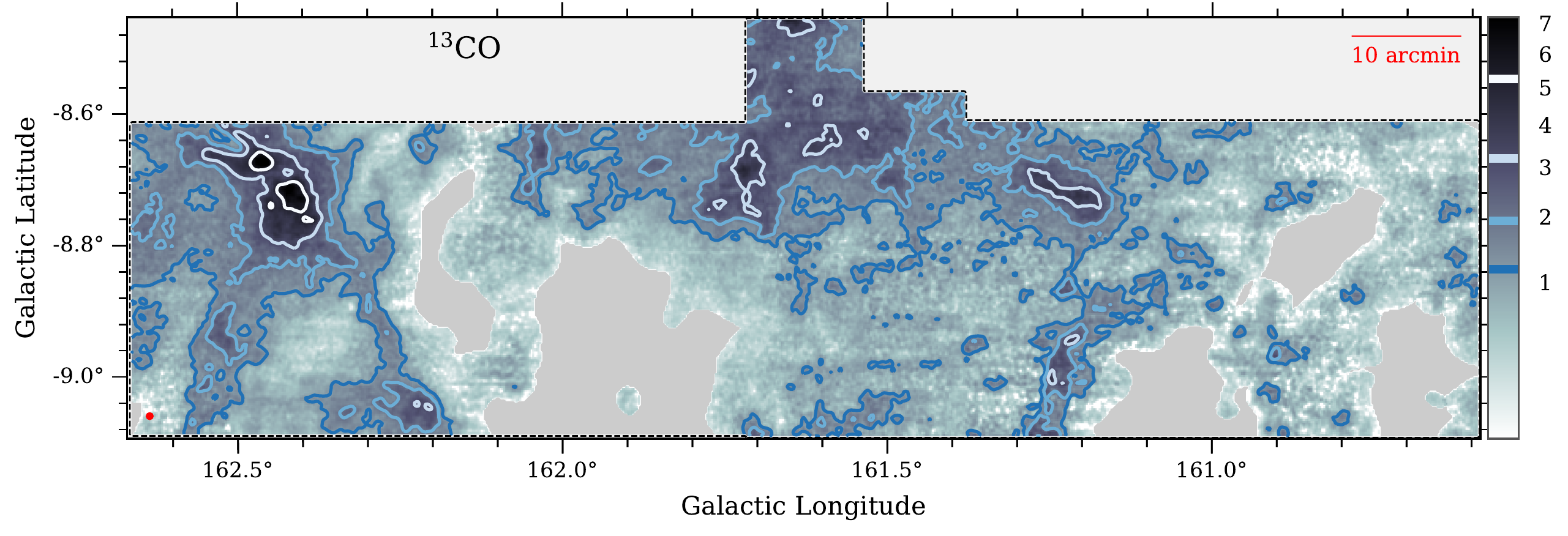}
    \caption{CO integrated intensity maps (Top:\twelveco, Bottom:\thirteenco) for the West region in the CMC. The black dashed line indicates the survey boundaries. West was not observed in \ceighteeno. The survey resolution (38\arcsec) is shown as a red circle in the lower left hand corner. The grayscale is the total integrated intensity in K km/s. The contours from light to dark are for - \twelveco: (7.9, 10.1, 12.2, 17.2, 23.1) K km/s; \thirteenco: (0.8, 1.4, 2.1, 3.7, 6.8 ) K km/s.}
    \label{fig:wcowest}
\end{figure}

\begin{figure}[ht]
    \includegraphics[height=4.25in]{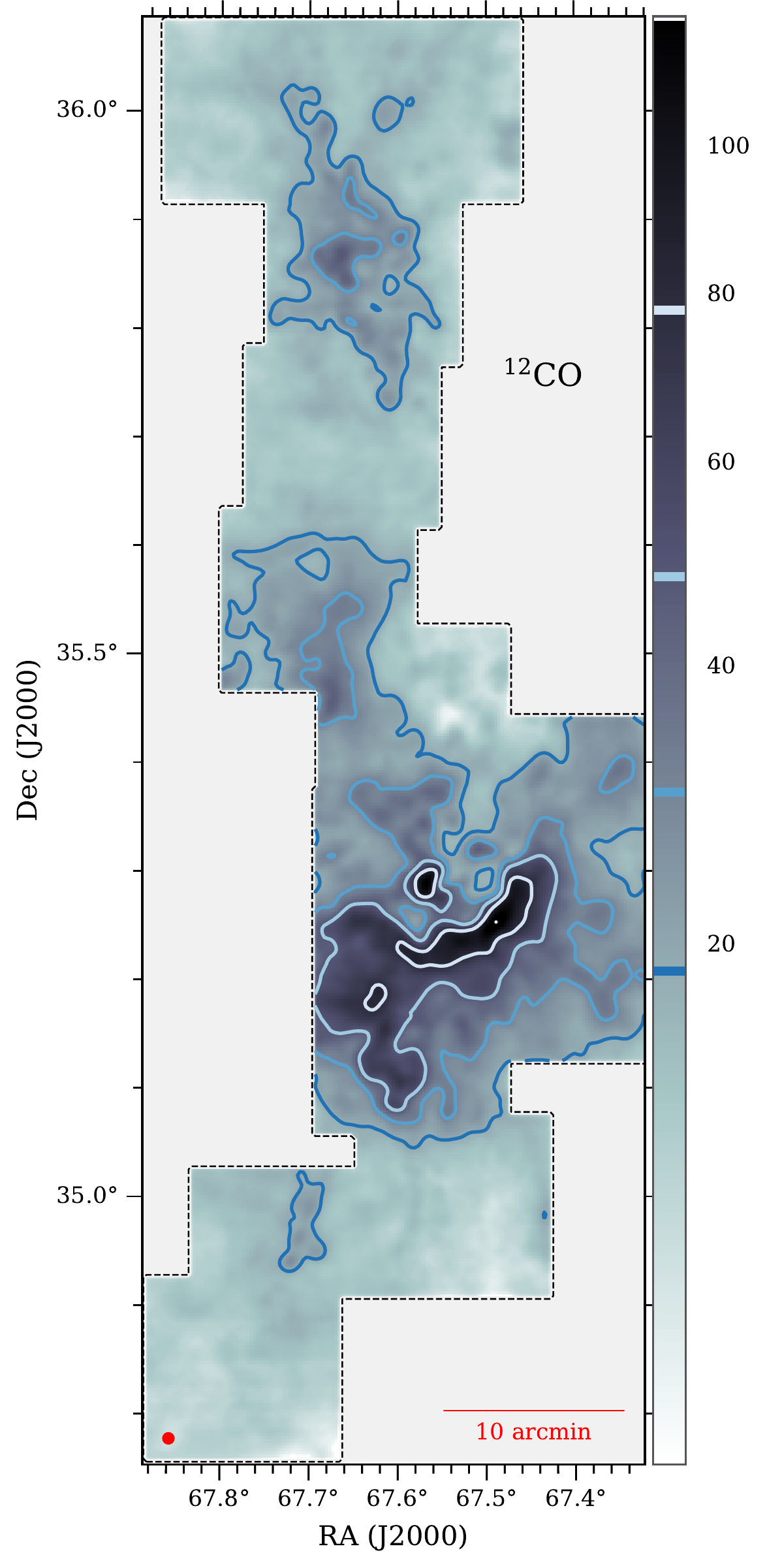}
    \includegraphics[height=4.25in]{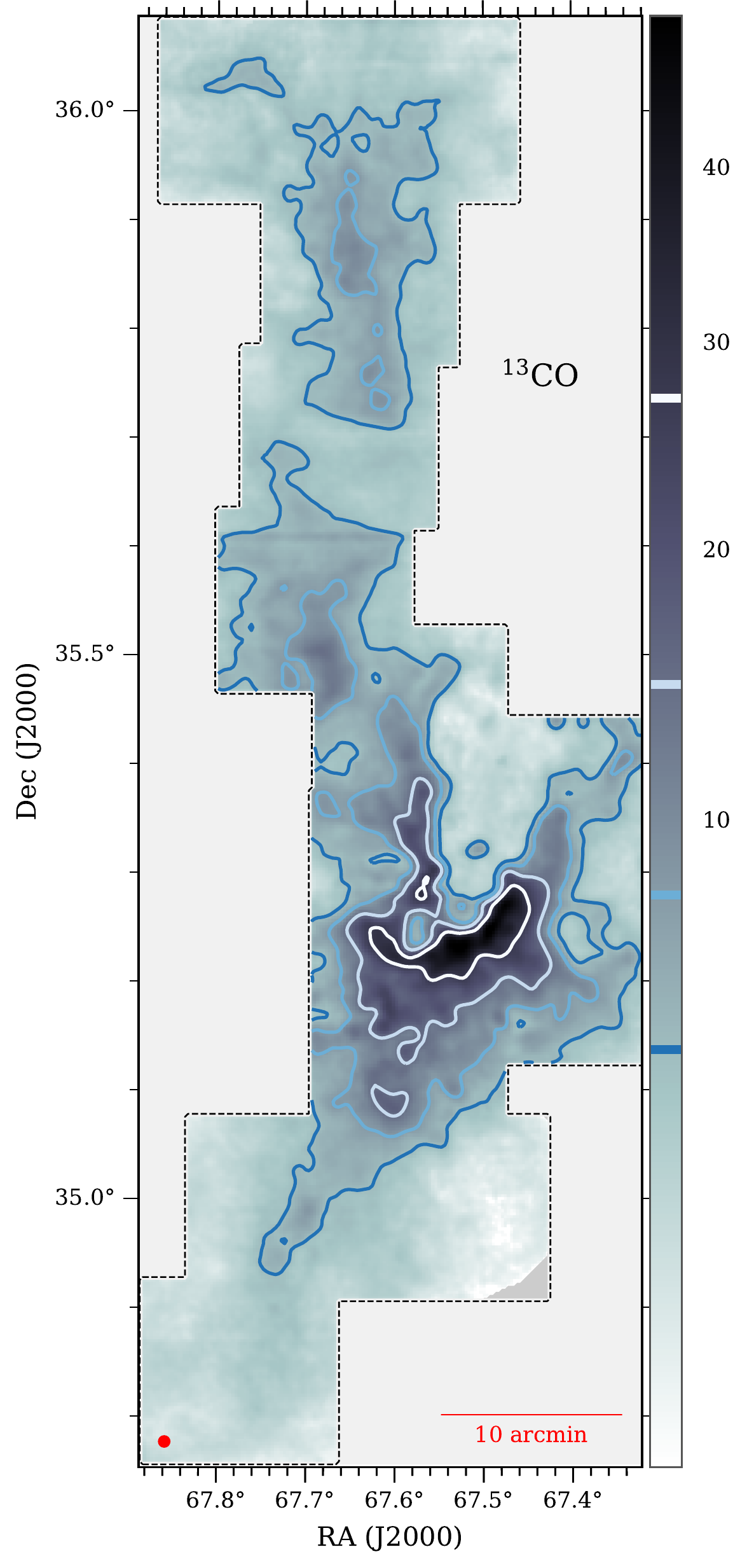}
    \includegraphics[height=4.25in]{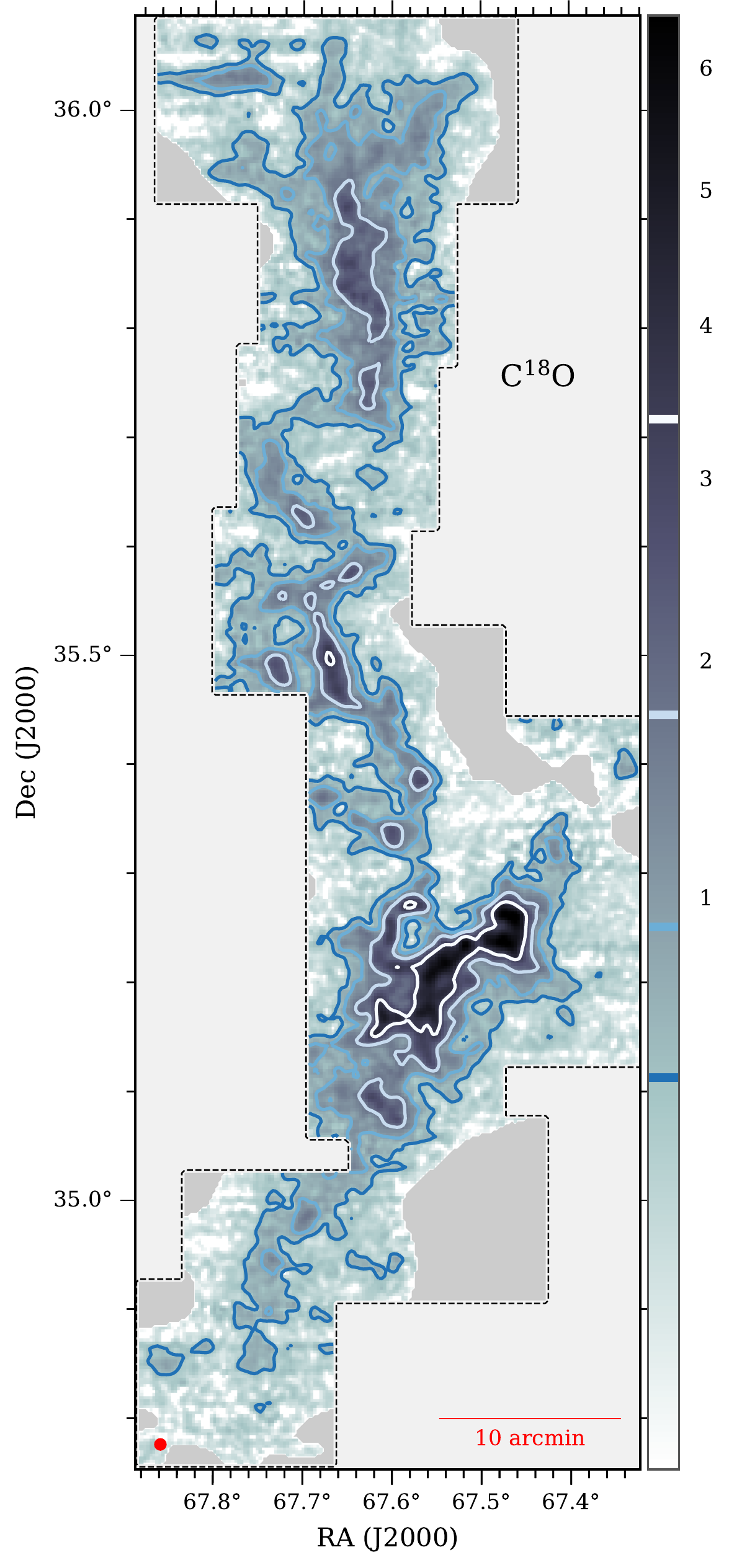}
    \caption{CO integrated intensity maps (Left:\twelveco, Middle:\thirteenco, Right:\ceighteeno) for the L1482 in the CMC. The black dashed line indicates the survey boundaries. The survey resolution (38\arcsec) is shown as a red circle in the lower left hand corner. The grayscale is the total integrated intensity in K km/s. The contours from light to dark are for - \twelveco: (14.1, 21.3, 32.8, 48.5, 75.7, 112.6) K km/s; \thirteenco:(2.9, 4.9, 8.7, 13.6, 27.2, 46.1) K km/s;  \ceighteeno (0.3, 0.6, 1.15, 1.7, 3.5, 5.8) K km/s.}
    \label{fig:wcolkha}
\end{figure}

\clearpage

\section{CO Column Density}\label{sec:methnco} \label{sec:appnco}

\parnum  We follow \cite{2015ApJ...805...58K} and  \cite{2010ApJ...721..686P}\footnote{We apply the correction to their equation 18 as noted in \cite{2013MNRAS.431.1296R}} to calculate the total column densities of \thirteenco and \ceighteeno assuming {\it local thermodynamic equilibrium} (LTE). In LTE, the main beam brightness temperature ($T_{\rm mb}$) is related to optical depth and excitation temperature (\tex) by
\begin{equation} \label{eq:Tmb}
    T_{\rm mb} = \left(\frac{h\nu/k}{e^{h\nu/k\tex}-1} - \frac{h\nu/k}{e^{h\nu/k T_{\rm bg}}-1}\right) (1 - e^{-\tau})
\end{equation}
Under LTE, the excitation temperature can be derived from the peak temperature of an optically thick line. We use the peak main beam temperature of \twelveco (2-1), $T_{\rm mb, p}$, as the line is optically thick throughout California.
\begin{equation} \label{eq:texeq}
 \tex = \frac{h\nu/k}{\ln{\!\left(1 + \frac{h\nu/k}{T_{\rm mb, p}+T_{\rm bg}}\right)}}
\end{equation}
where $T_{\rm bg} = 0.19~{\rm K}$ is the Rayleigh-Jeans equivalent temperature of the cosmic microwave background at $\nu_{\twelveco(2-1)} = 230.538~{\rm GHz}$. The excitation temperature varies from 4 - 50 K, with the distribution peaking around 7-10 K. As in \citetalias{2015ApJ...805...58K}, the highest temperatures are found in the \ion{H}{2} region around \lkhalpha. Additionally we find that the distribution of \tex in L1482 is generally hotter than in L1478 and West which are far removed from the cluster and have less star formation \citep{2017A&A...606A.100L}. \tex is assumed to be the same for all isotopologues. The optical depth $\tau$ of \thirteenco (2-1) and \ceighteeno (2-1) is,

\begin{equation}\label{eq:tau}
 \tau = - \ln{\left[1 - \frac{T_{\rm mb}}{h\nu/k}\left({\left( e^{\frac{h\nu}{k \tex}}-1 \right)^{-1} - \left( e^{\frac{h\nu}{k T_{\rm bg}}} -1 \right)^{-1}  }\right)^{-1} \right]}.
\end{equation}
This equation is only real valued when the argument of the logarithm is positive which corresponds approximately to when the \thirteenco or \ceighteeno brightness temperature ($T_{\rm mb}$) is less than the \twelveco peak brightness temperature. This is violated in some regions where \twelveco is self-absorbed. We mask these regions out of our analyses. Finally, the CO column density is given by,
\begin{equation}
\label{eq:columndensity}
N[\cdot]=C(\tex) \times \frac{\int \tau({\rm v})\ {\rm dv}}{\int [1-e^{-\tau({\rm v})}]\ {\rm d v}}\int\! T_{\rm mb}\ {\rm dv}~(cm^{-2})
\end{equation}
where, $T_{\rm mb}$ is the main beam brightness temperature of \thirteenco or \ceighteeno, $\nu$ is the frequency of the transition for which we want the column density, $C(\tex)$ is a function of the excitation temperature,

\begin{equation}\label{eq:ctex}
C(\tex) = \frac{8\pi k\nu^2}{hc^3A_{2\rightarrow1}}\frac{e^{\frac{h\nu}{k T_{\rm bg}}}-1}{e^{\frac{h\nu}{k T_{\rm bg}}}-e^{\frac{h\nu}{k\tex}}}\frac{Q}{(2J+1)e^{\frac{-hB_0J(J+1)}{k\tex}}},
\end{equation}

\begin{equation*}
Q \equiv \sum_J (2J+1) e^{\frac{-h B_0 J(J+1)}{k \tex}},
\end{equation*}

\begin{figure}[ht]
    \centering
    \includegraphics[height=2.5in]{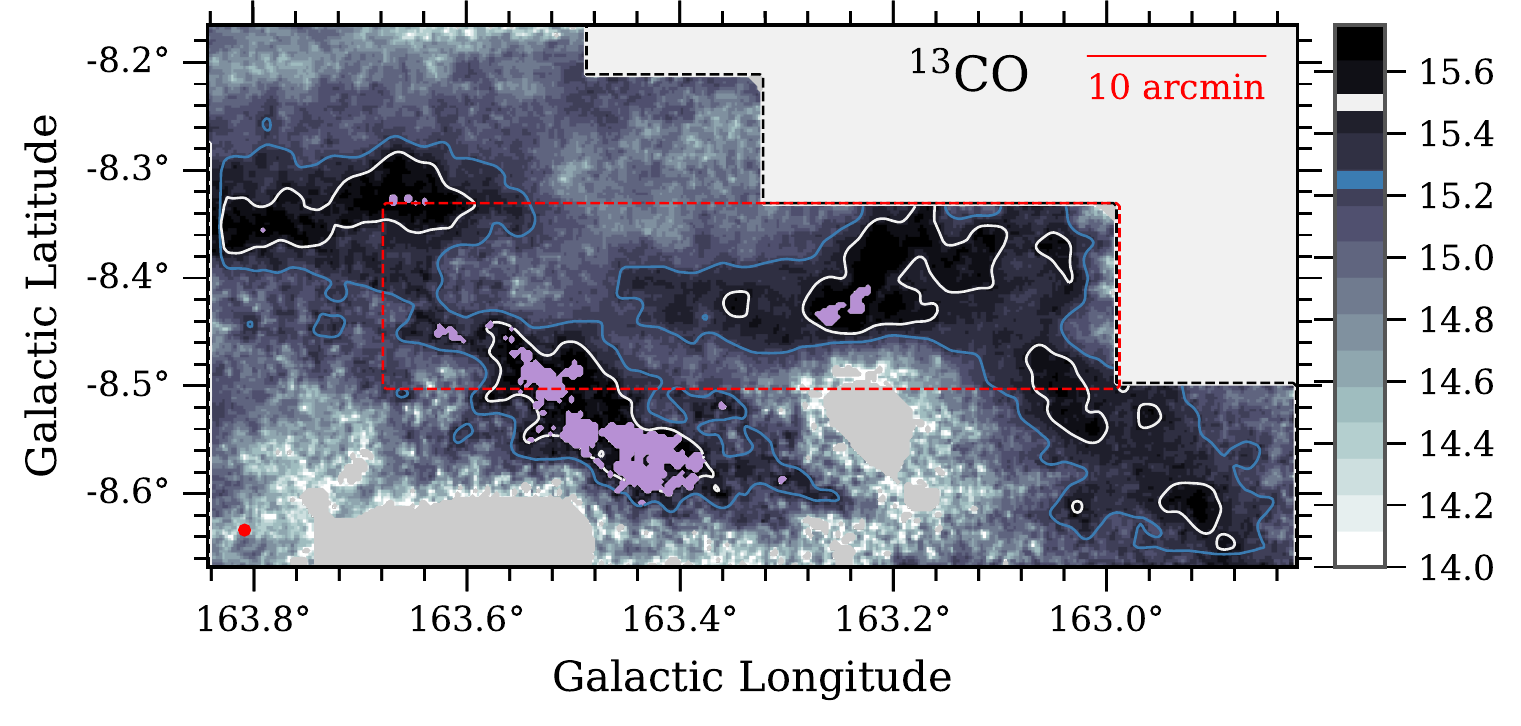}
    \includegraphics[height=2.5in]{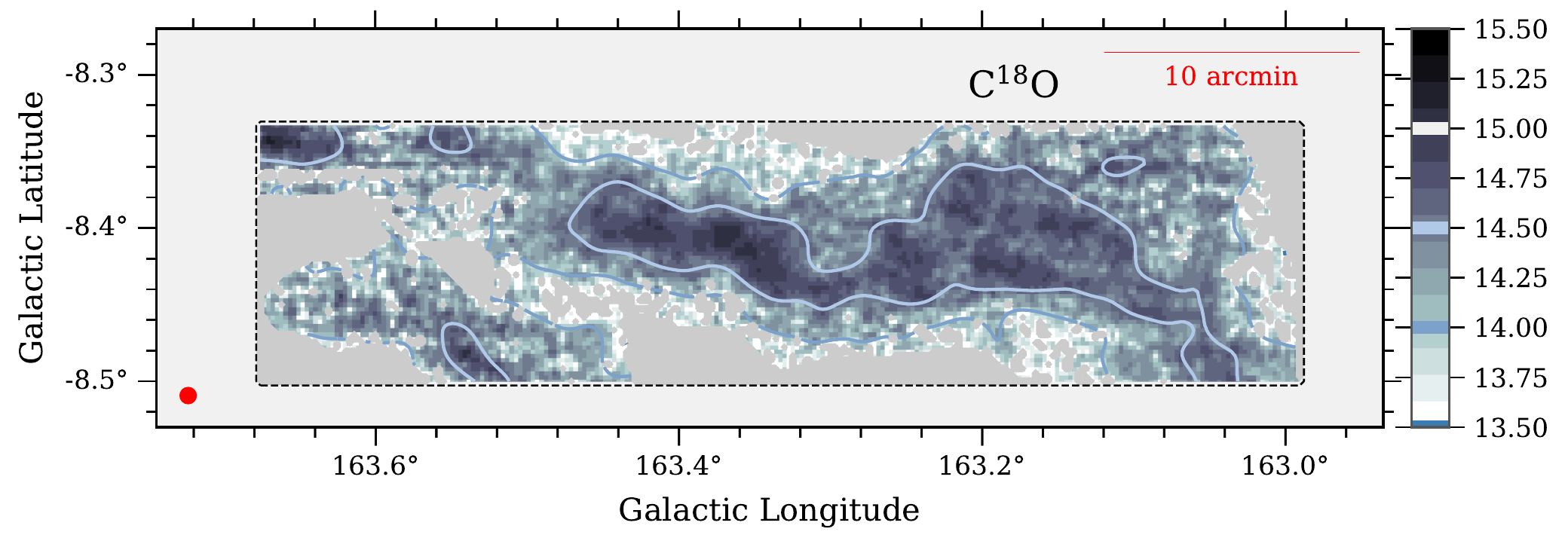}
    \caption{CO log(column density) maps (Top: \thirteenco, Bottom: \ceighteeno) for the L1478 region in the CMC. The black dashed line indicates the survey boundaries, and the red dashed line visible in the \thirteenco indicates the extent of the \ceighteeno survey. The survey resolution (38\arcsec) is shown as a red circle in the lower left hand corner. We mask (red with hatches) the regions of the cloud where a valid column density could not be calculated as described in Appendix \ref{sec:methnco}. The grayscale is the log CO column density in $\rm cm^{-2}$. The contours are - \thirteenco: (15.2, 15.5); \ceighteeno (13.5, 14.0, 14.5, 15.0) }
    \label{fig:ncol1478}
\end{figure}

\begin{figure}[ht]
    \centering
    \includegraphics[height=2.5in]{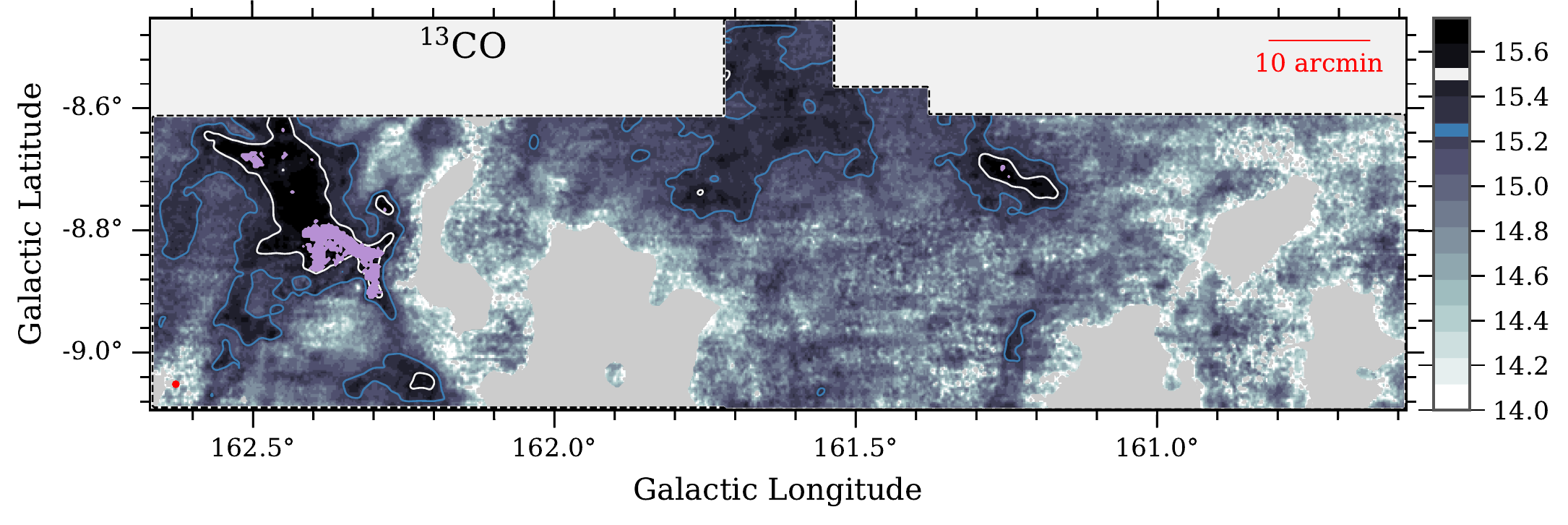}
    \caption{\thirteenco log(column density) map for the West region in the CMC. The black dashed line indicates the survey boundaries. The survey resolution (38\arcsec) is shown as a red circle in the lower left hand corner. We mask (in the filled purple regions) the regions of the cloud where a valid column density could not be calculated as described in Appendix \ref{sec:methnco}. The grayscale is the log CO column density in  in $\rm cm^{-2}$. The contours are (15.2, 15.5).}
    \label{fig:ncowest}
\end{figure}

\begin{figure}[ht]
    \centering
    \includegraphics[height=4.25in]{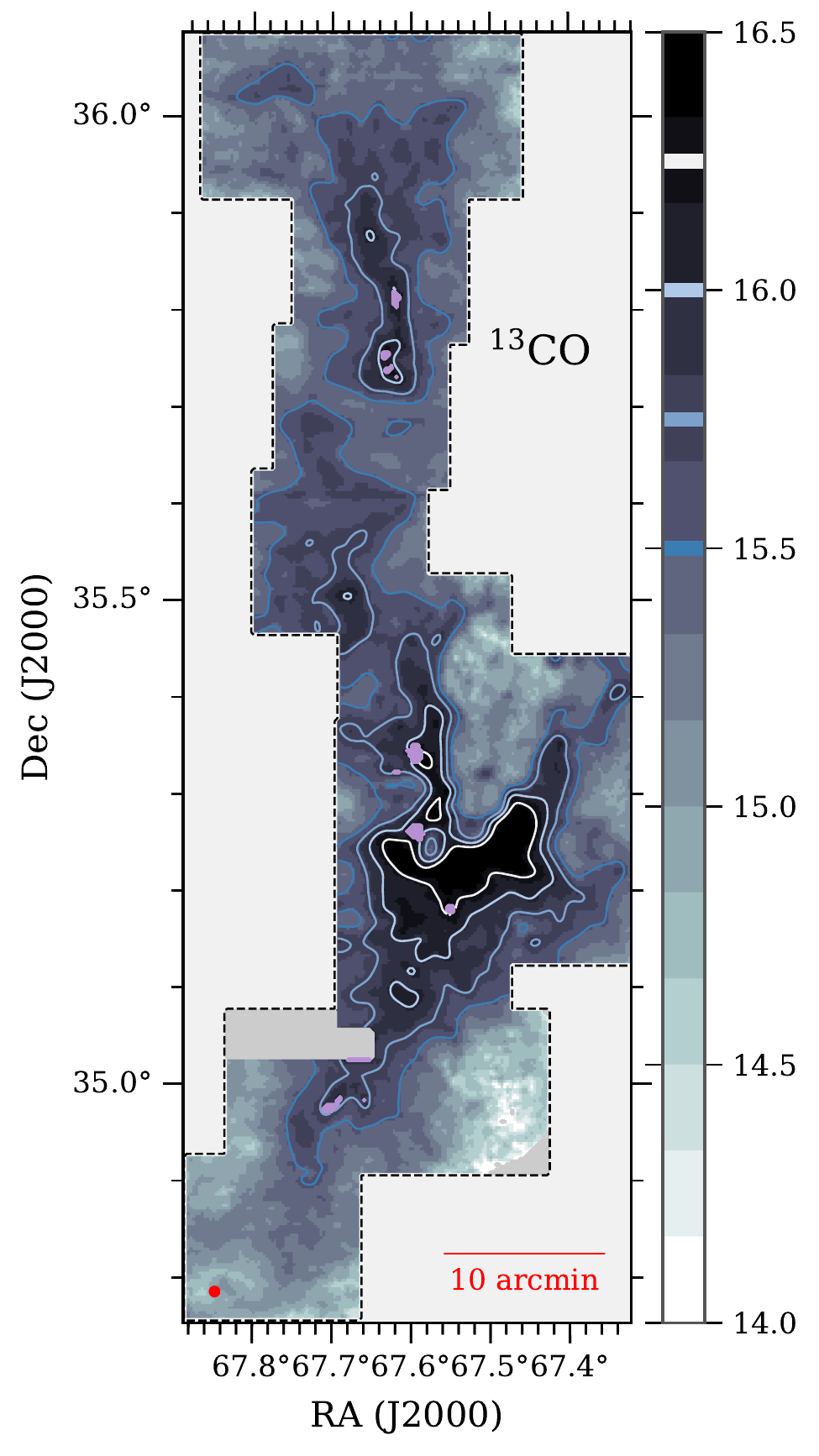}
    \includegraphics[height=4.25in]{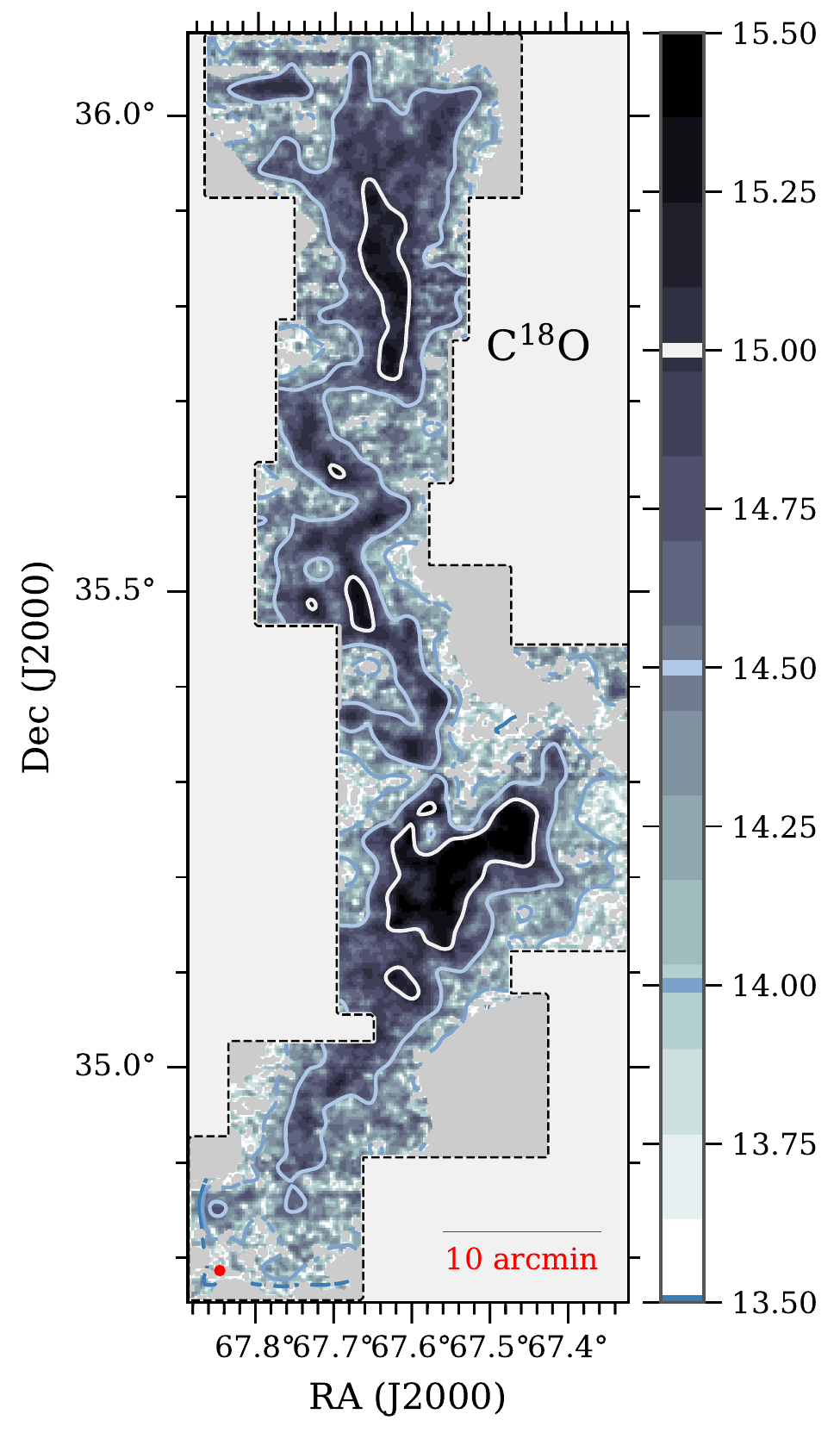}
    \caption{CO log(column density) maps (Top: \thirteenco, Bottom: \ceighteeno) for the L1482 region in the CMC. The black dashed line indicates the survey boundaries, and the red dashed line visible in the \thirteenco indicates the extent of the \ceighteeno survey. The survey resolution (38\arcsec) is shown as a red circle in the lower left hand corner.We mask (in the filled purple regions) the regions of the cloud where a valid column density could not be calculated as described in Appendix \ref{sec:methnco}.
    The grayscale is log CO column density in  in $\rm cm^{-2}$. The contours from dark to light are for - \thirteenco: (15.5, 15.8, 16.0, 16.2); \ceighteeno (13.5, 14.0, 14.5, 15.0)}
    \label{fig:ncolkha}
\end{figure}

\parnum where $Q$ is the partition function which we calculate numerically to the J=100 term, at which point $Q$ has converged to within machine precision (better than 1 part in $10^{15}$) for the range of \tex in our data, instead of using an approximation. The error associated with using the LTE approximation relative to a non-LTE determination was estimated to be a factor of $\sim2$ in column density by \citetalias{2015ApJ...805...58K}. $\rm N_{non-LTE}/N_{LTE}$ is larger for lower column densities and is does not have a significant dependence on \tex \citep{2000ApJ...529..259P}, for \thirteenco). For the range of column densities in our data $\rm N_{non-LTE}/N_{LTE}$  ranges from $\sim 1 - 2$. Figs. \ref{fig:ncol1478}, \ref{fig:ncowest} and \ref{fig:ncolkha} show the maps of N[\thirteenco] and N[\ceighteeno] with areas where $\tau$ is not real-valued masked.

\clearpage

\section{\thirteenco Depletion Core Maps}\label{sec:appcores}

In Fig. \ref{fig:fullcoremap}, we showed the depletion core contours in the context of the entire cloud, overlaid on the \herschel extinction. Here we focus on each region and show the depletion core positions overlaid on the \herschel extinction map. We include the position of the protostars from \citet{2017A&A...606A.100L} in blue.
\vskip 1.0in
\begin{figure}
    \centering
    \includegraphics[height=2.4in]{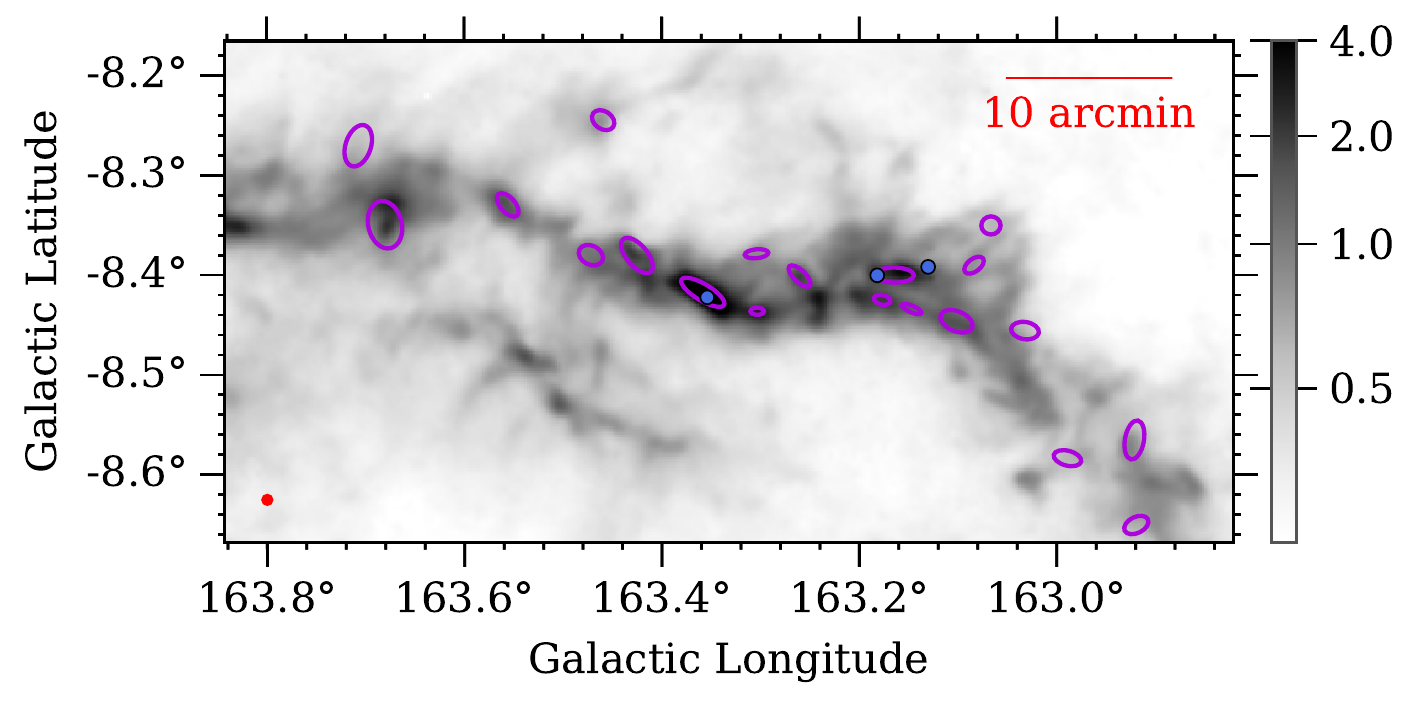}
    \includegraphics[height=2.4in]{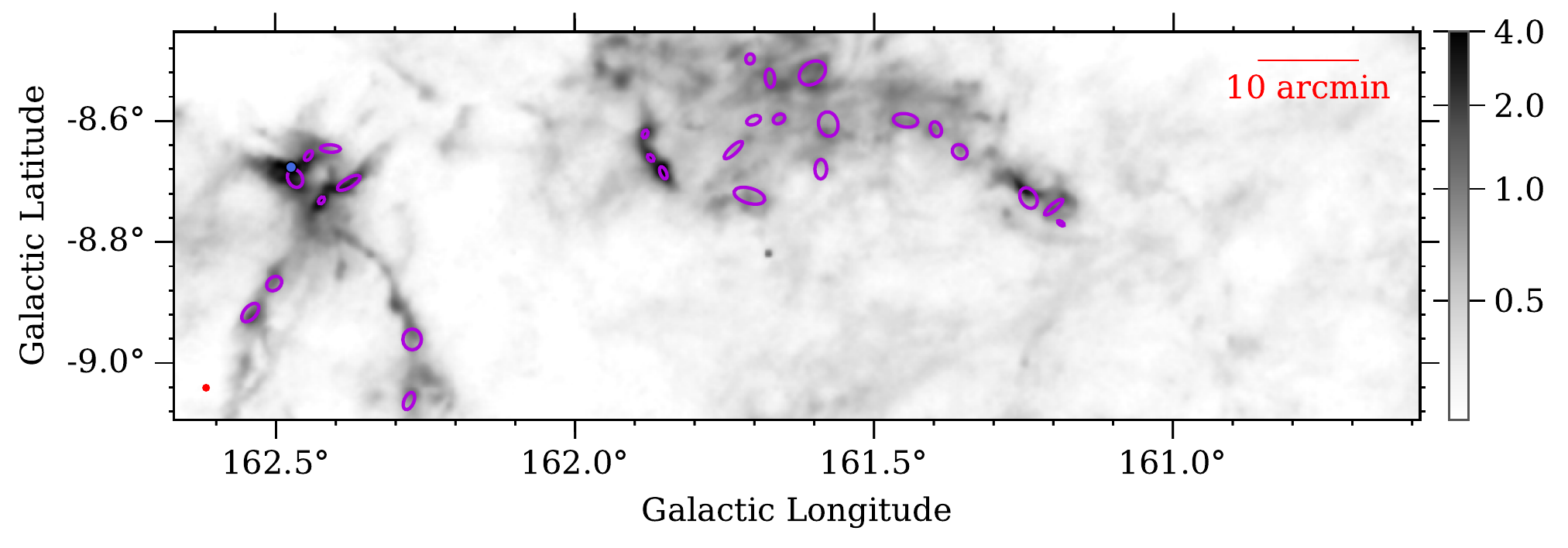}
    \caption{Map of depletion cores (ellipses) overlaid on \herschel extinction. The core's size and locations is represented as the best fit ellipse derived using the moment method \citep{2006PASP..118..590R}. Protostars from \citet{2017A&A...606A.100L} are in blue. from The top shows L1478, and the bottom CMC-West. Our $38\arcsec$ beam is located in the lower left corner of both plots in red. }
    \label{fig:coremaps}
\end{figure}

\begin{figure}
    \centering
    \includegraphics[height=7.2in]{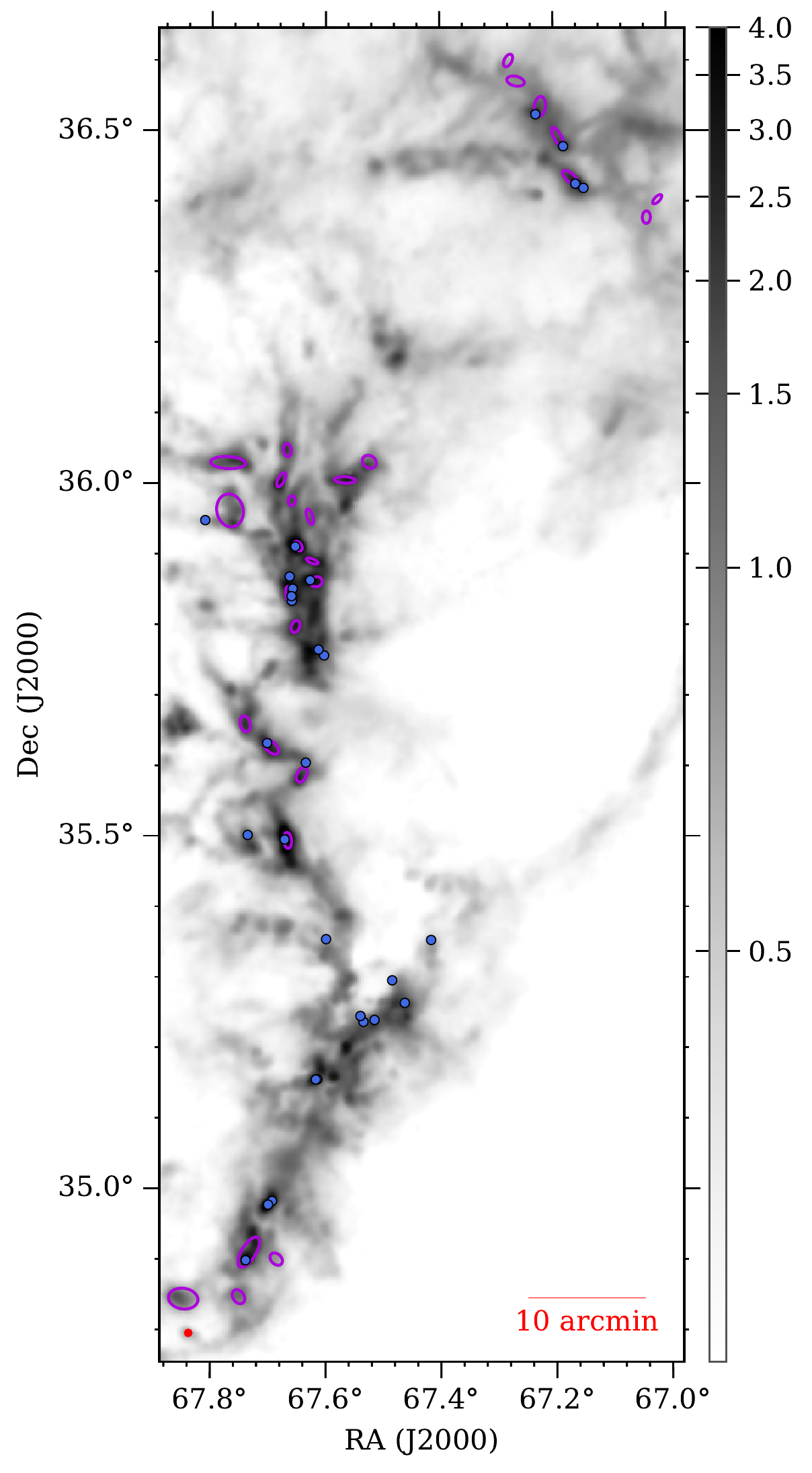}
    \caption{Same as Fig. \ref{fig:coremaps} but for the L1482 region in the CMC.}
    \label{fig:coremaps2}
\end{figure}

\clearpage

\clearpage
\startlongtable\label{tab:cores}
\begin{deluxetable}{cccccccccccccc}
\tablecaption{Catalog of Depletion Cores in the California Molecular Clouds}\label{tab:cores}
\tablehead{\colhead{NAME} & \colhead{Right Ascension} & \colhead{Declination} & \colhead{Radius [pc]} & \colhead{$M_{\rm core}$} & \colhead{$\tdust$} & \colhead{$\ak$} & \colhead{V$_{\rm cen}$} & \colhead{$\sigma_{\rm 13}$} & \colhead{$\sigma_{\rm 18}$} & \colhead{$\langle\eta_{13}\rangle$} & \colhead{$\langle\eta_{18}\rangle$} & \colhead{{\it Zhang et al.}} \\
&  &  & \colhead{pc} & \colhead{$M_\odot$} & \colhead{K} & \colhead{$\magn_{\rm K}$} & \colhead{km/s} & \colhead{km/s} & \colhead{km/s} &  &  & }
\startdata
West-001 & 4:19:36.34 & 37:26:00.91 & 0.13 & 8.1 & 15.3 & 0.76 & -0.70 & 0.44 & \omit & 5.18 & \omit & 103 \\
West-002 & 4:19:57.06 & 37:30:32.60 & 0.19 & 14.9 & 15.4 & 0.71 & -0.63 & 0.41 & \omit & 6.26 & \omit & 101 \\
West-003 & 4:21:03.28 & 37:20:59.46 & 0.18 & 17.2 & 14.7 & 0.89 & 0.08 & 0.36 & \omit & 5.31 & \omit & 106 \\
West-004 & 4:21:05.24 & 37:24:41.75 & 0.15 & 12.3 & 15.0 & 0.84 & -0.06 & 0.54 & \omit & 4.85 & \omit & 104 \\
West-005 & 4:16:47.84 & 38:24:08.92 & 0.05 & 1.3 & 15.7 & 0.62 & -3.40 & 0.46 & \omit & 5.24 & \omit &  \\
West-006 & 4:17:08.44 & 38:23:39.91 & 0.18 & 25.4 & 14.0 & 1.27 & -3.77 & 0.50 & \omit & 4.26 & \omit & 13 \\
West-007 & 4:16:56.39 & 38:24:14.09 & 0.08 & 5.3 & 14.2 & 1.04 & -3.78 & 0.49 & \omit & 4.17 & \omit & 12 \\
West-008 & 4:18:48.99 & 38:04:18.17 & 0.21 & 18.7 & 15.6 & 0.70 & -2.23 & 1.04 & \omit & 3.79 & \omit &  \\
West-009 & 4:21:18.31 & 37:33:49.77 & 0.06 & 9.3 & 12.1 & 3.25 & -0.62 & 0.86 & \omit & 4.79 & \omit & 90 \\
West-010 & 4:21:15.01 & 37:36:58.80 & 0.16 & 30.3 & 13.1 & 1.86 & -0.65 & 0.85 & \omit & 6.15 & \omit & 85 \\
West-011 & 4:21:35.40 & 37:33:29.75 & 0.17 & 49.1 & 13.3 & 2.70 & -1.06 & 0.80 & \omit & 6.96 & \omit & 94 \\
West-012 & 4:18:33.06 & 38:11:10.45 & 0.15 & 7.8 & 15.5 & 0.59 & -2.63 & 0.62 & \omit & 4.13 & \omit &  \\
West-013 & 4:19:27.74 & 37:59:51.66 & 0.09 & 20.1 & 12.6 & 3.34 & -2.37 & 0.61 & \omit & 21.72 & \omit & 75 \\
West-014 & 4:21:38.86 & 37:36:01.17 & 0.07 & 7.9 & 13.8 & 1.98 & -1.42 & 0.70 & \omit & 5.24 & \omit & 87 \\
West-015 & 4:19:37.54 & 38:00:01.09 & 0.06 & 7.5 & 13.2 & 2.40 & -2.31 & 0.52 & \omit & 18.71 & \omit & 76 \\
West-016 & 4:19:11.01 & 38:06:22.36 & 0.12 & 6.2 & 15.8 & 0.64 & -2.43 & 1.16 & \omit & 4.15 & \omit &  \\
West-017 & 4:17:49.60 & 38:22:06.79 & 0.14 & 8.4 & 15.1 & 0.67 & -3.12 & 0.49 & \omit & 5.61 & \omit &  \\
West-018 & 4:21:33.69 & 37:38:03.80 & 0.11 & 7.1 & 14.9 & 0.86 & -1.53 & 0.43 & \omit & 4.42 & \omit &  \\
West-019 & 4:18:46.50 & 38:14:51.00 & 0.22 & 23.4 & 15.2 & 0.78 & -2.79 & 0.73 & \omit & 4.34 & \omit & 29 \\
West-020 & 4:19:48.07 & 38:01:17.52 & 0.05 & 5.2 & 13.5 & 1.92 & -2.28 & 0.53 & \omit & 11.83 & \omit & 73 \\
West-021 & 4:18:06.23 & 38:22:01.59 & 0.13 & 8.0 & 15.0 & 0.79 & -2.96 & 0.45 & \omit & 6.91 & \omit &  \\
West-022 & 4:18:20.22 & 38:20:31.81 & 0.18 & 16.6 & 15.0 & 0.81 & -3.08 & 0.48 & \omit & 6.87 & \omit &  \\
West-023 & 4:19:14.46 & 38:09:52.89 & 0.10 & 4.0 & 15.7 & 0.60 & -3.18 & 0.69 & \omit & 4.85 & \omit &  \\
West-024 & 4:19:05.90 & 38:11:45.63 & 0.11 & 6.2 & 14.9 & 0.81 & -3.00 & 0.65 & \omit & 5.04 & \omit &  \\
West-025 & 4:19:10.45 & 38:17:18.70 & 0.26 & 42.3 & 14.2 & 1.07 & -3.10 & 0.61 & \omit & 7.02 & \omit & 21 \\
West-026 & 4:19:23.70 & 38:13:57.82 & 0.14 & 12.6 & 14.5 & 1.09 & -2.95 & 0.59 & \omit & 6.07 & \omit &  \\
West-027 & 4:19:37.65 & 38:13:55.32 & 0.08 & 3.3 & 14.8 & 0.73 & -2.73 & 0.62 & \omit & 4.47 & \omit &  \\
L1478-001 & 4:23:19.19 & 37:16:08.27 & 0.11 & 5.1 & 15.7 & 0.67 & -1.28 & 0.57 & \omit & 3.63 & \omit &  \\
L1478-002 & 4:23:48.17 & 37:15:57.10 & 0.11 & 4.6 & 15.8 & 0.57 & -1.10 & 0.52 & \omit & 3.27 & \omit &  \\
L1478-003 & 4:23:38.08 & 37:19:36.42 & 0.14 & 8.7 & 15.6 & 0.70 & -1.33 & 0.56 & \omit & 3.37 & \omit &  \\
L1478-004 & 4:24:24.86 & 37:19:25.83 & 0.12 & 6.1 & 15.6 & 0.63 & -1.49 & 0.42 & 0.39 & 4.50 & 8.40 &  \\
L1478-005 & 4:24:41.24 & 37:16:51.04 & 0.16 & 21.4 & 14.2 & 1.32 & -1.58 & 0.53 & 0.31 & 6.44 & 3.82 & 110 \\
L1478-006 & 4:25:40.50 & 37:07:01.95 & 0.19 & 82.6 & 12.8 & 3.65 & -1.76 & 0.61 & 0.41 & 14.42 & 5.74 & 121 \\
L1478-007 & 4:25:25.06 & 37:08:36.60 & 0.05 & 5.9 & 12.9 & 2.44 & -1.74 & 0.54 & 0.22 & 10.29 & 5.01 & 120 \\
L1478-008 & 4:24:53.30 & 37:15:24.82 & 0.05 & 3.7 & 13.8 & 1.36 & -1.77 & 0.49 & 0.35 & 5.70 & 3.03 & 112 \\
L1478-009 & 4:25:01.45 & 37:14:31.03 & 0.07 & 7.7 & 13.2 & 1.84 & -1.92 & 0.50 & 0.29 & 6.27 & 3.30 & 115 \\
L1478-010 & 4:25:23.86 & 37:11:54.88 & 0.11 & 11.3 & 13.6 & 1.42 & -1.94 & 0.70 & 0.38 & 4.80 & 3.81 & 118 \\
L1478-011 & 4:26:02.20 & 37:05:41.68 & 0.17 & 24.9 & 13.7 & 1.38 & -1.46 & 0.43 & 0.28 & 10.50 & 4.22 & 122 \\
L1478-012 & 4:25:04.29 & 37:16:05.02 & 0.16 & 32.5 & 13.2 & 2.05 & -1.72 & 0.62 & 0.34 & 7.33 & 4.42 & 111 \\
L1478-013 & 4:24:49.80 & 37:19:56.73 & 0.09 & 4.8 & 15.3 & 0.77 & -1.37 & 0.51 & 0.36 & 4.09 & 6.53 &  \\
L1478-014 & 4:26:11.85 & 37:03:41.96 & 0.13 & 11.3 & 14.3 & 1.02 & -1.41 & 0.40 & 0.27 & 10.49 & 4.35 &  \\
L1478-015 & 4:27:01.14 & 36:55:58.79 & 0.25 & 47.7 & 14.4 & 1.30 & -1.66 & 0.54 & 0.31 & 4.54 & 7.36 & 131 \\
L1478-016 & 4:25:37.73 & 37:10:58.15 & 0.08 & 3.9 & 14.9 & 0.77 & -1.54 & 0.58 & 0.28 & 5.84 & 4.84 &  \\
L1478-017 & 4:24:54.94 & 37:22:19.91 & 0.09 & 3.5 & 15.7 & 0.58 & -1.52 & 0.41 & 0.32 & 3.07 & 3.30 &  \\
L1478-018 & 4:26:40.09 & 37:02:09.39 & 0.12 & 10.6 & 13.9 & 1.22 & -1.65 & 0.45 & 0.26 & 6.62 & 3.59 & 124 \\
L1478-019 & 4:27:23.92 & 36:58:04.32 & 0.19 & 13.0 & 15.5 & 0.59 & -1.25 & 0.48 & \omit & 4.73 & \omit &  \\
L1478-020 & 4:26:38.82 & 37:09:49.67 & 0.12 & 5.5 & 15.4 & 0.62 & -1.66 & 0.51 & \omit & 5.85 & \omit &  \\
L1482-001 & 4:31:23.75 & 34:50:35.33 & 0.22 & 21.2 & 15.4 & 0.71 & -0.68 & 0.37 & 0.28 & 5.30 & 4.55 & 299 \\
L1482-002 & 4:31:00.11 & 34:50:47.39 & 0.11 & 7.7 & 14.8 & 0.92 & -0.77 & 0.40 & 0.36 & 4.80 & 4.82 & 298 \\
L1482-003 & 4:30:55.89 & 34:54:34.78 & 0.21 & 44.5 & 13.9 & 1.65 & -0.42 & 0.56 & 0.43 & 5.92 & 5.03 & 296 \\
L1482-004 & 4:30:44.47 & 34:53:59.43 & 0.10 & 5.0 & 15.5 & 0.70 & -0.44 & 0.42 & 0.96 & 4.44 & 3.64 &  \\
L1482-005 & 4:30:39.93 & 35:29:36.34 & 0.10 & 31.3 & 14.7 & 4.44 & -0.61 & 0.65 & 0.40 & 5.86 & 2.70 & 232 \\
L1482-006 & 4:30:33.98 & 35:35:11.62 & 0.11 & 13.2 & 15.1 & 1.54 & -0.60 & 0.68 & 0.44 & 3.48 & 2.01 & 229 \\
L1482-007 & 4:30:46.89 & 35:37:33.38 & 0.13 & 18.7 & 14.8 & 1.87 & -0.87 & 0.61 & 0.36 & 4.28 & 1.75 & 225 \\
L1482-008 & 4:30:57.75 & 35:39:31.13 & 0.12 & 13.4 & 14.1 & 1.55 & -1.14 & 0.47 & 0.33 & 4.40 & 1.93 & 222 \\
L1482-009 & 4:30:36.69 & 35:47:47.82 & 0.09 & 12.2 & 12.8 & 2.09 & -1.15 & 0.73 & 0.42 & 5.54 & 3.02 & 210 \\
L1482-010 & 4:30:39.00 & 35:50:26.19 & 0.11 & 23.6 & 13.6 & 2.83 & -1.14 & 0.79 & 0.53 & 5.51 & 2.94 & 205 \\
L1482-011 & 4:30:27.99 & 35:51:36.16 & 0.10 & 18.2 & 12.5 & 2.75 & -0.79 & 0.77 & 0.54 & 6.02 & 3.05 & 203 \\
L1482-012 & 4:30:29.72 & 35:53:21.97 & 0.06 & 7.5 & 12.8 & 2.68 & -1.02 & 0.73 & 0.37 & 6.40 & 2.15 & 201 \\
L1482-013 & 4:30:35.65 & 35:54:37.89 & 0.08 & 19.2 & 12.9 & 4.40 & -1.14 & 0.89 & 0.54 & 7.42 & 3.07 & 197 \\
L1482-014 & 4:31:04.28 & 35:57:40.08 & 0.24 & 30.6 & 14.4 & 0.86 & -0.86 & 0.52 & 0.30 & 3.89 & 5.28 & 188 \\
L1482-015 & 4:30:30.71 & 35:57:08.77 & 0.08 & 6.3 & 13.8 & 1.25 & -0.97 & 0.59 & 0.32 & 3.56 & 1.80 & 192 \\
L1482-016 & 4:30:38.33 & 35:58:29.51 & 0.07 & 5.9 & 13.5 & 1.53 & -1.05 & 0.60 & 0.55 & 4.43 & 2.21 & 185 \\
L1482-017 & 4:30:42.69 & 36:00:17.49 & 0.08 & 10.1 & 13.4 & 1.97 & -1.11 & 0.53 & 0.38 & 6.80 & 3.72 & 183 \\
L1482-018 & 4:30:15.92 & 36:00:15.72 & 0.10 & 13.0 & 13.3 & 1.96 & -0.70 & 0.62 & 0.36 & 5.57 & 3.03 & 181 \\
L1482-019 & 4:31:05.13 & 36:01:44.21 & 0.18 & 25.1 & 14.2 & 1.26 & -0.68 & 0.88 & 0.41 & 3.73 & 2.15 & 177 \\
L1482-020 & 4:30:05.72 & 36:01:48.07 & 0.12 & 10.3 & 14.6 & 1.13 & -0.78 & 0.43 & 0.31 & 7.24 & 4.56 & 179 \\
L1482-021 & 4:30:40.16 & 36:02:47.95 & 0.09 & 8.9 & 14.0 & 1.46 & -1.07 & 0.70 & 0.15 & 7.56 & 3.01 & 176 \\
L1482-022 & 4:28:08.56 & 36:22:36.88 & 0.08 & 3.5 & 15.9 & 0.68 & 0.03 & 0.65 & \omit & 6.34 & \omit &  \\
L1482-023 & 4:28:03.94 & 36:24:08.44 & 0.06 & 1.8 & 15.8 & 0.56 & 0.07 & 0.73 & \omit & 6.17 & \omit &  \\
L1482-024 & 4:28:40.59 & 36:25:58.25 & 0.12 & 16.4 & 13.7 & 1.79 & -1.07 & 0.66 & 0.40 & 5.21 & 2.56 & 160 \\
L1482-025 & 4:28:45.85 & 36:29:23.96 & 0.12 & 14.2 & 14.1 & 1.63 & -1.07 & 0.53 & 0.35 & 7.35 & 2.61 & 155 \\
L1482-026 & 4:28:53.39 & 36:32:02.45 & 0.13 & 15.4 & 14.5 & 1.43 & -1.21 & 0.56 & 0.42 & 6.94 & 2.72 & 151 \\
L1482-027 & 4:29:03.68 & 36:34:11.18 & 0.11 & 6.6 & 15.3 & 0.81 & -1.28 & 1.11 & 0.41 & 5.79 & 2.36 &  \\
L1482-028 & 4:29:06.82 & 36:35:56.29 & 0.07 & 2.2 & 15.4 & 0.64 & -1.35 & 1.25 & 0.31 & 5.08 & 5.88 &
\enddata
\tablecomments{Catalog of CMC depletion cores.  The radius is the beam-deconvolved radius of the source, $r_d$, as defined in the text. \tdust and \ak are the average \herschel dust temperature and extinction respectively. V$_{\rm cen}$ is the central velocity of the average \thirteenco line in the core. The last column is the core number for the best-matched dust core from Zhang et al. (2018, Table 3)}
\end{deluxetable}

\newpage

\bibliographystyle{aasjournal}

\begin{thebibliography}{}
\expandafter\ifx\csname natexlab\endcsname\relax\def\natexlab#1{#1}\fi
\providecommand{\url}[1]{\href{#1}{#1}}
\providecommand{\dodoi}[1]{doi:~\href{http://doi.org/#1}{\nolinkurl{#1}}}
\providecommand{\doeprint}[1]{\href{http://ascl.net/#1}{\nolinkurl{http://ascl.net/#1}}}
\providecommand{\doarXiv}[1]{\href{https://arxiv.org/abs/#1}{\nolinkurl{https://arxiv.org/abs/#1}}}

\bibitem[{{Astropy Collaboration} {et~al.}(2013){Astropy Collaboration},
  {Robitaille}, {Tollerud}, {Greenfield}, {Droettboom}, {Bray}, {Aldcroft},
  {Davis}, {Ginsburg}, {Price-Whelan}, {Kerzendorf}, {Conley}, {Crighton},
  {Barbary}, {Muna}, {Ferguson}, {Grollier}, {Parikh}, {Nair}, {Unther},
  {Deil}, {Woillez}, {Conseil}, {Kramer}, {Turner}, {Singer}, {Fox}, {Weaver},
  {Zabalza}, {Edwards}, {Azalee Bostroem}, {Burke}, {Casey}, {Crawford},
  {Dencheva}, {Ely}, {Jenness}, {Labrie}, {Lim}, {Pierfederici}, {Pontzen},
  {Ptak}, {Refsdal}, {Servillat}, \& {Streicher}}]{2013A&A...558A..33A}
{Astropy Collaboration}, {Robitaille}, T.~P., {Tollerud}, E.~J., {et~al.} 2013,
  \aap, 558, A33, \dodoi{10.1051/0004-6361/201322068}

\bibitem[{{Astropy Collaboration} {et~al.}(2018){Astropy Collaboration},
  {Price-Whelan}, {Sip{\H{o}}cz}, {G{\"u}nther}, {Lim}, {Crawford}, {Conseil},
  {Shupe}, {Craig}, {Dencheva}, {Ginsburg}, {Vand erPlas}, {Bradley},
  {P{\'e}rez-Su{\'a}rez}, {de Val-Borro}, {Aldcroft}, {Cruz}, {Robitaille},
  {Tollerud}, {Ardelean}, {Babej}, {Bach}, {Bachetti}, {Bakanov}, {Bamford},
  {Barentsen}, {Barmby}, {Baumbach}, {Berry}, {Biscani}, {Boquien}, {Bostroem},
  {Bouma}, {Brammer}, {Bray}, {Breytenbach}, {Buddelmeijer}, {Burke},
  {Calderone}, {Cano Rodr{\'\i}guez}, {Cara}, {Cardoso}, {Cheedella}, {Copin},
  {Corrales}, {Crichton}, {D'Avella}, {Deil}, {Depagne}, {Dietrich}, {Donath},
  {Droettboom}, {Earl}, {Erben}, {Fabbro}, {Ferreira}, {Finethy}, {Fox},
  {Garrison}, {Gibbons}, {Goldstein}, {Gommers}, {Greco}, {Greenfield},
  {Groener}, {Grollier}, {Hagen}, {Hirst}, {Homeier}, {Horton}, {Hosseinzadeh},
  {Hu}, {Hunkeler}, {Ivezi{\'c}}, {Jain}, {Jenness}, {Kanarek}, {Kendrew},
  {Kern}, {Kerzendorf}, {Khvalko}, {King}, {Kirkby}, {Kulkarni}, {Kumar},
  {Lee}, {Lenz}, {Littlefair}, {Ma}, {Macleod}, {Mastropietro}, {McCully},
  {Montagnac}, {Morris}, {Mueller}, {Mumford}, {Muna}, {Murphy}, {Nelson},
  {Nguyen}, {Ninan}, {N{\"o}the}, {Ogaz}, {Oh}, {Parejko}, {Parley}, {Pascual},
  {Patil}, {Patil}, {Plunkett}, {Prochaska}, {Rastogi}, {Reddy Janga},
  {Sabater}, {Sakurikar}, {Seifert}, {Sherbert}, {Sherwood-Taylor}, {Shih},
  {Sick}, {Silbiger}, {Singanamalla}, {Singer}, {Sladen}, {Sooley},
  {Sornarajah}, {Streicher}, {Teuben}, {Thomas}, {Tremblay}, {Turner},
  {Terr{\'o}n}, {van Kerkwijk}, {de la Vega}, {Watkins}, {Weaver}, {Whitmore},
  {Woillez}, {Zabalza}, \& {Astropy Contributors}}]{2018AJ....156..123A}
{Astropy Collaboration}, {Price-Whelan}, A.~M., {Sip{\H{o}}cz}, B.~M., {et~al.}
  2018, \aj, 156, 123, \dodoi{10.3847/1538-3881/aabc4f}

\bibitem[{{Bergin} {et~al.}(1995){Bergin}, {Langer}, \&
  {Goldsmith}}]{1995ApJ...441..222B}
{Bergin}, E.~A., {Langer}, W.~D., \& {Goldsmith}, P.~F. 1995, \apj, 441, 222,
  \dodoi{10.1086/175351}

\bibitem[{{Berry} {et~al.}(2007){Berry}, {Reinhold}, {Jenness}, \&
  {Economou}}]{2007ASPC..376..425B}
{Berry}, D.~S., {Reinhold}, K., {Jenness}, T., \& {Economou}, F. 2007, in
  Astronomical Society of the Pacific Conference Series, Vol. 376, Astronomical
  Data Analysis Software and Systems XVI, ed. R.~A. {Shaw}, F.~{Hill}, \& D.~J.
  {Bell}, 425

\bibitem[{{Bertoldi} \& {McKee}(1992)}]{1992ApJ...395..140B}
{Bertoldi}, F., \& {McKee}, C.~F. 1992, \apj, 395, 140, \dodoi{10.1086/171638}

\bibitem[{{Bieging} {et~al.}(2010){Bieging}, {Peters}, \&
  {Kang}}]{2010ApJS..191..232B}
{Bieging}, J.~H., {Peters}, W.~L., \& {Kang}, M. 2010, \apjs, 191, 232,
  \dodoi{10.1088/0067-0049/191/2/232}

\bibitem[{{Bisschop} {et~al.}(2006){Bisschop}, {Fraser}, {{\"O}berg}, {van
  Dishoeck}, \& {Schlemmer}}]{2006A&A...449.1297B}
{Bisschop}, S.~E., {Fraser}, H.~J., {{\"O}berg}, K.~I., {van Dishoeck}, E.~F.,
  \& {Schlemmer}, S. 2006, \aap, 449, 1297, \dodoi{10.1051/0004-6361:20054051}

\bibitem[{{Bohlin} {et~al.}(1978){Bohlin}, {Savage}, \&
  {Drake}}]{1978ApJ...224..132B}
{Bohlin}, R.~C., {Savage}, B.~D., \& {Drake}, J.~F. 1978, \apj, 224, 132,
  \dodoi{10.1086/156357}

\bibitem[{{Bolatto} {et~al.}(2013){Bolatto}, {Wolfire}, \&
  {Leroy}}]{2013ARA&A..51..207B}
{Bolatto}, A.~D., {Wolfire}, M., \& {Leroy}, A.~K. 2013, \araa, 51, 207,
  \dodoi{10.1146/annurev-astro-082812-140944}

\bibitem[{{Broekhoven-Fiene} {et~al.}(2014){Broekhoven-Fiene}, {Matthews},
  {Harvey}, {Gutermuth}, {Huard}, {Tothill}, {Nutter}, {Bourke}, {DiFrancesco},
  {J{\o}rgensen}, {Allen}, {Chapman}, {Dunham}, {Mer{\'\i}n}, {Miller},
  {Terebey}, {Peterson}, \& {Stapelfeldt}}]{2014ApJ...786...37B}
{Broekhoven-Fiene}, H., {Matthews}, B.~C., {Harvey}, P.~M., {et~al.} 2014,
  \apj, 786, 37, \dodoi{10.1088/0004-637X/786/1/37}

\bibitem[{{Burgh} {et~al.}(2007){Burgh}, {France}, \&
  {McCandliss}}]{2007ApJ...658..446B}
{Burgh}, E.~B., {France}, K., \& {McCandliss}, S.~R. 2007, \apj, 658, 446,
  \dodoi{10.1086/511259}

\bibitem[{{Frerking} {et~al.}(1982){Frerking}, {Langer}, \&
  {Wilson}}]{1982ApJ...262..590F}
{Frerking}, M.~A., {Langer}, W.~D., \& {Wilson}, R.~W. 1982, \apj, 262, 590,
  \dodoi{10.1086/160451}

\bibitem[{{Gildas Team}(2013)}]{2013ascl.soft05010G}
{Gildas Team}. 2013, {GILDAS: Grenoble Image and Line Data Analysis Software}.
\newblock \doeprint{1305.010}

\bibitem[{{Glover} \& {Clark}(2016)}]{2016MNRAS.456.3596G}
{Glover}, S. C.~O., \& {Clark}, P.~C. 2016, \mnras, 456, 3596,
  \dodoi{10.1093/mnras/stv2863}

\bibitem[{{Harvey} {et~al.}(2013){Harvey}, {Fallscheer}, {Ginsburg}, {Terebey},
  {Andr{\'e}}, {Bourke}, {Di Francesco}, {K{\"o}nyves}, {Matthews}, \&
  {Peterson}}]{2013ApJ...764..133H}
{Harvey}, P.~M., {Fallscheer}, C., {Ginsburg}, A., {et~al.} 2013, \apj, 764,
  133, \dodoi{10.1088/0004-637X/764/2/133}

\bibitem[{{Hernandez} {et~al.}(2011){Hernandez}, {Tan}, {Caselli}, {Butler},
  {Jim{\'e}nez-Serra}, {Fontani}, \& {Barnes}}]{2011ApJ...738...11H}
{Hernandez}, A.~K., {Tan}, J.~C., {Caselli}, P., {et~al.} 2011, \apj, 738, 11,
  \dodoi{10.1088/0004-637X/738/1/11}

\bibitem[{{Imara} {et~al.}(2017){Imara}, {Lada}, {Lewis}, {Bieging}, {Kong},
  {Lombardi}, \& {Alves}}]{2017ApJ...840..119I}
{Imara}, N., {Lada}, C., {Lewis}, J., {et~al.} 2017, \apj, 840, 119,
  \dodoi{10.3847/1538-4357/aa6d74}

\bibitem[{{Kirk} {et~al.}(2017){Kirk}, {Friesen}, {Pineda}, {Rosolowsky},
  {Offner}, {Matzner}, {Myers}, {Di Francesco}, {Caselli}, {Alves},
  {Chac{\'o}n-Tanarro}, {Chen}, {Chun-Yuan Chen}, {Keown}, {Punanova}, {Seo},
  {Shirley}, {Ginsburg}, {Hall}, {Singh}, {Arce}, {Goodman}, {Martin}, \&
  {Redaelli}}]{2017ApJ...846..144K}
{Kirk}, H., {Friesen}, R.~K., {Pineda}, J.~E., {et~al.} 2017, \apj, 846, 144,
  \dodoi{10.3847/1538-4357/aa8631}

\bibitem[{{Kong} {et~al.}(2015){Kong}, {Lada}, {Lada},
  {Rom{\'a}n-Z{\'u}{\~n}iga}, {Bieging}, {Lombardi}, {Forbrich}, \&
  {Alves}}]{2015ApJ...805...58K}
{Kong}, S., {Lada}, C.~J., {Lada}, E.~A., {et~al.} 2015, \apj, 805, 58,
  \dodoi{10.1088/0004-637X/805/1/58}

\bibitem[{{Kramer} {et~al.}(1999){Kramer}, {Alves}, {Lada}, {Lada}, {Sievers},
  {Ungerechts}, \& {Walmsley}}]{1999A&A...342..257K}
{Kramer}, C., {Alves}, J., {Lada}, C.~J., {et~al.} 1999, \aap, 342, 257

\bibitem[{{Lada} {et~al.}(1994){Lada}, {Lada}, {Clemens}, \&
  {Bally}}]{1994ApJ...429..694L}
{Lada}, C.~J., {Lada}, E.~A., {Clemens}, D.~P., \& {Bally}, J. 1994, \apj, 429,
  694, \dodoi{10.1086/174354}

\bibitem[{{Lada} {et~al.}(2017){Lada}, {Lewis}, {Lombardi}, \&
  {Alves}}]{2017A&A...606A.100L}
{Lada}, C.~J., {Lewis}, J.~A., {Lombardi}, M., \& {Alves}, J. 2017, \aap, 606,
  A100, \dodoi{10.1051/0004-6361/201731221}

\bibitem[{{Lada} {et~al.}(2009){Lada}, {Lombardi}, \&
  {Alves}}]{2009ApJ...703...52L}
{Lada}, C.~J., {Lombardi}, M., \& {Alves}, J.~F. 2009, \apj, 703, 52,
  \dodoi{10.1088/0004-637X/703/1/52}

\bibitem[{{Lada} {et~al.}(2008){Lada}, {Muench}, {Rathborne}, {Alves}, \&
  {Lombardi}}]{2008ApJ...672..410L}
{Lada}, C.~J., {Muench}, A.~A., {Rathborne}, J., {Alves}, J.~F., \& {Lombardi},
  M. 2008, \apj, 672, 410, \dodoi{10.1086/523837}

\bibitem[{{Lee} {et~al.}(2018){Lee}, {Leroy}, {Bolatto}, {Glover},
  {Indebetouw}, {Sandstrom}, \& {Schruba}}]{2018MNRAS.474.4672L}
{Lee}, C., {Leroy}, A.~K., {Bolatto}, A.~D., {et~al.} 2018, \mnras, 474, 4672,
  \dodoi{10.1093/mnras/stx2760}

\bibitem[{{Lee} {et~al.}(2014){Lee}, {Stanimirovi{\'c}}, {Wolfire}, {Shetty},
  {Glover}, {Molina}, \& {Klessen}}]{2014ApJ...784...80L}
{Lee}, M.-Y., {Stanimirovi{\'c}}, S., {Wolfire}, M.~G., {et~al.} 2014, \apj,
  784, 80, \dodoi{10.1088/0004-637X/784/1/80}

\bibitem[{Lewis(2020)}]{DVN/FTOHSO_2020}
Lewis, J.~A. 2020, {ARO SMT Survey of the California Molecular Cloud}, V1,
  Harvard Dataverse, \dodoi{10.7910/DVN/FTOHSO}

\bibitem[{{Lombardi}(2009)}]{2009A&A...493..735L}
{Lombardi}, M. 2009, \aap, 493, 735, \dodoi{10.1051/0004-6361:200810519}

\bibitem[{{Lombardi} {et~al.}(2006){Lombardi}, {Alves}, \&
  {Lada}}]{2006A&A...454..781L}
{Lombardi}, M., {Alves}, J., \& {Lada}, C.~J. 2006, \aap, 454, 781,
  \dodoi{10.1051/0004-6361:20042474}

\bibitem[{{Lombardi} {et~al.}(2014){Lombardi}, {Bouy}, {Alves}, \&
  {Lada}}]{2014A&A...566A..45L}
{Lombardi}, M., {Bouy}, H., {Alves}, J., \& {Lada}, C.~J. 2014, \aap, 566, A45,
  \dodoi{10.1051/0004-6361/201323293}

\bibitem[{{Lynds}(1962)}]{1962ApJS....7....1L}
{Lynds}, B.~T. 1962, \apjs, 7, 1, \dodoi{10.1086/190072}

\bibitem[{{Mangum} {et~al.}(2007){Mangum}, {Emerson}, \&
  {Greisen}}]{2007A&A...474..679M}
{Mangum}, J.~G., {Emerson}, D.~T., \& {Greisen}, E.~W. 2007, \aap, 474, 679,
  \dodoi{10.1051/0004-6361:20077811}

\bibitem[{{Nittler} \& {Gaidos}(2012)}]{2012M&PS...47.2031N}
{Nittler}, L.~R., \& {Gaidos}, E. 2012, Meteoritics and Planetary Science, 47,
  2031, \dodoi{10.1111/j.1945-5100.2012.01410.x}

\bibitem[{{Padoan} {et~al.}(2000){Padoan}, {Juvela}, {Bally}, \&
  {Nordlund}}]{2000ApJ...529..259P}
{Padoan}, P., {Juvela}, M., {Bally}, J., \& {Nordlund}, {\r{A}}. 2000, \apj,
  529, 259, \dodoi{10.1086/308229}

\bibitem[{{Pety}(2005)}]{2005sf2a.conf..721P}
{Pety}, J. 2005, in SF2A-2005: Semaine de l'Astrophysique Francaise, ed.
  F.~{Casoli}, T.~{Contini}, J.~M. {Hameury}, \& L.~{Pagani}, 721

\bibitem[{{Pety}(2018)}]{2018ssdd.confE..11P}
{Pety}, J. 2018, in Submillimetre Single-dish Data Reduction and Array
  Combination Techniques, 11, \dodoi{10.5281/zenodo.1205423}

\bibitem[{{Pineda} {et~al.}(2008){Pineda}, {Caselli}, \&
  {Goodman}}]{2008ApJ...679..481P}
{Pineda}, J.~E., {Caselli}, P., \& {Goodman}, A.~A. 2008, \apj, 679, 481,
  \dodoi{10.1086/586883}

\bibitem[{{Pineda} {et~al.}(2010){Pineda}, {Goldsmith}, {Chapman}, {Snell},
  {Li}, {Cambr{\'e}sy}, \& {Brunt}}]{2010ApJ...721..686P}
{Pineda}, J.~L., {Goldsmith}, P.~F., {Chapman}, N., {et~al.} 2010, \apj, 721,
  686, \dodoi{10.1088/0004-637X/721/1/686}

\bibitem[{{Polehampton} {et~al.}(2005){Polehampton}, {Baluteau}, \&
  {Swinyard}}]{2005A&A...437..957P}
{Polehampton}, E.~T., {Baluteau}, J.~P., \& {Swinyard}, B.~M. 2005, \aap, 437,
  957, \dodoi{10.1051/0004-6361:20052737}

\bibitem[{{Rachford} {et~al.}(2002){Rachford}, {Snow}, {Tumlinson}, {Shull},
  {Blair}, {Ferlet}, {Friedman}, {Gry}, {Jenkins}, {Morton}, {Savage},
  {Sonnentrucker}, {Vidal-Madjar}, {Welty}, \& {York}}]{2002ApJ...577..221R}
{Rachford}, B.~L., {Snow}, T.~P., {Tumlinson}, J., {et~al.} 2002, \apj, 577,
  221, \dodoi{10.1086/342146}

\bibitem[{{Ripple} {et~al.}(2013){Ripple}, {Heyer}, {Gutermuth}, {Snell}, \&
  {Brunt}}]{2013MNRAS.431.1296R}
{Ripple}, F., {Heyer}, M.~H., {Gutermuth}, R., {Snell}, R.~L., \& {Brunt},
  C.~M. 2013, \mnras, 431, 1296, \dodoi{10.1093/mnras/stt247}

\bibitem[{{Robitaille} {et~al.}(2019){Robitaille}, {Rice}, {Beaumont},
  {Ginsburg}, {MacDonald}, \& {Rosolowsky}}]{2019ascl.soft07016R}
{Robitaille}, T., {Rice}, T., {Beaumont}, C., {et~al.} 2019, {astrodendro:
  Astronomical data dendrogram creator}.
\newblock \doeprint{1907.016}

\bibitem[{{Rosolowsky} \& {Leroy}(2006)}]{2006PASP..118..590R}
{Rosolowsky}, E., \& {Leroy}, A. 2006, \pasp, 118, 590, \dodoi{10.1086/502982}

\bibitem[{{Rosolowsky} {et~al.}(2008){Rosolowsky}, {Pineda}, {Kauffmann}, \&
  {Goodman}}]{2008ApJ...679.1338R}
{Rosolowsky}, E.~W., {Pineda}, J.~E., {Kauffmann}, J., \& {Goodman}, A.~A.
  2008, \apj, 679, 1338, \dodoi{10.1086/587685}

\bibitem[{{Roy} {et~al.}(2014){Roy}, {Andr{\'e}}, {Palmeirim}, {Attard},
  {K{\"o}nyves}, {Schneider}, {Peretto}, {Men'shchikov}, {Ward-Thompson},
  {Kirk}, {Griffin}, {Marsh}, {Abergel}, {Arzoumanian}, {Benedettini}, {Hill},
  {Motte}, {Nguyen Luong}, {Pezzuto}, {Rivera-Ingraham}, {Roussel}, {Rygl},
  {Spinoglio}, {Stamatellos}, \& {White}}]{2014A&A...562A.138R}
{Roy}, A., {Andr{\'e}}, P., {Palmeirim}, P., {et~al.} 2014, \aap, 562, A138,
  \dodoi{10.1051/0004-6361/201322236}

\bibitem[{{Sakamoto} {et~al.}(1995){Sakamoto}, {Hasegawa}, {Hayashi}, {Handa},
  \& {Oka}}]{1995ApJS..100..125S}
{Sakamoto}, S., {Hasegawa}, T., {Hayashi}, M., {Handa}, T., \& {Oka}, T. 1995,
  \apjs, 100, 125, \dodoi{10.1086/192210}

\bibitem[{{Sault} {et~al.}(1995){Sault}, {Teuben}, \&
  {Wright}}]{1995ASPC...77..433S}
{Sault}, R.~J., {Teuben}, P.~J., \& {Wright}, M.~C.~H. 1995, in Astronomical
  Society of the Pacific Conference Series, Vol.~77, Astronomical Data Analysis
  Software and Systems IV, ed. R.~A. {Shaw}, H.~E. {Payne}, \& J.~J.~E.
  {Hayes}, 433.
\newblock \doarXiv{astro-ph/0612759}

\bibitem[{{Savage} \& {Mathis}(1979)}]{1979ARA&A..17...73S}
{Savage}, B.~D., \& {Mathis}, J.~S. 1979, \araa, 17, 73,
  \dodoi{10.1146/annurev.aa.17.090179.000445}

\bibitem[{{Schneider} {et~al.}(2016){Schneider}, {Bontemps}, {Motte},
  {Ossenkopf}, {Klessen}, {Simon}, {Fechtenbaum}, {Herpin}, {Tremblin},
  {Csengeri}, {Myers}, {Hill}, {Cunningham}, \&
  {Federrath}}]{2016A&A...587A..74S}
{Schneider}, N., {Bontemps}, S., {Motte}, F., {et~al.} 2016, \aap, 587, A74,
  \dodoi{10.1051/0004-6361/201527144}

\bibitem[{{Shimajiri} {et~al.}(2014){Shimajiri}, {Kitamura}, {Saito}, {Momose},
  {Nakamura}, {Dobashi}, {Shimoikura}, {Nishitani}, {Yamabi}, {Hara},
  {Katakura}, {Tsukagoshi}, {Tanaka}, \& {Kawabe}}]{2014A&A...564A..68S}
{Shimajiri}, Y., {Kitamura}, Y., {Saito}, M., {et~al.} 2014, \aap, 564, A68,
  \dodoi{10.1051/0004-6361/201322912}

\bibitem[{{Taylor}(2005)}]{2005ASPC..347...29T}
{Taylor}, M.~B. 2005, in Astronomical Society of the Pacific Conference Series,
  Vol. 347, Astronomical Data Analysis Software and Systems XIV, ed.
  P.~{Shopbell}, M.~{Britton}, \& R.~{Ebert}, 29

\bibitem[{{Whittet} {et~al.}(2007){Whittet}, {Shenoy}, {Bergin}, {Chiar},
  {Gerakines}, {Gibb}, {Melnick}, \& {Neufeld}}]{2007ApJ...655..332W}
{Whittet}, D.~C.~B., {Shenoy}, S.~S., {Bergin}, E.~A., {et~al.} 2007, \apj,
  655, 332, \dodoi{10.1086/509772}

\bibitem[{{Wilson}(1999)}]{1999RPPh...62..143W}
{Wilson}, T.~L. 1999, Reports on Progress in Physics, 62, 143,
  \dodoi{10.1088/0034-4885/62/2/002}

\bibitem[{{Yoda} {et~al.}(2010){Yoda}, {Handa}, {Kohno}, {Nakajima}, {Kaiden},
  {Yonekura}, {Ogawa}, {Morino}, \& {Dobashi}}]{2010PASJ...62.1277Y}
{Yoda}, T., {Handa}, T., {Kohno}, K., {et~al.} 2010, \pasj, 62, 1277,
  \dodoi{10.1093/pasj/62.5.1277}

\bibitem[{{Zari} {et~al.}(2016){Zari}, {Lombardi}, {Alves}, {Lada}, \&
  {Bouy}}]{2016A&A...587A.106Z}
{Zari}, E., {Lombardi}, M., {Alves}, J., {Lada}, C.~J., \& {Bouy}, H. 2016,
  \aap, 587, A106, \dodoi{10.1051/0004-6361/201526597}

\bibitem[{{Zhang} {et~al.}(2018){Zhang}, {Xu}, {Vasyunin}, {Semenov}, {Wang},
  {Dib}, {Liu}, {Liu}, {Zhang}, {Liu}, {Wang}, {Li}, {Wu}, {Yuan}, {Li}, \&
  {Gao}}]{2018A&A...620A.163Z}
{Zhang}, G.-Y., {Xu}, J.-L., {Vasyunin}, A.~I., {et~al.} 2018, \aap, 620, A163,
  \dodoi{10.1051/0004-6361/201833622}

\end{thebibliography}

\end{document}